\documentclass{article}
\usepackage{booktabs}
\usepackage{makecell}
\usepackage[english]{babel}

\usepackage[authoryear,round]{natbib}
\usepackage[letterpaper,top=2cm,bottom=2cm,left=3cm,right=3cm,marginparwidth=1.75cm]{geometry}

\usepackage{bm}

\usepackage{amsmath}
\usepackage{graphicx}
\usepackage{xcolor,soul}
\usepackage{mathptmx}   
\usepackage{amssymb}
\usepackage{textgreek}
\usepackage{multicol}
\usepackage{longtable}
\usepackage{multirow}
\usepackage{tabularx}
\usepackage{siunitx}
\usepackage[normalem]{ulem} 
\usepackage{tikz}
\usepackage[mathscr]{euscript}

\usepackage{ulem}  

\newcommand{\Cout}{$\mathscr{C}a_{out}$}
\newcommand{\Win}{$\mathscr{W}_{in}$}

\definecolor{mygreen}{RGB}{100, 10, 50}

\newcolumntype{Y}{>{\centering\arraybackslash}X}

\makeatletter
\renewcommand\st[1]{\@bsphack\@esphack}
\makeatother

\usepackage[colorlinks=true, allcolors=blue]{hyperref}
\date{}
\author{
David Gabriel Harispe$^{1,2}$
 Matías Gerard$^{3}$ 
Pablo A. Kler$^{1,2}$\thanks{Corresponding author: 
\texttt{kler@cimec.unl.edu.ar}}
\\[1em]
\small $^1$ Centro de Investigación en Métodos Computacionales
(CIMEC, UNL--CONICET)
\\
\small Predio CCT CONICET Santa Fe, RN 168,
Santa Fe, 3000, Argentina
\\[0.5em]
\small $^2$ Departamento de Ingeniería en Sistemas de Información,
FRSF-UTN
\\
\small Lavaise 610, Santa Fe, 3000, Argentina
\\[0.5em]
\small $^3$ Research Institute for Signals, Systems and Computational Intelligence (sınc(i)),
FICH–UNL/CONICET.\\
\small Ciudad Universitaria UNL, Santa Fe, Argentina.
\\
}

\title{Microdroplet Generator Surrogate Enhancement and Inverse Design Through an Uncertainty-Guided {Framework}}

\begin{document}

\maketitle

\begin{abstract}
Microfluidic droplet generation is critical for many applications, yet efficient device design in high-dimensional parameter spaces remains computationally prohibitive due to complex multiphase flow dynamics. This work presents a unified methodology that couples adaptive surrogate refinement with simulation-driven inverse design. High-fidelity Volume-of-Fluid simulations provide ground-truth data, while neural network ensembles predict flow regimes and droplet diameters. Ensemble-based uncertainty quantification identifies epistemic uncertainty using consensus levels for regime classification and the coefficient of variation for diameter prediction, revealing that uncertainty concentrates near regime transition boundaries where physics is highly sensitive to parameter perturbations.

A physics-informed enrichment strategy targets these critical zones and addresses class imbalances by deploying additional targeted simulations and retraining the surrogates forming the ensembles. A framework that integrates Particle Swarm Optimization for constrained inverse design, balancing accuracy with predictive confidence and physical consistency is therefore proposed. Validation on independent test datasets demonstrates that the enriched ensembles achieve higher accuracy, with the fraction of predictions within the 0–3\% error range on droplet diameter nearly doubling, and high-error cases (above 10\%) dropping substantially. The inverse design procedure reliably identifies physically realizable configurations meeting target droplet sizes. This approach significantly reduces computational costs while ensuring reliable predictions, offering a transferable framework for optimizing complex microfluidic systems. 
\end{abstract}




\section{Introduction}
\label{intro}

The precise control and characterization of droplet formation in microfluidic systems are fundamental to a wide array of applications, ranging from single-cell analysis and drug delivery to the synthesis of advanced materials \citep{Li2025, Wang2024, Wu2024}. 
In these microscale devices, the physics of multiphase flows is governed by complex interfacial dynamics, where the viscous forces, inertial effects, and surface tension interact to produce known formation patterns, such as dripping or jetting \citep{SHENG2022, BOUREGA2023}.

A central challenge in modern microfluidic engineering is the ability to efficiently navigate vast design spaces and operating conditions to optimize droplet formation. While semi-empirical correlations can provide rapid means of estimating droplet and flow regime characteristics \citep{Sontti2023, Mu2021-int, Sang2017}, they often lack adaptability to new architectures and fluid properties resulting in costly and repetitive recalibration processes \citep{Wang2025, DaLing2022, YU2022}. High-fidelity numerical simulations using Computational Fluid Dynamics (CFD) methods have been widely adopted for bypassing these empirical limitations, facilitating the study of these multi-phase systems while reducing manpower and material costs \citep{Tsitouridou2025, Wu2023, Soroor2021}. Specifically, the Volume of Fluid (VOF) approach provides high precision in capturing the interface dynamics inherent to multiphase flows \citep{Mardani2023, daSilva2025, Yin2022, Malakshah2024}. However, resolving critical phenomena like neck thinning and pinch-off demands extremely fine spatial and temporal resolutions, making extensive parameter sweeps over large design spaces computationally prohibitive.

To address this, machine learning (ML)-based surrogate modeling has emerged as an innovative tool, offering the ability to approximate complex, non-linear mappings between operational parameters (such as flow rates, viscosities, and capillary radius) and physical outputs like droplet diameter and regime types \citep{Harispe2026}. However, many ML models struggle to maintain accuracy in unexplored parameter regions or near regime transitions where the underlying physics becomes highly sensitive \citep{Lu2021}. A key challenge in building reliable models is distinguishing between two types of uncertainty: epistemic uncertainty, arising from limited data and model-inherent flaws, and aleatoric uncertainty, arising from the system's intrinsic, complex stochastic behavior. Without a rigorous uncertainty quantification strategy, design decisions risk relying on erroneous surrogate predictions, particularly in regions exhibiting sharp non-linearities or abrupt behavioral shifts \citep{hullermeier2021aleatoric}.

In microfluidic applications, such inaccuracies may result in failed experimental campaigns or economic consequences, such as the inefficient use of expensive chemical or biological reagents, the propagation of critical downstream errors within the microfluidic system, and challenges in scaling these architectures for industrial production. While existing models provide valuable insights, they often lack the inherent mechanisms, like an unified error quantification or uncertainty-guided refinement, required to identify and correct inaccuracies in challenging regions before they lead to such failures \citep{Harispe2024}.

While recent studies have advanced individual components of such frameworks, be it adaptive surrogate enhancement, ML-assisted design, or ML-assisted predictive modeling in microfluidic devices \citep{bae2024cfd, LIU2025, SOLANKI2022, Hong2020, Manekar2025, Lashkaripour2024, Lashkaripour2021}, the present work introduces a unified, simulation-driven methodology.
Prior works, such as \citep{Lashkaripour2024, Lashkaripour2021}, have achieved impressive design automation for microdroplet generators by leveraging large experimental databases. However, these approaches are inherently constrained to the parameter space and device configurations covered by prior experiments. Additionally, existing methods lack a unified framework for prediction error detection. The methodology proposed here integrates ensemble-based uncertainty quantification with physics-guided error detection, facilitating the continuous enrichment of the training dataset through targeted VOF simulations. 

In this work, a fully adaptive surrogate-optimization pipeline, that is both uncertainty- and physics-guided, is proposed and validated. The pipeline is specifically tailored for computationally expensive CFD problems. Our methodology moves beyond simple statistical uncertainty estimation by employing a dual-layered approach to identify and improve predictions on regions of limited confidence.


First, ensemble learning is used to extract specific variance-based uncertainty metrics, and then, those metrics are complemented by statistical diagnostic tools. By projecting these estimates onto the operational parameter space, genuine epistemic uncertainty can be isolated from the system’s intrinsic stochastic behavior. Second, our approach employs a physics-informed enrichment strategy. The decision of where to deploy additional high-fidelity VOF simulations is not driven by statistics alone, but by engineering knowledge regarding the system's operational behavior and flow regime transition boundaries, to ensure robustness where these flow regimes change \citep{mao2020physics}.

\begin{figure}[ht]
\centering
\includegraphics[scale=0.421]{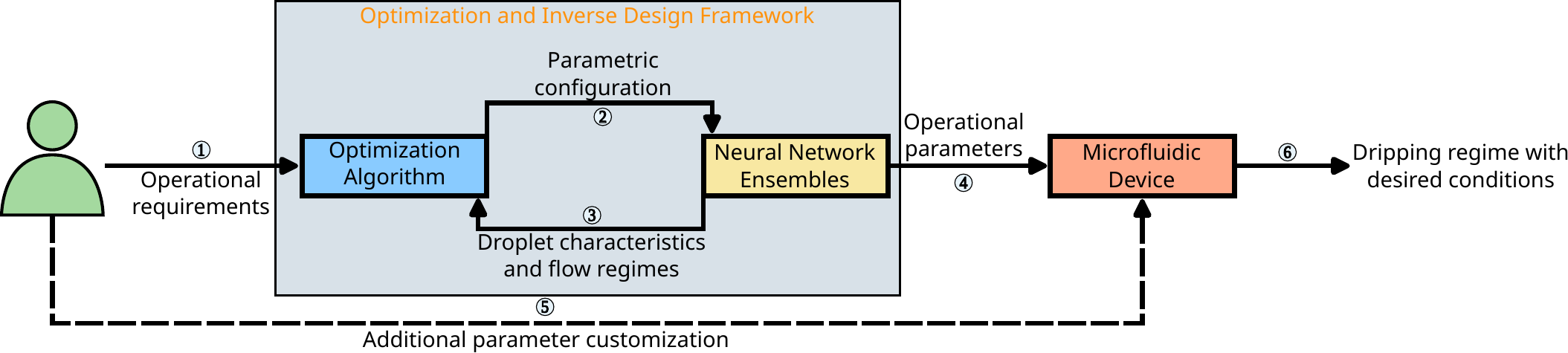}
\caption{Schematic of the proposed optimization and inverse design framework. Operational requirements (1) are processed via an optimization algorithm to generate parametric configurations (2). Neural network ensembles evaluate these to provide feedback on droplet and flow regimes (3) to guide the optimization loop, to output eventually the operational parameters (4) for the microfluidic device. The framework also enables parameter customization (5) to achieve the target dripping regime (6).}
\label{fig:workflow0}
\end{figure}

The true value of this high-fidelity surrogate lies in its application to complex design problems. For many end-users, manual optimization of microfluidic devices is a daunting task because even subtle fluctuations in operational parameters can drastically alter droplet morphology. To address this, the model refinement was included into an user-oriented workflow by integrating it with a constrained optimization algorithm for reliable inverse device design. This allows for the identification of specific operational settings that yield droplets with desired physical properties while exploring the parameter space under user-defined constraints, such as fixed values or allowable ranges for fluid viscosity, velocity, and capillary radius. By leveraging neural surrogates for instantaneous evaluation, our framework can efficiently explore the parameter space under user-defined constraints, such as maintaining a strictly dripping regime while simultaneously targeting specific droplet diameter requirements. Such workflow is illustrated in Fig. \ref{fig:workflow0}.

The primary contribution of this work is thus a unified computational methodology that couples adaptive surrogate refinement with simulation-based optimization. This framework enables efficient, constrained, and simulation-driven design of microfluidic droplet generators while significantly reducing the computational cost associated with exhaustive high-fidelity simulations. 

Although the methodology is demonstrated using a specific type of micro-droplet generator, its underlying principles are broadly applicable to other microfluidic systems and, more generally, to computationally intensive multiphase CFD problems. Despite variations in governing physics, geometry, or numerical solvers across applications, the same methodology can be adopted to systematically enhance surrogate fidelity and enable efficient design exploration.

The remainder of this paper is organized as follows. Section~\ref{sec:system_description} describes the coaxial micro-droplet generator and its governing equations. Section~\ref{sec:methods} presents the computational methodology. This methodology couples high-fidelity volume-of-fluid simulations with baseline neural network surrogates. It also introduces the optimization algorithm used for inverse design. Section~\ref{adapt_enh_models} details the adaptive surrogate enhancement pipeline. This pipeline combines ensemble-based uncertainty quantification with physics-informed data enrichment to improve predictive accuracy. Section~\ref{inverse_design} applies the refined surrogates to a constrained inverse design problem. The results are summarized in Section~\ref{sec:results} and show the impact of the uncertainty-guided optimization as well as that optimal microfluidic configurations can be identified at low computational cost. Section~\ref{sec:discussion} discusses the scope and limitations of the uncertainty diagnostics and of the inverse design method. Finally, Section~\ref{sec:conclusions} summarizes the main findings.

\section{Studied System Description and Governing Equations}
\label{sec:system_description}

\subsection{Target Microfluidic System: Coaxial Micro-droplet Generator}
\label{sec:device_conditions}

\begin{figure}[h!]
\centering
\includegraphics[scale=0.5]{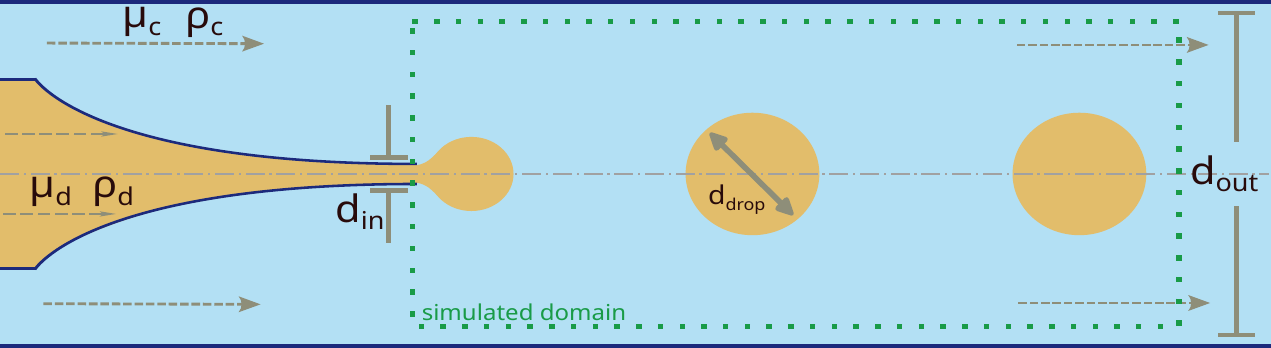}
\caption{Computational domain and coaxial droplet generator geometry used in the numerical simulations. The dispersed phase (orange) is injected through a cylindrical inner capillary into the continuous phase (light-blue), which advects the generated droplets through the surrounding concentric outer capillary.}
\label{fig:coaxDevice}
\end{figure}

In this work, a coaxial micro-droplet generator \citep{ma2024} is adopted as the target system for applying the adaptive surrogate enhancement pipeline and, ultimately, for leveraging the resulting surrogates in the efficient inverse design of microfluidic devices.
As illustrated in Fig. \ref{fig:coaxDevice}, the device consists of two aligned capillaries through which immiscible fluids are introduced coaxially, leading to droplet formation when the two phases interact at the inner capillary outlet. The fluid that forms the discrete droplets is identified as the dispersed phase ($d_p$), and the fluid that surrounds and carries those droplets constitutes the continuous phase ($c_p$).

Device's geometry is characterized by the outer capillary diameter ($d_{out}$) and an inner capillary diameter ($d_{in}$), which directly influence the shear rate and droplet breakup dynamics. Along with those geometric parameters, material parameters like the velocity ($u$), density ($\rho$), and dynamic viscosity ($\mu$) of both phases form the high-dimensional input space explored by the surrogate models that were used in this work. 

\begin{figure}[h!]
\centering
\includegraphics[scale=0.25]{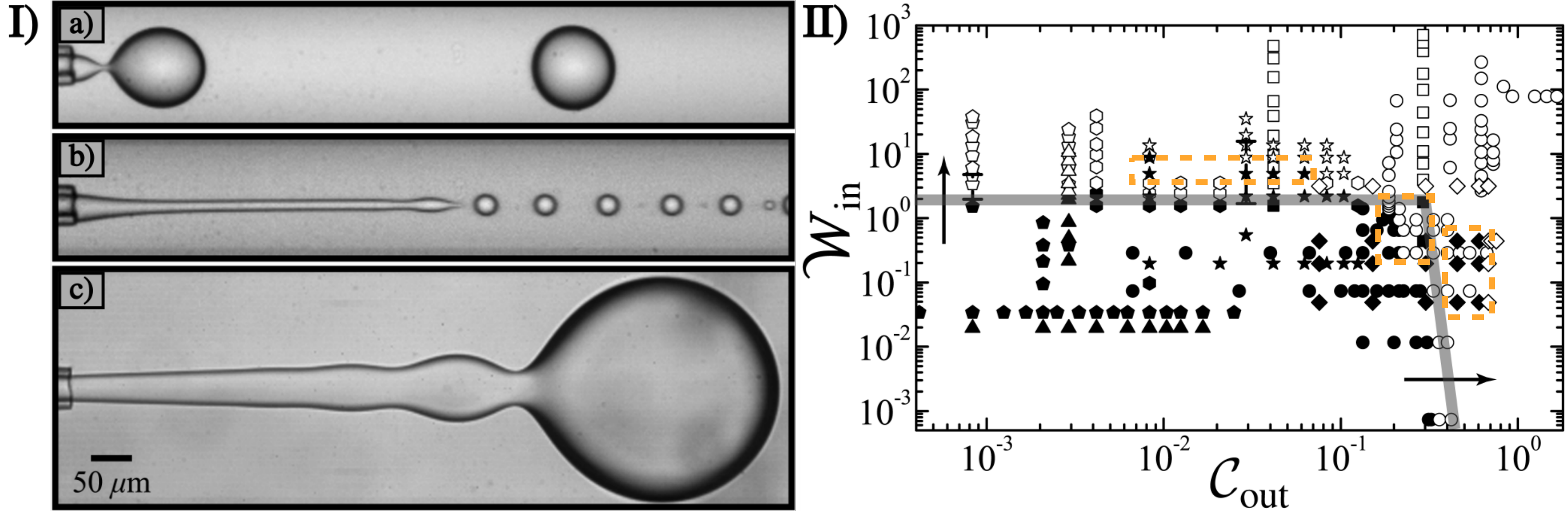}
\caption{Panel (I) shows three experimentally observed coaxial flow regimes: dripping, narrowing jet, and widening jet. Panel (II) summarizes the tested conditions from Utada \textit{et al.} Open symbols denote jetting regimes (\(\mathrm{Ca}_{\mathrm{out}} \gtrsim 0.3-0.5\) for the narrowing jet and \(\mathrm{W}_{\mathrm{in}} \gtrsim 2\) for the widening jet), while filled symbols denote the dripping regime. Dashed orange rectangles mark regions where the expected regime is not observed. Reprinted figures with permission from Utada et al., \url{http://dx.doi.org/10.1103/PhysRevLett.99.094502}. Copyright 2007 by the American Physical Society.}
\label{fig:UtadaFigure}
\end{figure}

This type of devices are characterized by two main operating regimes: dripping (Fig. \ref{fig:UtadaFigure} Ia)) and jetting . The jetting regime can be further subdivided into "narrowing" and "widening" jets. In a narrowing jet (Fig. \ref{fig:UtadaFigure} Ib)), the fluid form a jet that is broken downstream into droplets due to Rayleigh-Plateau instabilities \citep{rayleigh1879capillary}, whereas in a widening jet (Fig. \ref{fig:UtadaFigure} Ic)), the diameter of the jet increases with distance as the droplets grow while attached to the fluid through a neck before eventually detaching.

The behavior of the flow in this system can be characterized by the Capillary number ($\mathscr{C}a$) and the Weber number ($\mathscr{W}$), defined as follows:

\begin{align}
\mathscr{C}a&=\frac{\mu \mathbf{u}}{\sigma} \\
\mathscr{W}&=\frac{\rho u^2 L}{\sigma}
\end{align}

In these equations, $\mu$ denotes the dynamic viscosity, $\mathbf{u}$ represents the characteristic flow velocity, $\rho$ is the fluid density, $L$ denotes the characteristic length scale of the system (typically the capillary radius), and $\sigma$ is the interfacial surface tension.

In the work of \cite{Utada2007} a mapping between the capillary number of the continuous phase (\Cout) and the Weber number of the dispersed phase (\Win) was developed to define value ranges of these dimensionless parameters where the system operates in each flow regime (Fig. \ref{fig:UtadaFigure}).

While these results provide a convenient approximation framework for predicting the resulting flow regime, they do not fully capture the complexity of the underlying flow dynamics. In Utada's reported results and related  literature \citep{Harispe2026}, it was revealed that the same combination of \Cout{} and \Win{} can lead to different flow behaviors depending on the specific physical, geometrical, and operating conditions used. These limitations are particularly evident near regime transition regions, where small variations in almost any parameter may result in unstable behavior between dripping and jetting, or between jetting modes (dashed rectangles in Fig. \ref{fig:UtadaFigure} (II))

This limitation in predicting flow regimes provides strong motivation for using surrogate models. In addition to identifying the dominant flow regime, a crucial performance measure in experimental and industrial settings is the resulting droplet diameter ($d_{drop}$). The droplet size directly influences the performance and applicability of the system and therefore represents a key quantity of interest. Accordingly, $d_{drop}$ is defined as the primary target variable to be predicted by the surrogate models in this work.


\subsection{Two-phase flow}
\label{two_phase_flow}

This section describes the mathematical model for the two-phase flow in the coaxial droplet generator. The system is composed of two immiscible, incompressible Newtonian fluids, namely the dispersed and continuous phases. Each fluid has its own constant density and viscosity, while the effective mixture properties vary only in the interfacial region. The momentum conservation is expressed through the Navier–Stokes equations for incompressible Newtonian fluids written in conservative form:

\begin{equation}
\frac{\partial\rho\mathbf{u}}{\partial t} + \nabla \cdot (\rho \mathbf{u}\mathbf{u}) = -\nabla p + \nabla \cdot \left[\mu \left(\nabla \mathbf{u} + (\nabla \mathbf{u})^T\right)\right]  + \mathbf{F}_\sigma,
\label{eq:ns_found}
\end{equation}

\begin{equation}
\nabla \cdot \mathbf{u} = 0.
\label{eq:cont_found}
\end{equation}

\noindent Here, $\mathbf{u}$ is the velocity vector, $p$ is the pressure, $\rho$ and $\mu$ are the local density and viscosity, respectively. The last term, $\mathbf{F}_\sigma$, represents the volumetric surface-tension force. In this work, gravitational effects are omitted because capillary, viscous, and inertial forces dominate the flow at this micro-scales.

Using the continuum surface force (CSF) model \citep{BRACKBILL1992}, $\mathbf{F}_\sigma$ is expressed as:

\begin{equation}
\mathbf{F}_{\sigma} = \sigma \kappa_{(\alpha)} \,\mathbf{n}_{(\alpha)}\, \delta_{(\alpha)},
\label{eq:st_found}
\end{equation}

\noindent where $\sigma$ is the surface-tension coefficient, $\kappa_{(\alpha)}$ is the curvature of the interface, $\mathbf{n}_{(\alpha)}$ is the unit normal to the interfacial surface, and $\delta_{(\alpha)}$ is a regularized Dirac-delta distribution that is non-zero only at the interface. The last three factors depend on the volume fraction $\alpha$ (or on its numerical reconstruction).

The Volume of Fluid (VOF) method was used in the simulations to capture the evolution of the interface between the fluids. The volume fraction $\alpha$ is defined as the local proportion of the dispersed phase: $\alpha = 1$ in regions occupied only by the dispersed phase, $\alpha = 0$ in regions occupied only by the continuous phase, and $0 < \alpha < 1$ within the interfacial region. With this definition, the local fluid properties are obtained by:

\begin{equation}
\rho(\alpha) = \alpha \rho_d + (1-\alpha) \rho_c, \qquad
\mu(\alpha) = \alpha \mu_d + (1-\alpha) \mu_c,
\label{eq:props}
\end{equation}

\noindent where the subscripts $d$ and $c$ denote the dispersed and continuous phases, respectively.
The transport of the interface is governed by the advection equation:

\begin{equation}
\frac{\partial\alpha}{\partial t} + \mathbf{u}\cdot \nabla \alpha = 0.
\label{eq:vof_found}
\end{equation}


\section{Methodology and Computational Approach}
\label{sec:methods}

\subsection{High-fidelity VOF Simulations}
\label{hi-fi_CFD}

The governing equations presented in Section \ref{two_phase_flow} are discretized and solved using the Basilisk open-source software \citep{POPINET2009}. Basilisk employs a staggered-in-time discretization, ensuring second-order accuracy in both space and time, which is essential for capturing the interfacial deformations observed in droplet generation.

A key feature of this solver is its adaptive mesh refinement (AMR) capability. The mesh is dynamically refined and coarsened at each time step using a quadtree (2D) or octree (3D) structure, guided by a wavelet-estimated discretization error \citep{POPINET2015, SchneiderVasilyev2010}. For the present study, the mesh refinement error thresholds were set to $10^{-3}$ for $\alpha$, and $3\times10^{-2}$ for $\mathbf{u}$. This approach ensures the high spatial resolution required near the interface while significantly reducing the computational cost in other regions. For a detailed description of how Basilisk solves the governing equations for multiphase flows, with particular emphasis on the treatment of the surface tension force, the reader is referred to \citep{Harispe2024}.

The numerical domain replicates the coaxial droplet generator geometry, imposing a constant flow rate at the inlet for both phases, and a reference pressure with a natural outflow condition at the outlet. Initially, the computational domain is filled with the continuous phase, and the simulation begins when the dispersed phase is injected through the inner capillary. It is worth noting that due to the cylindrical and concentric configuration of the capillaries, the numerical model can be reduced to a two-dimensional, axisymmetric domain. The intrinsic symmetry allows for the reduction of the number of cells required and the associated computational cost when simulating. 


The accuracy and robustness of this numerical framework have been assessed against experimental benchmarks and existing numerical data \citep{Harispe2026}. This reliable, high-fidelity model provides the reference data used to train the neural-network surrogates in the following subsections.


\subsection{Baseline Neural-Network Surrogate Models}
\label{baselineSurrogates}

\begin{figure}[h!]
\centering
\includegraphics[scale=0.45]{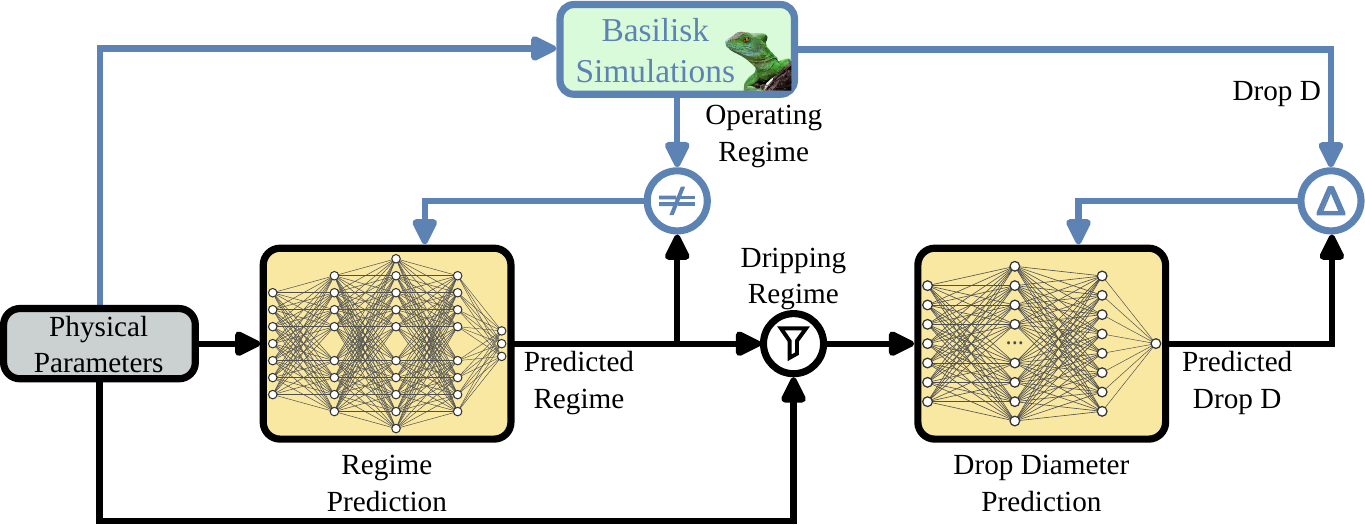}
\caption{Schematic of the surrogates training and inference procedure. The black arrows represent the inference flow, where the neural networks (NNs) map input physical parameters directly to the predicted flow regime and, conditionally, the droplet diameter. The blue arrows denote the training flow, which compares the predicted regimes and droplet diameters against the ground-truth Basilisk simulation results ($\neq$ and $\Delta$ respectively) to drive surrogates training. Note that the droplet diameter prediction module is specifically constrained to operate only when the regime prediction NN identifies a dripping regime}
\label{fig:dataFlowNNs}
\end{figure}

The high-fidelity VOF simulations performed in Basilisk serve as the ground-truth data generator for the surrogate models used in this work. As illustrated in Fig. \ref{fig:dataFlowNNs} (see blue lines), the solver is provided with a set of physical parameters, including phase velocities, viscosities, densities, surface tension, and capillary dimensions. These simulations resolve the two-phase flow and droplet generation dynamics, yielding both a flow regime and, specifically in the dripping regime, the corresponding droplet diameter.

The surrogate models used in this study were first introduced in \cite{Harispe2026} and are adopted here as the baseline for the enhancement framework, uncertainty quantification, and the support of the subsequent inverse design process. To train and utilize these neural network (NN) surrogates $\phi$, a 7-dimensional input vector is employed, consisting of phase velocities, viscosities, the inner capillary radius, \Cout, and \Win. While the Basilisk simulations utilize the full set of physical parameters to generate ground truth, the NNs are designed to map a subset of these parameters, along with \Cout{} and \Win{}, to key system behaviors and droplet characteristics. 

All the simulations utilized and generated throughout this work correspond to a device configuration with inner fluid density $\rho_{\mathrm{in}} = 997 \mathrm{kg/m^3}$, outer fluid density $\rho_{\mathrm{out}} = 965 \mathrm{kg/m^3}$, outer capillary radius $r_{\mathrm{out}} = 100 \mu\mathrm{m}$, and interfacial surface tension $\sigma = 40 \mathrm{mN/m}$. 
The dataset adopted for the different stages of this work closely resembles the one used to train and evaluate the baseline models. Approximately 40 cases were excluded from the literature-derived dataset due to ambiguous flow regime classification or unreliable droplet diameter measurements. The resulting dataset comprises 96 cases in the dripping regime, 121 in the narrowing-jet regime, and 179 in the widening-jet regime. A comprehensive description of how this baseline dataset was used is provided in Section \ref{adapt_enh_models}.

Based on the simulation-generated dataset, two different NNs were trained to provide instantaneous evaluations of the system's performance: one dedicated to predicting the flow regime (dripping, widening jet, or narrowing jet) denoted as $\phi_r$ and another tasked with predicting the resulting droplet diameter whenever a dripping regime is identified, denoted as $\phi_d$. As shown in Fig. \ref{fig:dataFlowNNs}, the droplet diameter predicting NN specifically infers values for those parameter combinations that produce a dripping regime. During the training phase, the discrepancies between the simulation results and the predicted values are used to adjust the internal weights of the NNs, thereby "teaching" the surrogates to align with the ground truth.

By leveraging these surrogates, it is possible to efficiently approximate the complex, non-linear mappings between operational parameters and physical outputs. This dual-output approach provides a robust foundation for subsequent stages of this work, such as targeted adaptive surrogate enhancement and optimization. However, while these baseline surrogates provide the necessary instantaneous-evaluation capability for the constrained inverse-design and adaptive-refinement stages, they possess inherent limitations. Specifically, as will be shown in Section \ref{ensemble_unc_quant}, it was identified that these models exhibit degraded fidelity in challenging parameter regions which were insufficiently sampled during training or near flow regime transitions. 
These areas lead to high epistemic uncertainty where the mapping becomes highly sensitive to small perturbations. Combined with the potential for further optimization of the underlying architecture, these factors necessitate the hyperparameter/architecture optimization and the physics-informed enrichment strategy described in the following sections.

\subsection{Particle Swarm Optimization}
\label{ParticleSwarmOpt}

Particle swarm optimization (PSO) is a population-based stochastic search algorithm inspired by the social behavior of bird flocks \citep{engelbrecht2007}. Each particle represents a candidate solution and moves through the search space, guided by the best position it has found and by the best position found by the swarm. Despite its simplicity, PSO remains one of the state-of-the-art metaheuristics for continuous single-objective optimization, and its variants are still actively developed and benchmarked \citep{bonyadi2017,shami2022}.

This work uses the global-best (gbest) PSO, in which every particle is connected to all the others. An improvement found by any particle is therefore immediately available to the whole swarm, which favors fast convergence. This behavior is adequate for the problem addressed here, which involves a small number of free parameters. Let $\bm{x}_i(t)$ and $\bm{v}_i(t)$ be the position and velocity of the particle $i$ at iteration $t$, in a search space of dimension $n_x$. The initial positions are sampled uniformly between the bounds $\bm{x}_{\min}$ and $\bm{x}_{\max}$. At each iteration, the velocity of each particle is updated in each dimension $j = 1, \dots, n_x$ as

\begin{equation}
v_{ij}(t+1) = w\,v_{ij}(t)
            + c_1 r_{1j}(t)\left[y_{ij}(t) - x_{ij}(t)\right]
            + c_2 r_{2j}(t)\left[\hat{y}_{j}(t) - x_{ij}(t)\right],
\label{eq:velocity}
\end{equation}

\noindent and the position as

\begin{equation}
\bm{x}_i(t+1) = \bm{x}_i(t) + \bm{v}_i(t+1),
\label{eq:position}
\end{equation}
where $\bm{y}_i$ is the personal best of particle $i$, $\hat{\bm{y}}$ is the global best, $w$ is the inertia weight, $c_1$ and $c_2$ are positive acceleration constants, and $r_{1j}(t)$ and $r_{2j}(t)$ are random numbers sampled from $U(0,1)$. The first term of Eq. \eqref{eq:velocity} keeps part of the previous velocity and sustains the exploration of the search space. The second term, or cognitive component, attracts the particle toward its personal best. The third term, or social component, attracts it toward the global best. The constants $c_1$ and $c_2$ set the balance between these two components, and therefore between exploration and exploitation.

For the minimization of a fitness function $f$, the personal best of each particle is updated as
\begin{equation}
\bm{y}_i(t+1) =
\begin{cases}
\bm{y}_i(t)   & \text{if } f(\bm{x}_i(t+1)) \geq f(\bm{y}_i(t)), \\
\bm{x}_i(t+1) & \text{otherwise},
\end{cases}
\label{eq:pbest}
\end{equation}
and the global best $\hat{\bm{y}}(t+1)$ is the personal best with the lowest value of $f$.

The implementation used in this work modifies this basic scheme in four aspects. First, the inertia weight $w$ is sampled at random, independently for each particle, dimension, and iteration \citep{EberhartShi2001}, instead of being kept constant or decreased linearly during the search. The randomness of $w$ is intended to maintain the diversity of the swarm throughout the search. 
Second, each velocity component is limited to $|v_{ij}| \leq \delta\,\Delta x_j$, where $\Delta x_j = x_{\max,j} - x_{\min,j}$ and $\delta$ is the velocity clamping coefficient. This prevents a particle from crossing a large part of the search space in a single iteration. Third, when a particle leaves the search space in dimension $j$, its position is set to the nearest bound and $v_{ij}$ is multiplied by a negative rebound factor $\alpha_b$. The particle therefore moves back toward the interior of the search space. Fourth, in addition to a maximum number of iterations, the search stops when the global best does not improve significantly during a given number of iterations. The values of these parameters are given in Section \ref{pso_configuration}.

\subsection{Computational Setup}

\subsubsection{Software}
The fluid dynamics simulations are performed using the Basilisk open-source software \citep{PopinetBasilisk}, which utilizes a Finite Volume Method (FVM) combined with a VOF strategy to resolve the multiphase interface. Python language \citep{python3} was used for all the scripts used in the main stages of the workflow, including data collection and preprocessing, as well as the cleaning and organization of the final simulation metrics. For the development of the deep learning components, PyTorch \citep{Paszke2019} is employed to implement and train the Multi-Layer Perceptrons (MLP) architectures, leveraging CUDA for GPU acceleration during the training and validation phases.

\subsubsection{Hardware}
The computational requirements for both the large-scale simulations and the machine learning models were met using high-performance computing resources. The Basilisk fluid dynamics simulations were executed on a CPU-based cluster. It features a head node with an Intel Xeon E5-2650 and 7 calculation nodes based on AMD Epyc Turin 9255 processors with 192 GB of RAM. The training and testing of the NNs were conducted on a dedicated workstation utilizing GPU acceleration to significantly reduce training time compared to CPU-only computations. This workstation was equipped with an AMD Ryzen 7 9700X 8-core CPU, 32 GB of DDR5 RAM, and an NVIDIA GeForce RTX 5070 GPU.

\section{Adaptive Enhancement of the Models}
\label{adapt_enh_models}

The evaluation of the baseline models showed that predictions are often inaccurate near the transitions between regimes and for certain parameter combinations of interest. This behavior is consistent with epistemic uncertainty, that is, uncertainty caused by insufficient training data, which can be reduced by adding samples in the affected regions. A single network, however, gives no indication of where its predictions are unreliable, and therefore of where new data would be most valuable.

Ensembles of NNs offer a practical way to locate these regions \citep{Lakshminarayanan2017}. An ensemble is a set of networks trained for the same task, whose predictions are combined into a single prediction. In this work, the regime ensemble is $E_{r} = \{\phi_{r,1}, \dots, \phi_{r,|E_{r}|}\}$ and the diameter ensemble is $E_{d} = \{\phi_{d,1}, \dots, \phi_{d,|E_{d}|}\}$, where each member $\phi_{r,k}$ or $\phi_{d,k}$ is a network with the architecture of $\phi_{r}$ or $\phi_{d}$, respectively. The number of members of an ensemble is denoted by $|E_{r}|$ or $|E_{d}|$. The predicted regime is the class that receives the most votes in $E_{r}$, and the predicted diameter is the mean of the predictions of the members of $E_{d}$. Members trained with different initializations and data partitions agree where the data constrain the prediction. They disagree in sparsely sampled regions and near regime boundaries, where they extrapolate differently or place the boundaries at slightly different locations. This disagreement can be measured using different metrics, depending on the task (as will be shown in Sections \ref{regime_pred_unc} and \ref{diam_pred_unc}). Said metrics are therefore used as indicators of epistemic uncertainty, making it possible to select new training samples in regions where the models are least reliable, rather than sampling the input space uniformly.
This follows the principle of active learning and, in particular, of query-by-committee strategies \citep{Seung1992,Settles2009}. It is especially relevant here because each new sample requires a VOF simulation, which is computationally expensive, so the points to simulate must be chosen carefully. In this work, the new points are selected within the regions of highest uncertainty identified in Section \ref{ensemble_unc_quant}, guided by physical criteria, as detailed in Section \ref{section:manual_enrich}.

\begin{figure}[h!]
\centering
\includegraphics[scale=0.4]{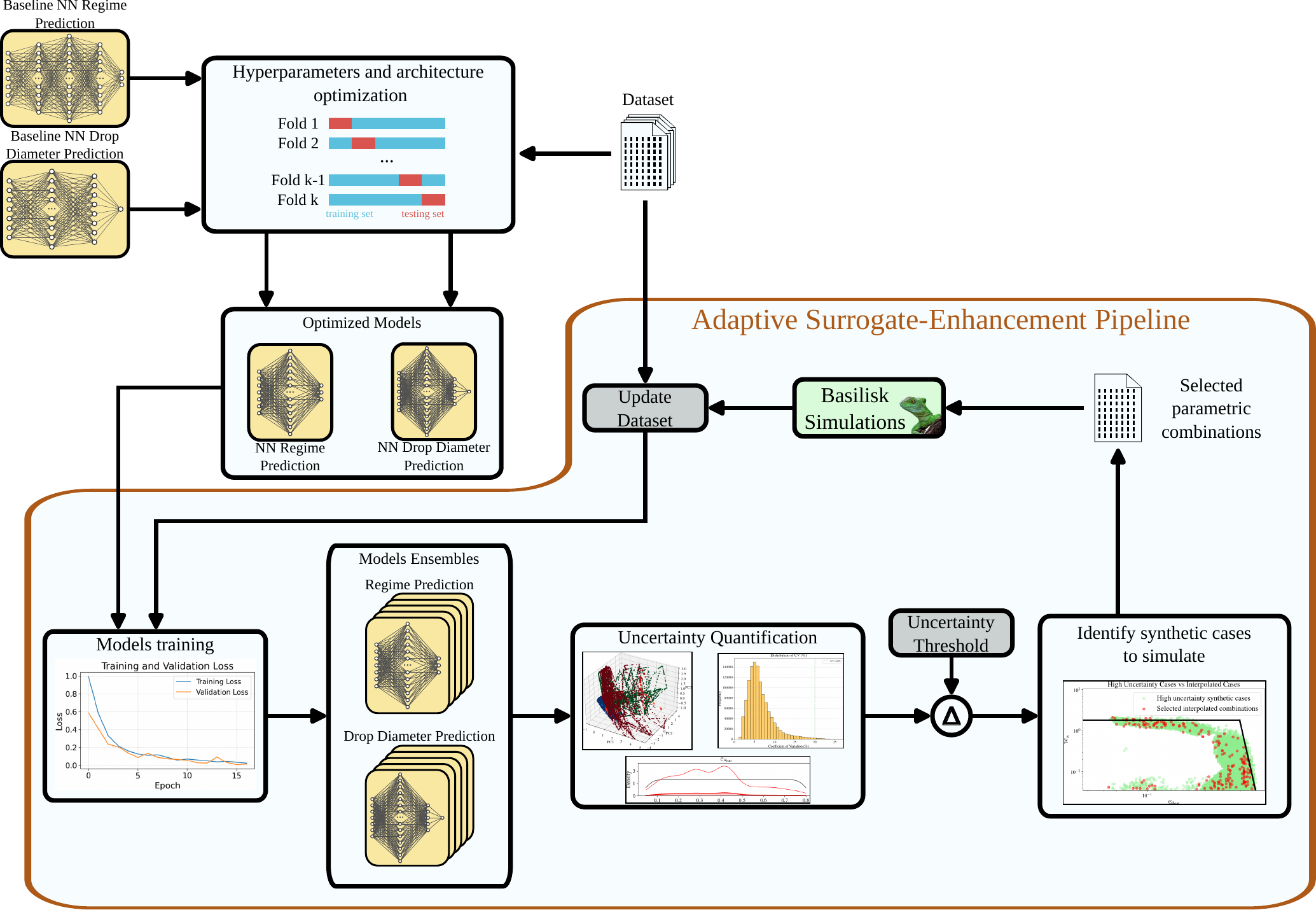}
\caption{Schematic overview of the proposed Adaptive Surrogate-Refinement Pipeline, whose closed loop couples ensemble-based uncertainty quantification with targeted Basilisk VOF simulations and dataset-driven model retraining, to progressively improve surrogate fidelity. The upstream optimization of baseline architectures and hyperparameters lies outside the loop and supplies the optimized models used to build the ensembles.}
\label{fig:bigPlot}
\end{figure}

Based on this idea, a framework for uncertainty quantification and model enrichment was implemented (Fig. \ref{fig:bigPlot}). It consists of an initial stage followed by an enhancement loop that progressively improves the surrogate models. In the initial stage, the NN architecture and hyperparameters are selected for each task (Section \ref{kFoldOpt}), and the baseline ensembles are trained with the selected configuration to initialize the enhancement loop. This loop comprises three steps:
(i) ensembles disagreement, on synthetic datasets, is used to quantify and localize uncertainty (Section \ref{ensemble_unc_quant}); (ii) new parameter combinations are selected within high-uncertainty regions integrating a physics-guided strategy, simulated, and added to the training dataset (Section \ref{section:manual_enrich}); and finally, {(iii) the ensemble models are retrained with the enriched data and the results of this process are compared and reported (Section }\ref{post_enrich}{).}

\subsection{Datasets}
\label{Datasets}

\begin{figure}[h!]
\centering
\includegraphics[scale=0.35]{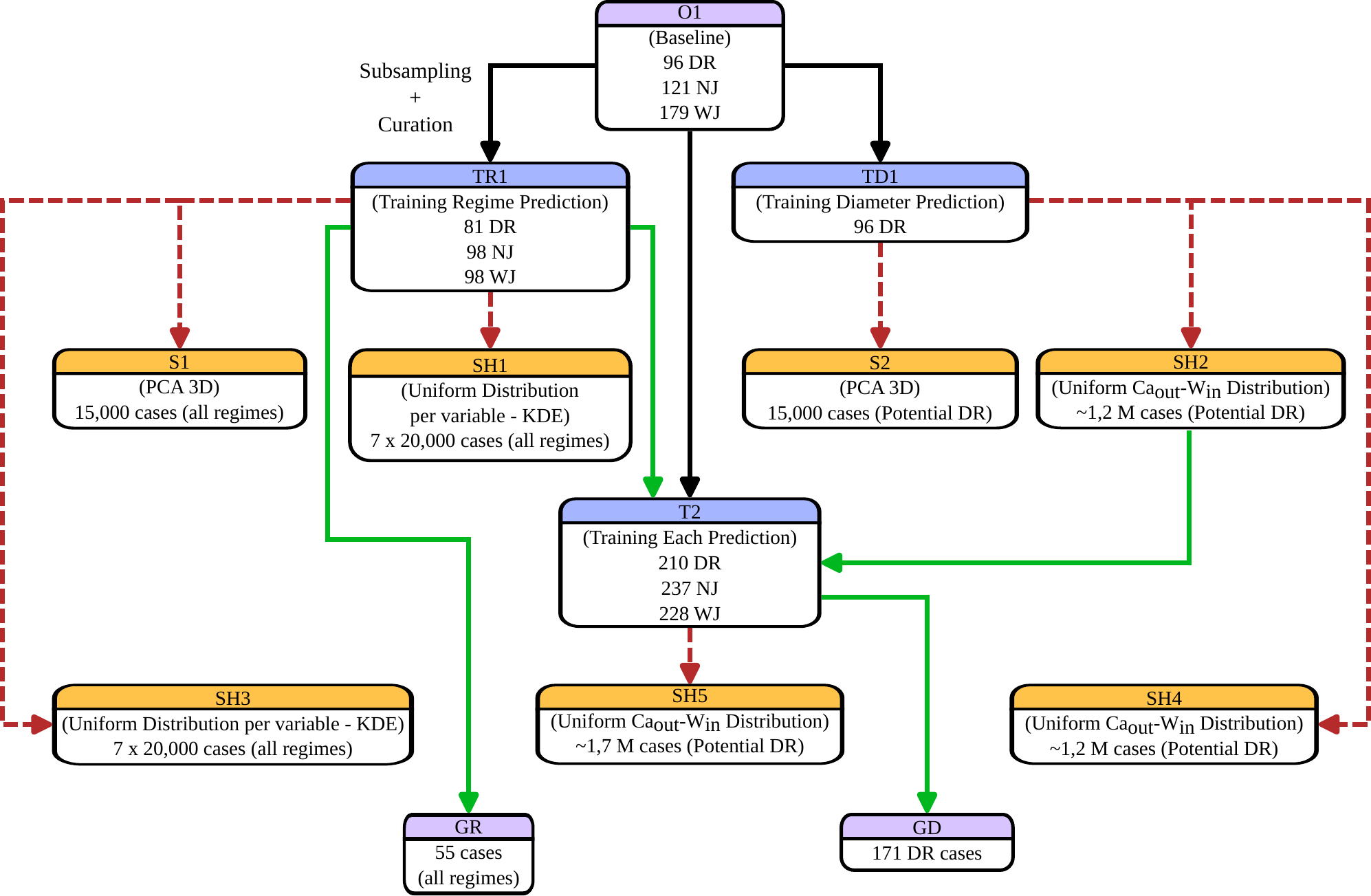}
\caption{Overview of the dataset generation and enrichment pipeline. The flow goes from literature-based and training datasets through balancing, synthetic sampling, training dataset enrichment, and uncertainty reduction verification, as well as final physical predictions test datasets. Dashed red arrows indicate interpolation procedures, whereas solid green arrows denote both interpolation procedures and FVM-VOF simulations. DR: Dripping Regime, NJ: Narrowing-Jet regime, WJ: Widening-Jet regime.}
\label{fig:dataset_diag}
\end{figure}

Several datasets were constructed throughout the pipeline to support various stages of the process. Figure \ref{fig:dataset_diag} outlines the generation procedures for these datasets, highlighting their origins, characteristics, and roles. The red dashed arrows in the figure indicate interpolation steps, while the green solid arrows denote procedures that combine interpolation with VOF simulations. The process starts with the literature-based dataset ($O1$) described in Section \ref{baselineSurrogates}. Following subsampling and curation
, $O1$ produces two training subsets. The subset $TR1$ supports regime prediction training, and the subset $TD1$ supports droplet diameter prediction training. These initial datasets provide the foundation for generating synthetic samples and for the subsequent enrichment of the training data.

\subsubsection{VOF-Simulated Datasets}
\label{vof_datasets}

The proposed pipeline is directed towards producing an enriched training dataset ($T2$) and two test datasets ($GR$ and $GD$), one for each task. The enriched training set $T2$ contains new data from targeted VOF simulations. These simulations were performed on cases sampled from regions where the pre-enrichment ensembles showed the highest predictive disagreement (Section \ref{section:manual_enrich}). The new cases for regime prediction were drawn from zones with maximal disagreement, and VOF simulations determined their resulting flow regimes. The selection process for droplet diameter prediction cases followed a similar approach, measuring the droplet diameters also from VOF simulations.
Two test datasets with simulation ground truth validated the performance of the retrained ensembles ($GR$ and $GD$). The $GR$ dataset contains configurations across all flow regimes. It includes samples from parametric zones flagged for significant model disagreement in Section \ref{regime_pred_unc} to study the effects of enrichment. The $GD$ dataset contains dripping-regime configurations that target parametric zones with high predictive variance in Section \ref{diam_pred_unc}. Candidate configurations for $GD$ underwent pre-screening based on the predicted regime by the regime post-enrichment ensemble. The process preserved in $GD$ only those with a confident dripping-regime prediction to avoid non-dripping outcomes in the simulation (further details in Section \ref{physical_impact}).

\subsubsection{Synthetic Datasets}
\label{synth_datasets}

Synthetic datasets were constructed mathematically to allow efficient sampling of the parameter space without the computational cost of additional VOF simulations. Their labels are not independently verified because doing so would require a dedicated simulation for each case. These datasets help to study throughout this document how to enrich the baseline training dataset. Synthetic samples for both tasks were generated by interpolating between simulated cases using different strategies. The regime prediction constitutes a classification task in which the regimes should be equally represented in the data. In consequence, the Synthetic Minority Over-sampling Technique (SMOTE) \citep{Chawla2002} was applied to generate synthetic samples with a balanced class distribution using dataset $TR1$. SMOTE creates new samples of a class by interpolating between a sample and one of its $k$ nearest neighbors of the same class. The default configuration with $k=5$ was used throughout this work. The droplet diameter prediction task is a regression problem where SMOTE is not applicable. Synthetic samples were generated by linearly interpolating input parameters between two randomly selected dripping-regime instances from datasets $TD1$ and $T2$. Since interpolation does not guarantee that a new case remains in the dripping regime, candidate samples for synthetic datasets used in diameter prediction were pre-screened by the regime-prediction ensemble. Only samples predicted as the dripping regime by a certain number of ensemble members were retained (as mentioned for $GD$). The confidence threshold, per member, for retention was greater than 75\%. 
Because synthetic samples are obtained by interpolation, they lie within the region spanned by the original data, and the uncertainty analysis therefore identifies sparsely populated zones within this region rather than extrapolating beyond it.

Pre-enrichment synthetic datasets mapped the initial uncertainty landscape for each predicting task. A dataset $S1$ was generated with SMOTE from $TR1$ for regime prediction. Each synthetic sample received the regime label of the source samples. The samples of this dataset were projected onto the first three principal components (3D PCA). 
To investigate per-variable uncertainty in more detail, seven additional datasets were constructed for each input variable, collectively referred to as $SH1$. The full observed range of each of the seven input variables was discretized into 1,000 equally spaced values. For each value, 20 different combinations of the remaining six variables were generated via SMOTE interpolation using $TR1$ as the source. This construction ensures uniform distribution across each parameter axis while preserving the joint correlation structure of $O1$.



Dataset $S2$ consisted of a synthetic set of 15,000 parametric combinations, obtained by linearly interpolating between a randomly selected reference sample from $TD1$ and one of its five nearest geometric neighbors in the feature space. The interpolation was done using a weight drawn from a uniform distribution. This localized strategy constrained synthetic samples to the existing data distribution while allowing all seven input parameters to vary simultaneously. The pre-enrichment regime ensemble pre-screened the samples for $S2$. $S2$ was utilized to project its samples onto a 3D PCA space for initial uncertainty visualization.

A larger scale dataset named $SH2$ allowed the mapping of the parametric combinations onto the (\Cout, \Win) plane (Section \ref{sec:device_conditions}). $SH2$ was generated from 10,000 base combinations, using a $100\times100$ grid of equally spaced (\Cout,\Win) points over the dripping region. The grid extended slightly beyond dripping region boundaries to account for approximate experimental regime transitions and occasional dripping configurations just outside the delineated region. Because the remaining parameters are not uniquely determined by the two dimensionless numbers, each grid point represents a range of potentially valid dripping configurations. Thus, for each grid point, up to 200 parametric samples were generated by linearly interpolating the remaining model-input variables between randomly selected pairs of dripping cases from $TD1$. Fluid velocities were computed from the corresponding \Cout and \Win values. The actual number of samples per grid point depended on how many physically valid dripping configurations the regime-prediction ensemble could confirm within 100 attempts.

Finally, three post-enrichment synthetic datasets were constructed to perform new and unbiased uncertainty testing. For regime prediction, new synthetic datasets per input variable were generated via SMOTE. These utilized the same generation procedure as $SH1$ and are collectively referred to as dataset $SH3$. Each dataset uniformly samples distributed across the full range of the corresponding variable so that a per-variable uncertainty diagnostic could be applied under the same conditions as the pre-enrichment analysis. The last two datasets were constructed for droplet diameter prediction. Dataset $SH4$ was generated using the same (\Cout,\Win)-grid constrained interpolation strategy as $SH2$, and once more each candidate sample was pre-filtered for potential dripping regime. Dataset $SH5$ was generated by an identical grid-based interpolation strategy but drew source data from the enriched $T2$ dataset. Also, in this last case, candidate samples were validated for the dripping regime using the retrained regime-prediction ensemble. This approach differed from the method used for $SH4$, which used the previous version of the ensemble. 

\subsection{Optimization of the architectures and hyperparameters}
\label{kFoldOpt}

Before uncertainty quantification and model enrichment, the architectures and hyperparameters of the NNs used for regime and droplet diameter prediction were optimized. This is a critical step to ensure that the surrogate models generalize across the diversity of operational conditions and system dynamics encountered in microdroplet generators.

To compare model architectures and hyperparameter configurations and to select the most suitable model, $k$-fold cross-validation \citep{Hastie2009} was employed. Datasets are partitioned into $k$ approximately equal-sized, non-overlapping folds. In each of $k$ iterations, the model is trained from scratch on $k-1$ folds and evaluated on the remaining one (held-out fold) so that every fold is used exactly once for evaluation. The performance metrics are averaged across iterations, and the mean is used for model selection, which reduces the dependence of the selection on a single arbitrary partition of the data. For the regime network, stratified $k$-fold cross-validation was used to preserve the proportion of samples of each regime (dripping, narrowing-jet, and widening-jet) across the folds.

Both networks were optimized following the same two-stage protocol: an architecture search, followed by hyperparameter tuning of the selected architecture. The candidate configurations were defined manually rather than as a full cartesian product, so both searches are non-exhaustive. Their size, ranges, and outcomes are given in Table \ref{tab:regime_opt} for the regime network and Table \ref{tab:diameter_opt} for the droplet diameter network. All networks were fully connected and trained with the Adam optimizer. In the first stage, every candidate architecture was trained with the fixed hyperparameters listed in the tables so that differences in performance could be attributed solely to architectural choices.

\begin{table}[t]
\centering
\caption{Optimization settings and selected configuration for the regime classification network.}
\label{tab:regime_opt}
\small
\begin{tabularx}{\textwidth}{@{}>{\raggedright\arraybackslash}p{2.4cm}>{\raggedright\arraybackslash}X>{\raggedright\arraybackslash}X@{}}
\toprule
 & \textbf{Stage 1: Architecture} & \textbf{Stage 2: Hyperparameters} \\
\midrule
Cross-validation & Stratified 8-fold on 80\% of $TR1$; 15\% internal validation split & Stratified 8-fold (same folds); 15\% internal validation split \\
\addlinespace
Search space & 50 configurations; 1--3 hidden layers; 3--80 neurons/layer; dropout 0--0.3 (step 0.05) & 52 configurations; batch: 4--25; epochs: 400--5000; learning rate: $0.5\times10^{-4}$--$2\times10^{-4}$; early-stopping delta: 0.02--0.08; patience: 70 or off \\
\addlinespace
Fixed parameters & Batch $=16$; epochs $=600$; patience $=70$; learning rate $=1.1\times10^{-4}$; early-stopping delta $=0.02$ & Architecture from Stage 1 \\
\midrule
Selected & Layers 7-30-3; dropout of 0.2 after the 30-neuron layer & Batch $=25$; epochs $=3500$; early stopping off (best model retained); learning rate $=1\times10^{-4}$ \\
\bottomrule
\end{tabularx}
\end{table}

\begin{table}[t]
\centering
\caption{Optimization settings and selected configuration for the droplet diameter regression network.}
\label{tab:diameter_opt}
\small
\begin{tabularx}{\textwidth}{@{}>{\raggedright\arraybackslash}p{2.4cm}>{\raggedright\arraybackslash}X>{\raggedright\arraybackslash}X@{}}
\toprule
 & \textbf{Stage 1: Architecture} & \textbf{Stage 2: Hyperparameters} \\
\midrule
Cross-validation & 10-fold on all of $TD1$; 10\% internal validation split & 10-fold (same folds); 10\% internal validation split \\
\addlinespace
Search space & 40 configurations; 1--3 hidden layers; 4--128 neurons/layer; dropout 0--0.3 (1st hidden layer), 0--0.1 (subsequent hidden layers) & 
15 configurations; batch: 5--20; epochs: 200--1600; 
learning rate: $2.5\times10^{-4}$--$4\times10^{-3}$; early stopping off (best model retained) \\
\addlinespace
Fixed parameters & Batch $=16$; epochs $=1500$; 
learning rate $=1.3\times10^{-3}$; early stopping off (best model retained)
& Architecture from Stage 1 \\
\midrule
Selected & Layers 7-40-5-1; 10\% dropout after the 40-neuron layer & Batch $=16$; epochs $=1600$; early stopping off (best model retained); learning rate $=8\times10^{-4}$ \\
\bottomrule
\end{tabularx}
\end{table}

For the regime network, 20\% of $TR1$ was first set aside as a fixed test set, which was excluded from both the architecture and hyperparameter optimization processes.
The cross-validation was performed on the remaining 80\% of $TR1$. For the diameter network, the cross-validation was performed on the whole of $TD1$. In each cross-validation iteration, the data were divided into three sets. The held-out fold served as the test set and was used only to evaluate the trained model. The other folds were further split into a training set and an internal validation set, the latter containing 15\% of these data for the regime network and 10\% for the diameter network. The internal validation set was used to monitor training, determining when to stop training (early stopping) or, when early stopping was disabled, which checkpoint to retain. To prevent information leakage, the input normalization parameters were computed on the training set only and then applied to the validation and test sets, and the outputs of the diameter network were normalized in the same way.

Candidate configurations were compared across all folds using a composite criterion, applied in both stages. For the regime classifier, it prioritized the highest mean test accuracy, the lowest standard deviation, the smallest interquartile range (IQR) \citep{HAN2012}, and the smallest generalization gap between training and outer-test performance. For the diameter regressor, the criterion comprised primarily the lowest mean test percentage error, followed by the lowest standard deviation and the smallest IQR.

The regime network has 7 inputs and 3 outputs, one per flow regime probability. The candidate architectures comprised one to three hidden layers of 3 to 80 neurons, covering bottleneck (e.g., 7-40-10-3)
, uniform-width (e.g., 7-20-20-3), and narrow-wide-narrow (e.g., 7-10-30-10-3) designs. Dropout rates between 0.0 and 0.3, in steps of 0.05, were combined with these architectures to limit overfitting. With the selected architecture fixed, the training hyperparameters (batch size, maximum number of epochs, early-stopping patience, and learning rate) were then tuned using the same cross-validation scheme and partitioning strategy as in the first stage, including the option of disabling early stopping. This search was guided, varying one hyperparameter at a time around the best setting found so far.

The droplet diameter network has 7 inputs and a single output. The candidate architectures comprised one to three hidden layers, ranging from shallow networks (e.g., 7-64-1 and 7-128-1) to deeper ones with narrow hidden layers (e.g., 7-10-8-10-1), and with up to 128 neurons per hidden layer. Dropout rates of up to 0.3 were applied to the first hidden layer and of up to 0.1 to the subsequent ones. The training hyperparameters of the selected architecture were then tuned by a quasi-grid search over the ranges in Table \ref{tab:diameter_opt}, following the same scheme as for the regime network hyperparameters, including the option of disabling early stopping. The selected architecture and hyperparameters were then used to build the ensembles described in the next section.

\subsection{Ensemble-Based Uncertainty Quantification}
\label{ensemble_unc_quant}

An ensemble is a set of models trained for the same task, whose predictions are combined into a single prediction \citep{Lakshminarayanan2017}. In regression, the prediction of the ensemble is usually the mean of the predictions of its members. In classification, it is usually the class that receives the most votes. Since the members are trained independently, they tend to agree in the regions of the input space that are well covered by the training data and to disagree elsewhere. The disagreement among members therefore provides an estimate of the model (epistemic) uncertainty.

In this work, two ensembles of 20 NNs each were built, $E_{r} = \{\phi_{r,1}, \dots, \phi_{r,20}\}$ for regime prediction and $E_{d} = \{\phi_{d,1}, \dots, \phi_{d,20}\}$ for droplet diameter prediction, where $\phi_{r,k}$ and $\phi_{d,k}$ denote the $k$-th member of each ensemble. 
All members of an ensemble share the architecture and training hyperparameters selected in Section \ref{kFoldOpt} (Tables \ref{tab:regime_opt} and \ref{tab:diameter_opt}), and differ only in their random weight initialization and in the random partition of the data used for training. For a given sample, the predicted regime is the class that receives the most votes in $E_{r}$, and the predicted diameter is the mean of the predictions of the members of $E_{d}$. The uncertainty of $E_{r}$ is measured by the fraction $c$ of members that vote for the predicted class (Section \ref{regime_pred_unc}), and the uncertainty of $E_{d}$ by the coefficient of variation $c_v$ of the predictions of its members (Section \ref{diam_pred_unc}). These measures are used in the subsequent stages of this work to identify the regions of the input space where additional training data should be provided. The ensembles trained on the initial datasets are denoted by $E_{r}^{1}$ and $E_{d}^{1}$, and those retrained after the enrichment of Section \ref{section:manual_enrich} by $E_{r}^{2}$ and $E_{d}^{2}$.

Both ensembles were built following the same procedure. For each candidate network, the corresponding dataset was randomly repartitioned into training, validation, and test subsets, with fixed proportions of 0.65/0.15/0.20 for \textit{TR1} (regime ensemble) and 0.80/0.10/0.10 for \textit{TD1} (droplet diameter ensemble). Since independently trained networks do not always converge to solutions of similar quality, each candidate was assessed once its training was completed and retained as an ensemble member only if it met a minimum performance requirement on both its validation and test subsets. This requirement was an accuracy above 90\% for the regime task, and a mean absolute percentage error (MAPE) of at most 7\% for the droplet diameter task. Candidate networks were trained until 20 of them met this requirement for each ensemble. These thresholds act solely as a post-training acceptance criterion and play no role in the optimization of the networks. Because the test subsets take part in member selection, and because the samples held out for one member belong to the training subset of others, neither \textit{TR1} nor \textit{TD1} could provide an unbiased estimate of ensemble performance. The ensembles uncertainty were therefore evaluated on new data not used at any stage of training or member selection as will be explained in the following subsections.

\subsubsection{Regime Prediction Uncertainty}
\label{regime_pred_unc}

The uncertainty of the regime prediction was quantified through the consensus among ensemble members. For a given input $\mathbf{x}$, each member independently predicts a regime label, and the ensemble prediction is assigned by majority vote. The agreement among members is measured by the consensus level

\begin{equation}
c(\mathbf{x}) = \frac{1}{N}\max_{r}\sum_{i=1}^{N}\mathbb{I}\left[\hat{y}_i(\mathbf{x}) = r\right],
\label{eq:consensus}
\end{equation}

\noindent where $N=20$ is the number of members, $\hat{y}_i(\mathbf{x})$ is the regime predicted by member $i$, $r$ ranges over the three regimes (DR, NJ, and WJ), and $\mathbb{I}[\cdot]$ is the indicator function. That is, $c(\mathbf{x})$ is the fraction of members that vote for the most voted regime. A value of $c=1$ corresponds to a unanimous vote, whereas lower values indicate increasing disagreement, down to a minimum of $c=0.35$ (7 of 20 votes) when the votes are split as evenly as possible among the three regimes. Vote ties are resolved by summing the predicted regime probabilities across ensemble members and selecting the tied regime with the highest total.

\begin{figure}[h!]
\centering
\includegraphics[scale=0.5]{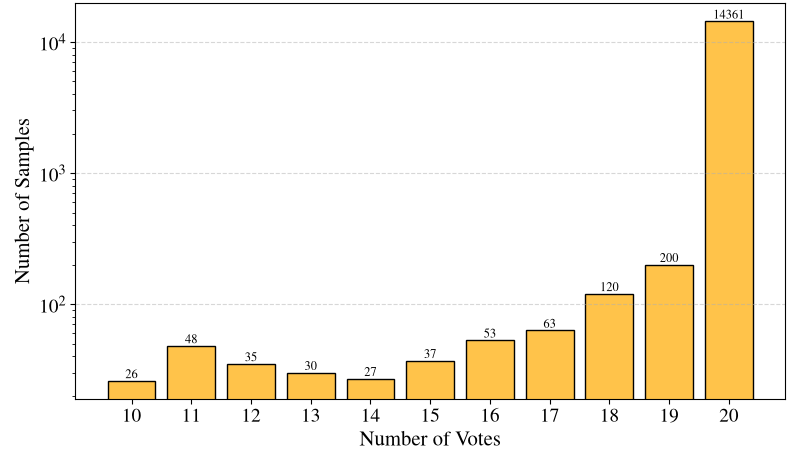}
\caption{Distribution of the consensus level over the samples in $S1$. The horizontal axis shows the number of ensemble members, out of $N=20$, that voted for the most voted regime, which corresponds to $N\,c(\mathbf{x})$ in Eq. \eqref{eq:consensus}. The vertical axis is in logarithmic scale, and the number above each bar indicates the sample count. Most samples (14,361 of 15,000) received a unanimous vote.}
\label{fig:histo_Reg}
\end{figure}

The regime ensemble was applied to dataset $S1$, and the consensus level of each sample was computed with Eq. \eqref{eq:consensus}. Figure \ref{fig:histo_Reg} shows the resulting distribution. Most samples ($\sim$96\%) received a unanimous vote, which indicates that the ensemble is confident over most of the sampled region. This is expected, since the samples in $S1$ were generated by interpolating between training samples of the same regime
. The remaining $\sim$4\% ($\sim$600 samples) showed some degree of disagreement. Although they are a minority, these are the samples for which the predicted regime is least reliable, and they can indicate the regions of the input space where data enrichment is needed.

The distribution of the consensus level over the input space is shown in Fig. \ref{fig:3DPCA_Reg} of the Supplementary Material, where the samples in $S1$ dataset are projected onto the first three principal components. While this projection provided a useful initial overview of the uncertainty landscape, the agreement patterns proved too diffuse to isolate specific regions of disagreement. This suggested that the localized parameter combinations driving ensemble divergence are not readily apparent in a three-dimensional reduction of the full input space. Consequently, a more granular, per-variable analysis was adopted to precisely identify the parametric sub-ranges responsible for the predictive uncertainty. The consensus level of each sample in $S1$ was plotted against each input parameter separately, over the full range covered by the training set. These plots encode only the consensus level and omit the predicted regime so that the reliability of the prediction can be assessed independently of the actual regime label. The complete set of plots is provided in the Supplementary Material (Fig. \ref{fig:VarRangeRegFull}).


The per-variable plots suggested that low-consensus samples cluster in certain sub-ranges of some parameters. This observation alone is not conclusive, however. The samples in $S1$ inherit the non-uniform marginal distributions of the training set, so the ensemble is evaluated on more samples in some sub-ranges than in others. A sub-range can therefore contain many low-consensus samples simply because it contains many samples, and not because the ensemble is less reliable there.

To remove this effect, the analysis was repeated on \textit{SH1} dataset (Section \ref{synth_datasets}). In each set, the values of the corresponding parameter are uniformly distributed over its range. With uniform sampling, every sub-range of each parameter is evaluated with the same number of samples, and any concentration of low-consensus samples can be attributed to the ensemble and not to the sampling density.

The distribution of the samples along each parameter was summarized with kernel density estimation (KDE) \citep{Silverman2018}, a non-parametric method that estimates the probability density function of a variable from a set of observations. Unlike histograms, which depend on the width and placement of the bins, KDE produces a smooth estimate by placing a kernel function $K$ on each observation. The estimated density at a point $x$ is

\begin{equation}
\hat{f}(x) = \frac{1}{nh} \sum_{i=1}^{n} K\left(\frac{x - x_i}{h}\right),
\label{eq:kde}
\end{equation}

\noindent where $n$ is the number of samples, $x_i$ are the observed values, $h$ is the bandwidth, which controls the smoothness of the estimate, and $K$ is the kernel, here a Gaussian. For the KDE curves in this work, the Gaussian kernel density estimator from SciPy library \citep{2020SciPy-NMeth} was used. The bandwidth $h$ was determined using the library's default value, which applies Scott's rule \citep{Scott1992}.

\begin{figure}[h!]
\centering
\includegraphics[scale=0.55]{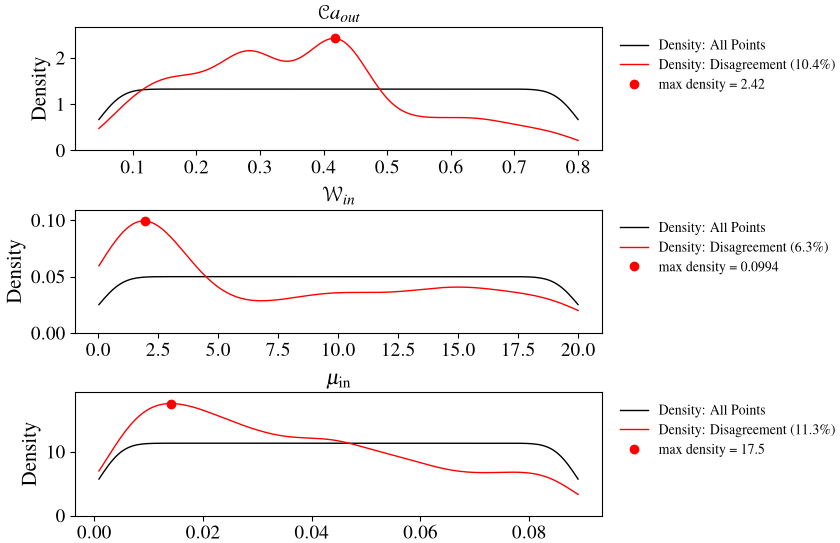}
\caption{Kernel density estimates of the samples in \textit{SH1} along three input parameters. For each parameter, the black curve corresponds to all samples in the set and the red curve to the samples with non-unanimous votes ($c(\mathbf{x})<1$), whose percentage is indicated in the legend. The red marker indicates the maximum of the red curve.}
\label{fig:KDE_Reg_3var}
\end{figure}

Figure \ref{fig:KDE_Reg_3var} shows the KDE curves for three of the input parameters (\Cout{}, \Win{}, and $\mu_{in}$). For each parameter, two curves were computed on the corresponding set of \textit{SH1}, one with all samples (black) and one restricted to the samples with non-unanimous votes ($c(\mathbf{x})<1$, in red). The black curve is approximately flat, as expected from the uniform sampling, and serves as a reference. The red curve is a conditional density. It describes how the samples with non-unanimous votes are distributed along the parameter, and it is normalized to unit area regardless of how many such samples exist. Its height alone therefore indicates where the disagreement is concentrated, not how frequent it is. The local frequency can be recovered by combining both curves with the overall fraction of non-unanimous samples, $\bar{p}$, reported in the legend. The fraction of samples with non-unanimous votes around a value $x$ is

\begin{equation}
p(x) = \bar{p}\,\frac{\hat{f}_{\mathrm{d}}(x)}{\hat{f}_{\mathrm{all}}(x)},
\label{eq:local_disagreement}
\end{equation}

\noindent where $\hat{f}_{\mathrm{d}}$ and $\hat{f}_{\mathrm{all}}$ are the red and black curves, respectively. Thus, where the red curve lies above the black one, disagreement is more frequent than the average over the range, and where it lies below, it is less frequent. Both curves decrease near the limits of each range, which is a known boundary effect of KDE. Near a limit, part of the kernel extends beyond the region where data exist, and the density is underestimated. Since this affects both curves, differences between them near the limits should be interpreted with caution.

The red curves of \Cout{}, \Win{}, and $\mu_{in}$ show clear peaks, located at approximately 0.42, 2, and 0.014, with maximum densities of 2.42, 0.0994, and 17.5, respectively. These maxima are about 1.8, 2.0, and 1.5 times the reference level of the corresponding black curves. According to Eq. \eqref{eq:local_disagreement}, the fraction of samples with non-unanimous votes at the peaks is approximately 19\%, 13\%, and 17\%, compared with overall fractions of 10.4\%, 6.3\%, and 11.3\%. Epistemic uncertainty is therefore concentrated in specific sub-ranges of these parameters. The same analysis was applied to the remaining input parameters, and no clear concentration of low-consensus samples was found. The corresponding curves are provided in the Supplementary Material (Fig. \ref{fig:KDE_Reg_AllVar_v2}).

The peaks for \Cout{} and \Win{} are located close to the regime transitions reported experimentally by \citet{Utada2007}. This is physically consistent, since these two parameters compare viscous and inertial forces, respectively, with capillary forces, and their balance determines the regime in which the system operates. Near a transition, the regime changes abruptly with small variations in the parameters, and the location of the boundary depends on several parameters jointly. With a limited number of training samples, each ensemble member places the boundary at a slightly different location, which gives rise to the disagreement. The agreement between the regions of low consensus and the known physical transitions supports the use of the consensus level as an indicator of where additional high-fidelity simulations are most informative.

\subsubsection{Diameter Prediction Uncertainty}
\label{diam_pred_unc}
Quantifying the uncertainty of the droplet diameter predictions requires a more refined strategy for generating synthetic samples than in the regime prediction case. The droplet diameter is only meaningful in the dripping regime, so the diameter predicting ensemble was trained exclusively on dripping-regime data, and the synthetic samples used to assess its uncertainty should also belong to this regime. Samples generated without this restriction would include non-dripping configurations, for which the predicted diameter has no physical relevance, and the resulting disagreement would not reflect the epistemic uncertainty of the models.

To avoid this, the synthetic samples were first classified with the regime ensemble (as explained in Section \ref{synth_datasets}), and only those for which at least 18 of the 20 members predicted the dripping regime ($c(\mathbf{x}) \geq 0.9$) were retained. Some of the retained samples may still be misclassified. However, the purpose of this analysis is to locate the regions where the diameter prediction disagrees, not to validate the regime labels, and the high consensus required limits these cases to a small fraction of the samples. 

The uncertainty of the diameter predictions was quantified with the coefficient of variation ($c_v$) of the predictions of the ensemble members,

\begin{equation}
c_v(\mathbf{x}) = \frac{s_d(\mathbf{x})}{\bar{d}(\mathbf{x})} \times 100\%,
\label{eq:cv}
\end{equation}

\noindent where $\bar{d}(\mathbf{x})$ and $s_d(\mathbf{x})$ are the mean and the standard deviation of the diameters predicted by the members for input $\mathbf{x}$. The standard deviation alone is not suitable for this purpose, because the predicted diameters span a wide range across the dripping configurations, and the same absolute spread represents a large disagreement for a small droplet but a small one for a large droplet. The $c_v$ expresses the spread relative to the mean prediction, so it can be compared across the whole dripping region regardless of the droplet size.

The $c_v$ measures the dispersion among the predictions of the ensemble members. It is neither the error with respect to the Basilisk simulation results nor the physical variability of different droplets produced by the simulated device. In particular, $c_v$ must not be read as a measure of how spread out the droplets themselves are. A high $c_v$ indicates that the ensemble is not confident, but it does not imply that the mean prediction is wrong. Conversely, a low $c_v$ does not guarantee an accurate measure, since all members may share the same bias, for example, in regions without training data where they extrapolate similarly. The $c_v$ therefore only identifies regions where the training data do not sufficiently constrain the model, which are the candidates for enrichment. 

Synthetic dataset $S2$ was used in order to study the distribution of $c_v$ values over the input space.
First, all synthetic samples were passed through the 20-member diameter ensemble to obtain droplet-diameter predictions. Then, the spatial distribution of the $c_v$ over the synthetic samples was first examined through a three-dimensional PCA projection. 

Figure \ref{fig:3DPCA_Dia} of the Supplementary Material shows a relatively localized cluster of high $c_v$ suggesting inter-member disagreement in droplet diameter prediction, though the projection cannot pinpoint the specific physical parameter range, within the dripping regime, where it occurs.

To quantify the uncertainty of the diameter ensemble, the synthetic dataset \textit{SH2} was built with randomly generated dripping cases (see Section \ref{synth_datasets}), uniformly distributed over the expected dripping region on the (\Cout, \Win) regime diagram of \citet{Utada2007}. 

A specific synthetic sample was considered to have high uncertainty when the standard deviation of the predicted diameters exceeds 20\% of their mean ($c_v > 20\%$). Because $c_v$ quantifies disagreement among ensemble members' predictions rather than physical variability between individual droplets (as mentioned above), this threshold is best interpreted as an upper bound on acceptable prediction variability. A prediction with this level of disagreement is not reliable enough for applications that require a stable droplet formation regime, 
where the diameter must be controlled within a narrow tolerance \citep{Mulligan2012}. This maximum value for $c_v$ was adopted in the first iteration of the refinement loop. The enrichment described in Section \ref{section:manual_enrich} aims to ensure that samples remain below this bound, and stricter bounds can be adopted in subsequent iterations to reduce the uncertainty further.

\begin{figure}[h!]
\centering
\includegraphics[scale=0.6]{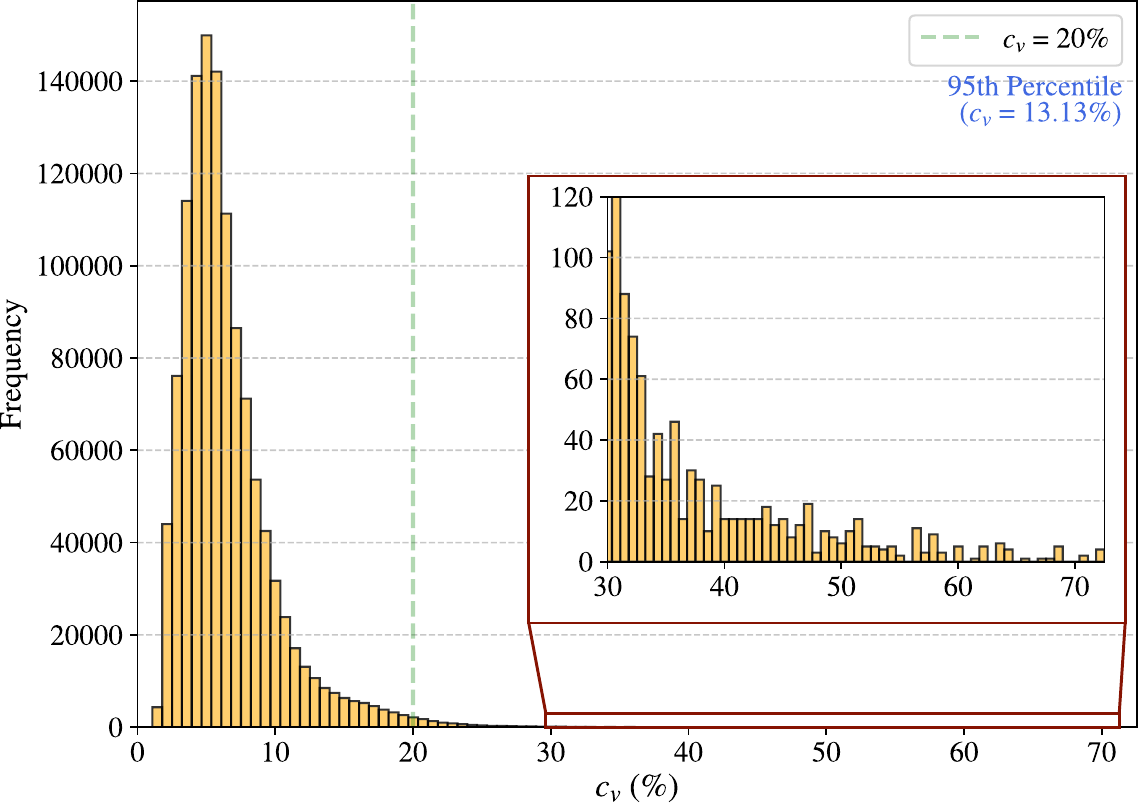}
\caption{Distribution of the $c_v$ for droplet diameter predictions. The dashed green line indicates the 20\% threshold used to define excessive uncertainty. The inset provides a magnified view of the distribution's tail, illustrating the presence of high-$c_v$ samples reaching up to 70\%, which characterize the pre-enrichment state of the ensemble.}
\label{fig:Histo_Dia_1}
\end{figure}

The distribution of $c_v$ over \textit{SH2} is shown in Figure \ref{fig:Histo_Dia_1}. Most samples have $c_v < 10\%$, but the distribution has a long tail that reaches values of up to 70\%. A total of 9,748 samples exceed the bound of 20\%, which represents less than 1\% of \textit{SH2}. Although this fraction is low, the uncertainty of these predictions is exceedingly high for many applications, as mentioned earlier, and these samples are therefore the starting point of the refinement process. However, as happened with the 3D PCA, the histogram does not show where they are located. If they are concentrated in specific regions rather than randomly scattered, their location would reveal where the model is least reliable and where new training data should be added.

\begin{figure}[h!]
\centering
\includegraphics[scale=0.65]{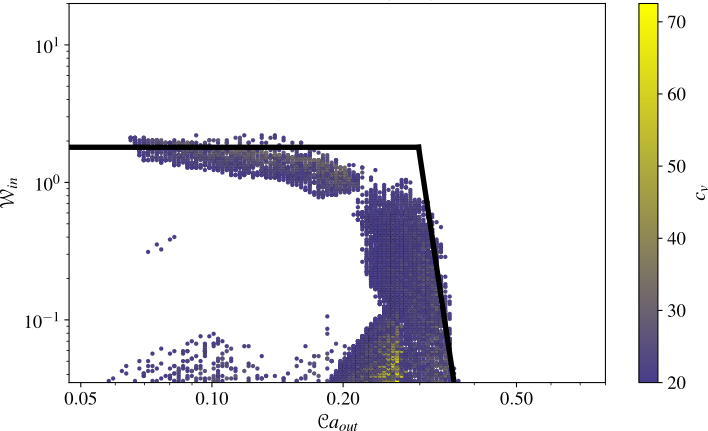}
\caption{Spatial distribution of high predictive uncertainty ($c_v> 20$\%) for droplet diameter predictions on the (\Cout, \Win) plane. High-uncertainty samples are primarily clustered along the regime transition boundaries, whereas the center of the dripping regime region exhibits mostly high ensemble confidence.}
\label{fig:HU_CV_0.2}
\end{figure}

Figure \ref{fig:HU_CV_0.2} shows the \textit{SH2} samples with $c_v > 20\%$ on Utada's regime diagram, colored according to their $c_v$ values. The samples are not randomly scattered, but rater concentrated near the boundaries between the dripping and widening jet regimes and between the dripping and narrowing jet regimes, whereas the interior of the dripping region is dominated by low $c_v$ values. This behavior is consistent with that observed for the regime ensemble. Near a transition, the droplet diameter changes rapidly with small variations in the input parameters, and if the training data available in these zones are scarce, each member will approximate the rapid variation differently, giving rise to the disagreement. 

The regions exhibiting disagreement are targeted by the enrichment step described in Section \ref{section:manual_enrich}, which runs new high-fidelity Basilisk VOF simulations to augment training data where the diameter ensemble is least reliable. The net effect of this enrichment on the distribution of $c_v$ is presented and analyzed in Section \ref{post_enrich}.

\subsection{Uncertainty-Guided Enrichment Strategy}
\label{section:manual_enrich}

The analyses in Sections \ref{regime_pred_unc} and \ref{diam_pred_unc} identified the regions of the input space where the predictions of both ensembles are least reliable. Since this uncertainty is mainly epistemic, it can be reduced by providing the models with new training data in these regions. The next step of the refinement loop was therefore to run new high-fidelity VOF simulations at selected parameter combinations and to add the resulting data to the training dataset. Because each simulation is computationally expensive, the combinations to be simulated were chosen according to two complementary criteria: the spatial distribution of the predictive uncertainty and the known physical behavior of the device. Additional simulations were therefore concentrated in the regions where the ensembles show low consensus or high $c_v$, and in the vicinity of the regime transitions reported by \citet{Utada2007}, namely the dripping-to-widening-jet and dripping-to-narrowing-jet transitions. Both criteria predominantly point to closely situated regions in the parameter space. Near a transition, the interplay of inertial, viscous, and capillary forces shifts, rendering the device highly sensitive to small parameter perturbations and causing the underlying flow mapping to change abruptly. In these zones the initial sampling strategy yielded insufficient data to resolve such sharp behavioral shifts, a scarcity that directly contributed to the observed ensemble disagreement.

The enrichment started with the dripping regime for three reasons. First, the regime transitions with the highest uncertainty in both ensembles, dripping-to-widening-jet and dripping-to-narrowing-jet, border this regime. Second, each simulated case near the dripping region provides a regime label for the regime ensemble and, if it is confirmed as dripping, also a diameter value for the diameter ensemble. Third, the diameter ensemble is trained only on dripping cases, so it has fewer training data than the regime ensemble and is more affected by their scarcity.

The starting point was the set of samples of $SH2$ with $c_v > 20\%$ (Section \ref{diam_pred_unc}). From this set, $\sim$210 base samples were drawn with probability proportional to their $c_v$, in order to preferentially select those with the highest uncertainty. For each base sample $\mathbf{x}$, its five nearest neighbors within this set were identified on the (\Cout{}, \Win{}) plane, and a new configuration was generated by linear interpolation in the five-dimensional parameter space (after discarding \Cout{} and \Win{} parameters) between $\mathbf{x}$ and a randomly selected neighbor $\mathbf{x}_k$, as $\mathbf{x}_{\mathrm{new}} = \mathbf{x} + \lambda (\mathbf{x}_k - \mathbf{x})$, with $\lambda$ drawn uniformly from $[0,1]$. 
Each new configuration produced was classified with the regime ensemble and retained only if it was predicted as dripping, with the same criterion used in Section \ref{diam_pred_unc} ($c(\mathbf{x}) \geq 0.9$). In case that the predicted regime was not dripping, the interpolation was retried.

\begin{figure}[h!]
\centering
\includegraphics[trim={0cm 0cm 0cm 0.5cm}, clip, scale=0.6]{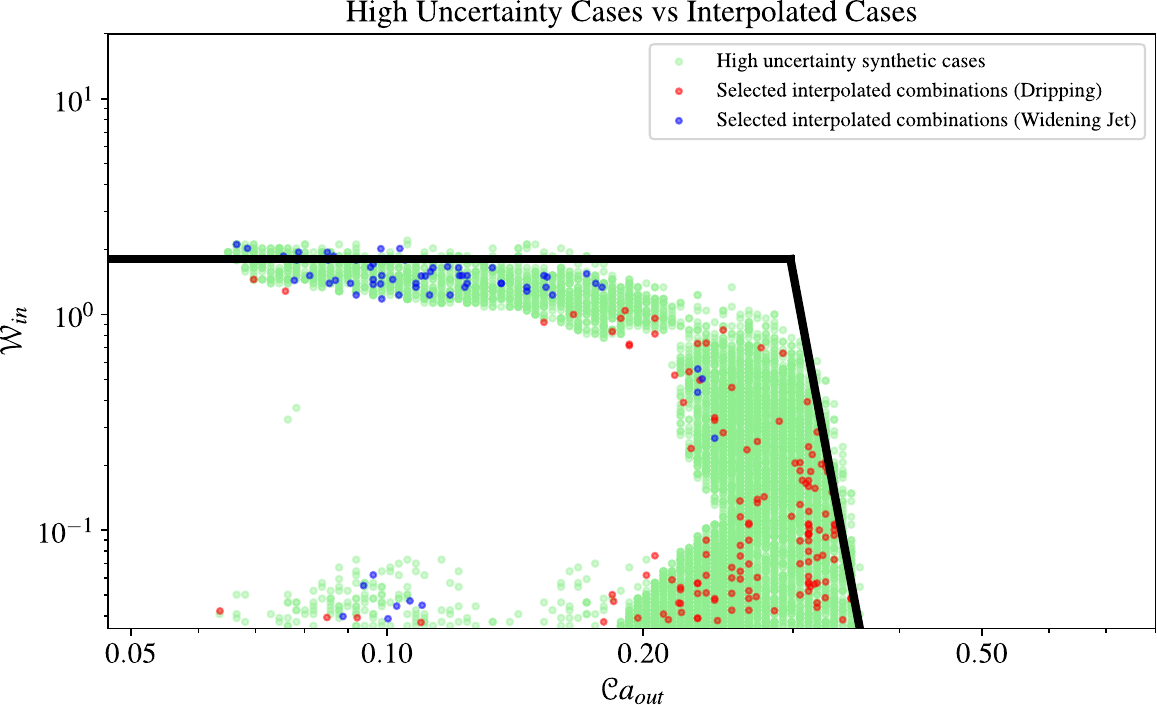}
\caption{Spatial distribution of $SH2$ synthetic cases with $c_v> 20$\% (light green) and the $\sim$200 interpolated combinations (red and blue) on the (\Cout, \Win) plane. The red dots indicate cases that produced dripping regime on the simulation, while blue dots are for widening jet. The black line represents the documented approximate boundary of the dripping regime. Interpolated points for the enrichment process are situated in the main regions of high epistemic uncertainty.}
\label{fig:newDripDist}
\end{figure}

After all the resulting samples were simulated in Basilisk, around 140 were confirmed as dripping and the remaining as widening jets. The dripping cases provided new training data for both the diameter and regime ensemble, while widening-jet cases were added to the training data of the regime ensemble. Figure \ref{fig:newDripDist} shows the set of samples of $SH2$ with $c_v > 20\%$ and the new simulated samples on the (\Cout{}, \Win{}) plane. The figure shows with blue dots those samples that resulted in a widening jet regime after simulation, and with red dots those that exhibited dripping regime. It must be noted that the majority of widening-jet cases lie close to the dripping-to-widening-jet transition defined in literature, which shows that the original regime ensemble itself is not reliable near this transition and confirms the need for new data in these regions. 


In addition, new narrowing-jet cases were generated to keep the regime classes balanced in the enriched dataset. The new cases were generated interpolating linearly between randomly sampled points from $TR1$, using a wight drawn from a uniform distribution. Additionally, \Cout and \Win values were restricted to sub-ranges covering partially the probable narrowing-jet region and its transition boundary (Section \ref{sec:device_conditions}).

When generating these new cases, candidate samples were pre-screened to retain only those that presented a high ensemble consensus, with 17 or more of the 20 members predicting narrowing-jet regime ($c(\mathbf{x}) \geq 0.85$). At the end, 250 new configurations were generated by interpolation and simulated in Basilisk, and around 130 of them were confirmed as narrowing jets from simulation. 
These new cases, together with the dripping and widening-jet cases previously generated, were added to the original \textit{O1} training data to form dataset \textit{T2} (see Section \ref{Datasets} for details). {This dataset was used to retrain the ensembles, as will be described in Section }\ref{post_enrich}. 
Fig. \ref{fig:datasetAddedCases} of the Supplementary Material shows \textit{T2} data distribution on the (\Cout{}, \Win{}) plane, together with the regime of each case.

\section{Inverse design of microfluidic droplet generators}
\label{inverse_design}

The main goal of this work, as outlined in the introduction, is to determine the operating parameters of a microfluidic droplet generator that produce droplets of a specified size. The previous sections developed the surrogate ensembles for this purpose and reduced their uncertainty in the dripping regime by enriching the training data. This section applies the retrained ensembles $E_{r}^{2}$ and $E_{d}^{2}$ to this inverse design problem. Given a target droplet diameter $d_{\mathrm{drop}}^{\mathrm{target}}$ and optional constraints on the operating parameters, the task is to find a configuration that operates in the dripping regime and produces droplets of the target size. Since the ensembles are much less expensive to evaluate than VOF simulations, they can be queried many times and used as the evaluation model within an iterative search algorithm. In this work, the search was performed with the PSO algorithm described in Section \ref{ParticleSwarmOpt}, which evaluates a population of candidate configurations at each iteration. {The following paragraphs} describe the formulation of the objective function and the settings of the optimization. Afterwards, several case studies are presented and discussed in Section~\ref{subsec:inverse_design_validation}, in which the configurations obtained were simulated with VOF to validate the method.

\subsection{PSO Configuration}
\label{pso_configuration}

The neural networks receive seven input parameters, but the search space has only five dimensions. The free parameters are \Cout{}, \Win{}, $\mu_{\mathrm{out}}$, $\mu_{\mathrm{in}}$, and $r_{\mathrm{in}}$. The phase velocities $u_{\mathrm{out}}$ and $u_{\mathrm{in}}$ are not independent variables, as they are derived from \Cout{} and \Win{} once the remaining physical and geometric parameters are fixed. Specifically, they are calculated using the definitions of the capillary and Weber numbers, and the same physical constants reported in \citet{Harispe2026} (see Section \ref{baselineSurrogates}). %
The user can fix any subset of the free parameters to prescribed values. In that case, the search is performed only over the remaining free parameters.

Table \ref{PSObounds} lists the range of each free parameter used in the inverse design problem. These intervals span the complete range of values present in the training data, and are not restricted to the observed dripping regime region reported in the literature \citep{Utada2007}. The boundary of the dripping regime depends jointly on several parameters and cannot be represented by independent bounds on each of them. The regime ensemble therefore decides whether a configuration operates in the dripping regime, through the fitness function detailed in Section \ref{fitnessFunct}.

\begin{table}[ht!]
\centering
\caption{Ranges of the free parameters in the inverse design problem.}
\label{PSObounds}
\begin{tabular}{@{}lcl@{}}
\toprule
Parameter & Symbol & Range \\
\midrule
Capillary number of the outer phase & \Cout{} & $[0.047,\;0.8]$ \\
Weber number of the inner phase & \Win{} & $[0.035,\;20.0]$ \\
Viscosity of the outer phase & $\mu_{\mathrm{out}}$ & $[8.9,\;95]$~mPa$\cdot$s \\
Viscosity of the inner phase & $\mu_{\mathrm{in}}$ & $[0.89,\;89]$~mPa$\cdot$s \\
Radius of the inner capillary & $r_{\mathrm{in}}$ & $[5.0,\;25]$~$\mu$m \\
\bottomrule
\end{tabular}
\end{table}

Table \ref{tab:pso_params} lists the values of the parameters of the PSO scheme described in Section \ref{ParticleSwarmOpt}. The personal best of each particle is its initial position, and the global best is the initial position with the lowest value of the fitness function. The initial velocities are sampled uniformly in $[-\Delta\bm{x},\;\Delta\bm{x}]$ and then clipped to $\pm\,0.5\,\Delta\bm{x}$, where $\Delta\bm{x}$ is the vector of the widths of these ranges. This conservative initialization mitigates the risk of large initial step sizes that would cause wild oscillation before cognitive and social components can guide particles. Furthermore, it prevents the swarm from overshooting promising regions while the stochastic inertia gradually facilitates broader search patterns in subsequent iterations.


\begin{table}[ht!]
\centering
\caption{Parameters of the PSO algorithm and termination criteria.}
\label{tab:pso_params}
\begin{tabular}{@{}lcl@{}}
\toprule
Quantity & Symbol & Value \\
\midrule
Number of particles & $N$ & 50 \\
Inertia weight & $w$ & $\mathcal{U}(0,1)$ \\
Cognitive acceleration constant & $c_1$ & 2 \\
Social acceleration constant & $c_2$ & 2 \\
Velocity clamping coefficient & $\delta$ & 0.9 \\
Boundary rebound factor & $\alpha_b$ & $-0.7$ \\
Maximum number of iterations & $T_{\max}$ & 500 \\
Patience limit & $P$ & 150 \\
Minimum relative improvement & $\varepsilon$ & 0.10 \\
\bottomrule
\end{tabular}
\end{table}


The inertia weight $w$ in Table \ref{tab:pso_params} is sampled from a uniform distribution $\mathcal{U}(0,1)$. The algorithm employs a patience criterion to prevent unnecessary iterations. A counter increments whenever the global best fails to improve. It resets only when the global best improves by at least $\varepsilon$ relative to the previous value. The search terminates once the counter reaches $P$. The counter remains fixed if a new solution yields a smaller fitness function value but the relative improvement falls below $\varepsilon$. This relative improvement is calculated by multiplying the previous best by $(1 - \varepsilon)$ instead of dividing by it. This formulation ensures numerical stability and prevents division by zero as the objective function values approach zero.

\subsection{Multi-Criteria Fitness Function}
\label{fitnessFunct}

The fitness function determines which configurations the PSO algorithm considers good solutions. Minimizing only the error of the predicted diameter is not sufficient. The optimizer could reach configurations with a predicted diameter close to the target in regions where the members of the ensembles disagree or where the configuration does not operate in the dripping regime. Such configurations would probably not produce the target diameter in a simulation or an experiment. To avoid these configurations, the fitness function was defined as the sum of the predicted diameter error ($t_0$) and four penalty terms,
\begin{equation}
F(\mathbf{x}) = t_0 + t_1 + t_2 + t_3 + t_4,
\label{eq:fitness}
\end{equation}
where $\mathbf{x}$ is the vector of free parameters. Each term takes values between 0 and 1, so no term dominates the sum because of its scale. The five terms are described below.

\paragraph{Diameter error ($t_0$).}
The first term measures the relative error between the target diameter $d_{\mathrm{drop}}^{\mathrm{target}}$ and the mean prediction $\bar{d}_{\mathrm{drop}}$ of the $|E_{d}^{2}|$ members of the ensemble, through a saturating function,
\begin{equation}
t_0 = \tanh\left(\frac{|d_{\mathrm{drop}}^{\mathrm{target}} - \bar{d}_{\mathrm{drop}}|}{0.1\, d_{\mathrm{drop}}^{\mathrm{target}}}\right).
\label{eq:t0}
\end{equation}
For small errors, $t_0$ grows approximately linearly with the relative error, while for large errors, it saturates at 1. A relative error of 1\% gives $t_0 \approx 0.10$, an error of 10\% gives $t_0 \approx 0.76$, and an error of 20\% gives $t_0 \approx 0.96$. The saturation keeps $t_0$ in the same range as the penalty terms. Far from the target diameter, the differences in fitness are therefore determined mainly by the penalty terms, which guide the swarm toward reliable parametric regions within the dripping regime. 

\paragraph{Regime confidence ($t_1$).}
The second term requires each member of $E_{r}^{2}$ to be confident in the dripping regime. It penalizes the fraction of the $|E_{r}^{2}|$ members that assign to the dripping class a probability lower than or equal to $p_{\mathrm{th}} = 75\%$,
\begin{equation}
t_1 = 1 - \frac{1}{|E_{r}^{2}|} \sum_{k=1}^{|E_{r}^{2}|} \mathbb{I}\left[p_k^{\mathrm{DR}} > p_{\mathrm{th}}\right],
\label{eq:t1}
\end{equation}
where $p_k^{\mathrm{DR}}$ is the softmax probability of the dripping class given by the \textit{k}-th member of the regime ensemble ($\phi_{r,k}$), and $\mathbb{I}[\cdot]$ equals 1 if its argument is true and 0 otherwise. The condition is applied to each member rather than to the mean probability of the ensemble. A high mean probability can result from a few members with very high confidence and many members with low confidence. The condition on each member penalizes these configurations. A value of $t_1 = 0$ indicates that every member assigns a probability higher than $p_{\mathrm{th}}$ to the dripping class. The value of $p_{\mathrm{th}} = 75\%$ was chosen to follow the same criteria used when selecting candidate dripping samples for the synthetic datasets (Section \ref{synth_datasets}). In this way, $t_1$ measures the confidence of each vote of the members. The considered criteria for this term is different than the number of votes for a regime ($c$), which is accounted on the next penalty term. 


\paragraph{Regime consensus ($t_2$).}
The third term penalizes the configurations for which the consensus of $E_{r}^{2}$ is not the dripping regime,
\begin{equation}
t_2 = \left(1 - \frac{N_{\mathrm{votes}}^{\mathrm{DR}}}{|E_r^2|}\right) \mathbb{I}\left[\hat{c} \neq \mathrm{DR}\right],
\label{eq:t2}
\end{equation}

\noindent where $N_{\mathrm{votes}}^{\mathrm{DR}}$ is the number of members that predict the dripping regime, and $\hat{c}$ is the consensus class. The consensus class is obtained by majority voting, and ties are resolved by the sum of the class probabilities over the members. If the consensus is dripping, $t_2 = 0$, and the agreement between members is penalized only by $t_1$. If the consensus is a jetting regime, at most half of the members vote for dripping, so $t_2$ is at least 0.5. A configuration in a jetting regime therefore cannot reach a low fitness value, even if its predicted diameter is close to the target. This prevents the optimizer from selecting configurations where the diameter prediction has no physical meaning.

\paragraph{Diameter uncertainty ($t_3$).}
The fourth term penalizes the spread of the predictions of the members of $E_{d}^{2}$, with a saturating function similar to that of Eq. \eqref{eq:t0},
\begin{equation}
t_3 = \tanh\left(\frac{c_v}{13\%}\right), \qquad c_v = \frac{s_d}{\bar{d}_{\mathrm{drop}}},
\label{eq:t3}
\end{equation}
where $s_d$ is the standard deviation of the $|E_{r}^{2}|$ predictions. A value of $c_v = 5\%$ gives $t_3 \approx 0.37$, $c_v = 13\%$ gives $t_3 \approx 0.76$, and $c_v = 20\%$ gives $t_3 \approx 0.91$. 
The value 13\% at the denominator was chosen based on the 95th percentile observed on $SH4$ for $E_{d}^{1}$. This term connects the optimization with the uncertainty quantification of Section \ref{ensemble_unc_quant}. A configuration with a mean prediction close to the target is penalized if the members of $E_{d}^{2}$ disagree on its diameter.

\paragraph{Boundary proximity ($t_4$).}
The fifth term penalizes the configurations approaching the bounds of \Cout{} and \Win{},

\begin{equation}
t_4 = \frac{1}{2} \left( |2\hat{x}_{\mathscr{C}a} - 1|^{3} + |2\hat{x}_{\mathscr{W}} - 1|^{3} \right),
\label{eq:t4}
\end{equation}

\noindent where $\hat{x}_{\mathscr{C}a}$ and $\hat{x}_{\mathscr{W}}$ represent the values of \Cout{} and \Win{} linearly normalized to $[0, 1]$ within the ranges of Table \ref{PSObounds}. The penalty is zero at the center of the ranges and equals 1 when both parameters are at a bound. With the cubic exponent, the penalty is small near the center and increases rapidly close to the bounds. The surrogates are less reliable near the bounds of the search space, where the training data starts to become scarce. {The choice of linear normalization ensures a symmetric penalty structure, which facilitates the exploration of broad parametric locations that extend beyond the traditional boundaries reported in literature }\citep{Utada2007}{. Furthermore, restricting $t_4$ exclusively to \Cout{} and \Win{} serves to reduce the complexity of the multivariable problem. These dimensionless groups encapsulate the dominant physical forces (viscous, inertial, and capillary) that govern the regime transition. Penalizing the extreme bounds of these variables can be used to maintain surrogate fidelity without unnecessarily over-constraining the secondary geometric and material parameters.}

The fitness function of Eq. \eqref{eq:fitness} favors configurations with a predicted diameter close to the target, a confident and consistent prediction of the dripping regime, a low spread of the diameter predictions, and parameters away from the bounds of the search space. Since the penalty terms use the uncertainty measures of Section \ref{ensemble_unc_quant}, the optimizer avoids the regions where the predictions of the ensembles are not reliable.


\section{Results}
\label{sec:results}
\subsection{Enrichment Impact Assessment}
\label{post_enrich}

Following the enrichment of the training dataset detailed in Section \ref{section:manual_enrich}, the neural networks comprising both ensembles were retrained on the expanded data. This section evaluates the impact of this enrichment on both ensembles, addressing two key questions: (i) whether the disagreement of the retrained regime ensemble decreases in regions associated with the dripping regime, and (ii) whether the uncertainty of the diameter predictions within this regime is reduced. The analysis focuses on the dripping regime because diameter predictions are only performed for selected candidate samples for which the regime ensemble had at least 85\% member consensus on dripping ($c(\mathbf{x}) \geq 0.85$). Disagreement within the jetting regimes can therefore affect the diameter predictions only through candidates misclassified as dripping, which are expected mainly near the regime boundaries.

The retrained ensembles were not evaluated on the synthetic datasets used to localize and quantify uncertainty or to select the enrichment cases. The new cases in \textit{T2} were generated by interpolation around high-uncertainty samples,  
so the retrained ensembles carry indirect information about those samples, and evaluating them on the same sets would overestimate the improvement. New synthetic datasets were therefore generated with the knowledge gathered in Sections \ref{regime_pred_unc} and \ref{diam_pred_unc}, and with the procedure described in Section \ref{synth_datasets}. For the regime task, \textit{SH3} was generated from \textit{TR1} using the same per-variable uniform sampling strategy as \textit{SH1}, in which only one variable is sampled uniformly over its range at a time. For the diameter task, datasets \textit{SH4} and \textit{SH5} were generated with the same procedure as \textit{SH2}. \textit{SH4} was obtained from \textit{TD1}, and \textit{SH5} from \textit{T2}. They differ in the regime ensemble used to filter the dripping candidates, as mentioned in \ref{synth_datasets} and detailed below.

In what follows, $E_{r}^{1}$ and $E_{d}^{1}$ denote the baseline regime and diameter ensembles, trained on \textit{TR1} and \textit{TD1}, respectively, and $E_{r}^{2}$ and $E_{d}^{2}$ denote the retrained ensembles, trained on \textit{T2} and its dripping subset. Both versions were evaluated on the new datasets with the same diagnostics used before. For the regime ensembles, the diagnostic is the disagreement curve, which is the per-variable KDE curve of the samples with non-unanimous votes ($c(\mathbf{x})<1$). For the diameter ensembles, it is the coefficient of variation $c_v$ of the predictions of the members on each sample. Because both versions are evaluated on the same datasets with the same diagnostics, any difference in the results can be attributed to the new training data rather than to the evaluation procedure.

Both diameter datasets differ in the population of samples on which $E_{d}^{1}$ and $E_{d}^{2}$ are compared. \textit{SH4} was generated from \textit{TD1} and filtered with $E_{r}^{1}$, so it contains the samples selected by the baseline pipeline. \textit{SH5} was generated from \textit{T2} and filtered with $E_{r}^{2}$, so it contains the samples that the updated pipeline would select in the inverse design workflow (Section \ref{inverse_design}), in which the corresponding retrained regime ensemble $E_{r}$ filters the candidates before the retrained diameter ensemble $E_{d}$ predicts their size. In each datasets, the two diameter ensembles predict for the same samples, so the change in $c_v$ is due only to the retraining of the diameter networks. While \textit{SH4} quantifies this change in the original population of synthetic samples, \textit{SH5} shows whether it holds on the samples selected by the complete updated pipeline, which is the relevant population in practice.

\subsubsection{Evaluation of uncertainty on regime ensembles}
\label{sec:uncertainty on regime ensembles}

\begin{figure}[ht]
\centering
\includegraphics[scale=0.55]{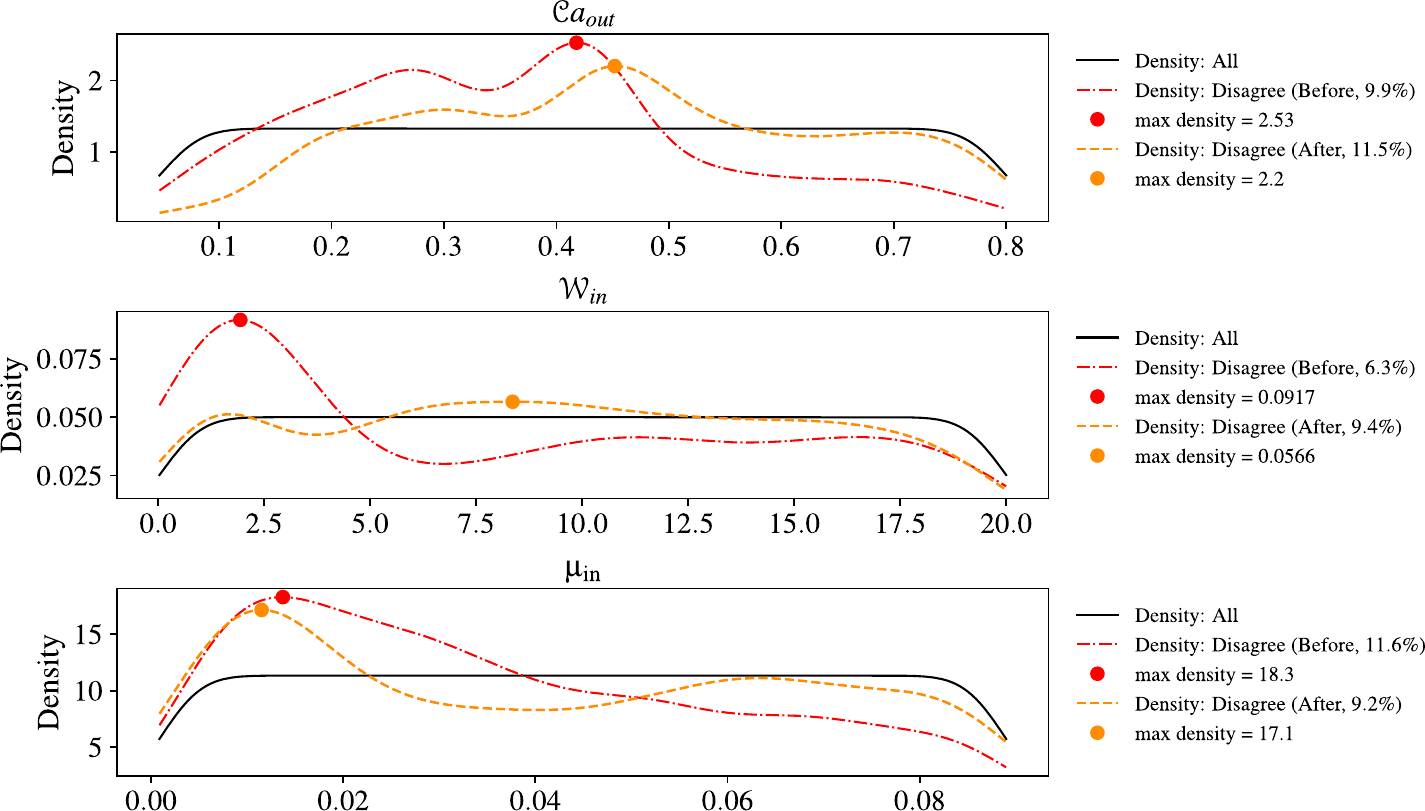}
\caption{Kernel density estimates of the samples in \textit{SH3} along three input parameters, before and after enrichment. Each panel corresponds to the subset of \textit{SH3} in which only the indicated parameter is sampled uniformly. The black curve corresponds to all samples in the subset, and the red and orange curves to the samples with non-unanimous votes ($c(\mathbf{x})<1$) for the baseline ($E_{r}^{1}$) and retrained ($E_{r}^{2}$) regime ensembles, respectively. The percentage of non-unanimous samples is indicated in the legend, and the markers indicate the maximum of each curve.}
\label{fig:newKDEReg}
\end{figure}

Figure \ref{fig:newKDEReg} compares the disagreement curves on \textit{SH3} of $E_{r}^{1}$ (red) and $E_{r}^{2}$ (orange). Each panel corresponds to the subset of \textit{SH3} in which the indicated parameter is sampled uniformly, so the overall fractions of non-unanimous samples given in the legends differ between panels. Only the disagreement curves for parameters \Cout{}, \Win{} and $\mu_{in}$ are shown on the figure since it is where the disagreement changed noticeably. The rest of the parameters' curves can be observed in Fig. \ref{fig:KDECurves_pre+post} of the Supplementary Material.

Since each KDE is normalized, the heights of the red and orange curves do not directly measure the amount of disagreement, so two quantities derived from them are also used to compare both ensembles. The first is the local fraction $p(x)$ of non-unanimous samples, given by Eq. \eqref{eq:local_disagreement}. For \Win{} and \Cout{}, it is evaluated in the sub-ranges associated with the dripping regime. These correspond to the low-\Win{} region around \Win{}~$\approx 2$ and the range of \Cout{} below the dripping-to-narrowing-jet transition, where the upper bound typically falls between $0.3$ and $0.5$ depending on the \Win{} value. For $\mu_{in}$, it is evaluated at the maximum of each disagreement curve, around $\mu_{in} \approx 0.01$. The second is the ratio $\max_x p(x)/\bar{p}$ between the maximum local fraction and the overall fraction. A value of 1 means that the disagreement is uniform along the parameter. Higher values mean that the disagreement is concentrated in a sub-range. In Figure \ref{fig:newKDEReg}, this ratio equals the maximum of the disagreement curve divided by the height of the flat part of the black curve. After enrichment, the ratio decreased from 1.92 to 1.67 for \Cout{}, from 1.84 to 1.14 for \Win{}, and from 1.62 to 1.51 for $\mu_{in}$. The disagreement of $E_{r}^{2}$ is therefore less concentrated in specific sub-ranges of the parameters than that of $E_{r}^{1}$.

The largest reduction in the concentration of disagreement took place for \Win{}. Before enrichment, the disagreement curve peaked at \Win{}~$\approx 2$, in the low-\Win{} range associated with the dripping regime, where $p(x)$ was almost 12\%, nearly twice the overall fraction $\bar{p}$ of 6.3\%. After enrichment, this peak disappears, and $p(x)$ in the same range is about  9.7\%, close to the overall fraction $\bar{p}$ of 9.4\%. The curve is mostly flat over the whole range, so no sub-range of \Win{} concentrates the disagreement. Consequently, the low-\Win{} range where dripping occurs is no longer the range with the highest disagreement of the regime ensemble, which is one of the main objectives. The increase in the overall fraction is examined later.

For \Cout{}, the dripping-to-narrowing-jet transition are expected at values between approximately 0.3 and 0.5, depending on \Win{} \citep{Utada2007}. After enrichment, the disagreement curve for \Cout decreased below the transition, for example from about 11\% to about 6\% at \Cout{}~$\approx 0.15$. The peak of the disagreement curve shifted from \Cout{}~$\approx 0.42$ to \Cout{}~$\approx 0.45$, toward the transition and the narrowing-jet regime regions, and became less pronounced relative to the reference. As a result, the disagreement of the retrained ensemble is lower within the range of \Cout{} where dripping is more frequent, and the remaining peak is located at the transition region or slightly above it. Above the transition, the disagreement increased, as discussed below.

For $\mu_{in}$, both the overall fraction of non-unanimous samples, from 11.6\% to 9.2\%, and the local fraction at the initial peak location, from about 19\% to about 14\%, decreased. In addition, the broad elevation of the disagreement curve that extended up to $\mu_{in} \approx 0.04$ before enrichment is confined after enrichment to a narrower peak that ends at about $\mu_{in} \approx 0.02$. The enrichment therefore reduced the regime disagreement over most of the inner-phase viscosity range, and the remaining concentration is limited to a narrow range of low viscosities that can be targeted directly in a subsequent iteration of the refinement loop. However, because the dripping regime is determined by geometric confinement and the balance of interfacial, viscous, and inertial forces, rather than $\mu_{in}$ alone \citep{Nunes2013, Utada2007}, the range where disagreement persists is not necessarily associated with dripping regime.

In contrast, the overall fraction of non-unanimous samples increased for \Cout{}, from 9.9\% to 11.5\%, and for \Win{}, from 6.3\% to 9.4\%. This increase is located at high \Cout{} values and at intermediate and high \Win{} values, away from the transition regions, which received most of the new training cases, most of them within the expected dripping regime region according to the regime map of \citet{Utada2007}. A plausible explanation is that the new cases near the transitions, including those initially predicted as dripping that turned out to be widening jets, forced the networks to learn more detailed regime boundaries around the dripping region. Since the ensemble members differ in the randomization of the training data and in their initialization, each of them fits these new boundaries in a slightly different way, and away from the new cases, where training data could remain scarce, they extrapolate them differently. This increases the disagreement within the jetting regimes. Since a sample in these regions reaches the diameter ensemble only if at least 85\% of the regime ensemble members vote for dripping, this increase has at most a limited effect on the diameter predictions. It does, however, identify the regions to be enriched if the regime ensemble is to be used disregarding the dripping regime.

The accuracy of the regime predictions cannot be evaluated on \textit{SH3}, which has no ground-truth labels, and is assessed on the simulated test dataset \textit{GR} in Section \ref{physical_impact}. In terms of disagreement, the enrichment achieved its purpose for the regime ensemble: the disagreement of $E_{r}^{2}$ decreased in the regions associated with the dripping regime, where its predictions are used to select the candidates to be evaluated by the diameter ensemble.

\subsubsection{Evaluation of uncertainty on diameter ensembles}
\label{sec:uncertainty on diameter ensembles}

\begin{figure}[h!]
\centering
\includegraphics[scale=0.55]{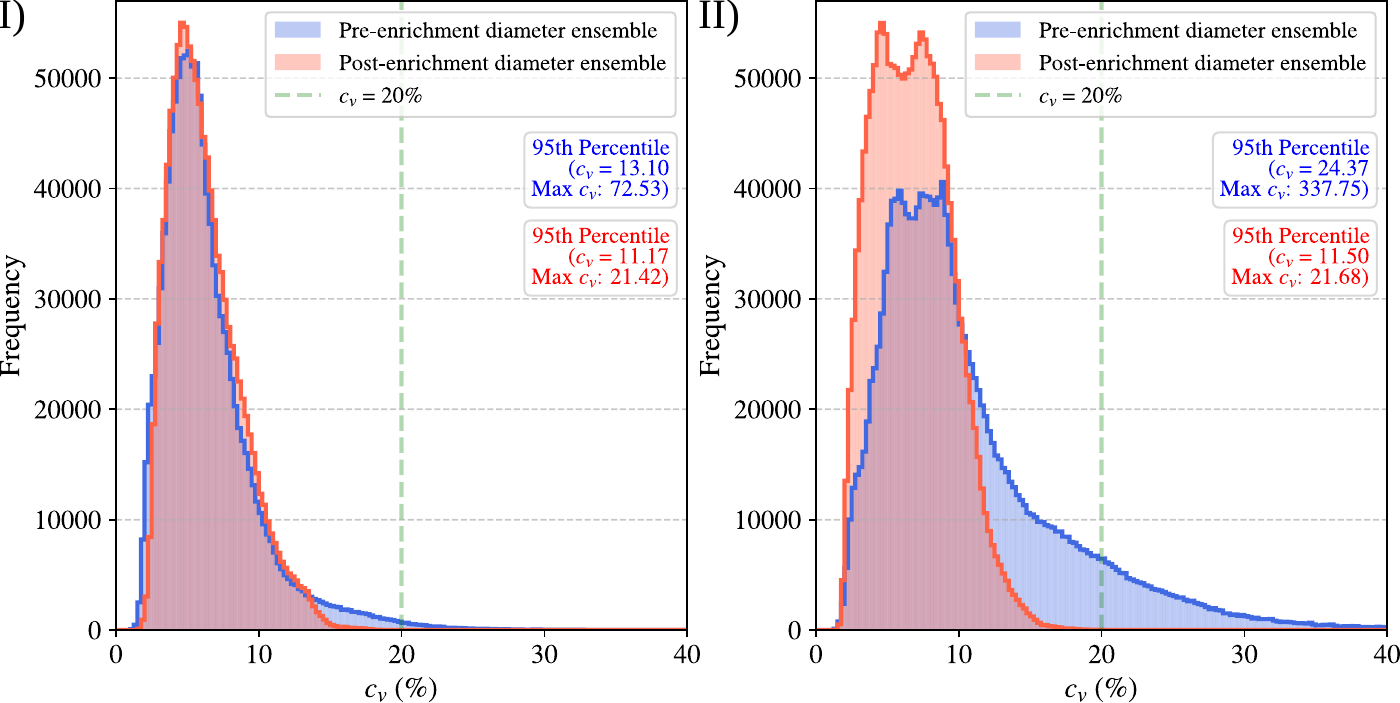}
\caption{Distribution of the coefficient of variation $c_v$ of the diameter predictions before ($E_{d}^{1}$, blue) and after ($E_{d}^{2}$, red) enrichment. I) Evaluation on \textit{SH4}, generated from \textit{TD1} and filtered with $E_{r}^{1}$. II) Evaluation on \textit{SH5}, generated from \textit{T2} and filtered with $E_{r}^{2}$. The dashed line marks the threshold of 20\%. The boxes give the 95th percentile and the maximum of $c_v$ for each ensemble.
}
\label{fig:newHistDiam}
\end{figure}

Figure \ref{fig:newHistDiam} compares the distributions of $c_v$ of $E_{d}^{1}$ and $E_{d}^{2}$ on \textit{SH4} (panel I) and \textit{SH5} (panel II). The dashed line marks the threshold of 20\% used in Section \ref{diam_pred_unc} to identify predictions with excessive uncertainty. On panel I, the two distributions are almost identical for $c_v$ below 10\%, with the mode close to 5\%, differing only in the tail to the right. The distribution of $E_{d}^{1}$ extends beyond the threshold up to a maximum of 72.5\%, and its 95th percentile is 13.1\%. Most samples therefore have a low $c_v$, but a small fraction exceeds the threshold. The distribution of $E_{d}^{2}$ ends at a maximum of around 21.4\%, and its 95th percentile is near to 11.2\%. The number of samples above the threshold decreased from 9526 to 15. Since both ensembles predict on the same samples, this reduction is due only to the retraining of the diameter networks.

The tail of $E_{d}^{1}$ on \textit{SH4} reaches high values but decreases much more rapidly than the distribution in \textit{SH5} (Panel II) because \textit{SH4} was generated from \textit{TD1}, the training data of $E_{d}^{1}$, and arguably does not contain many samples in regions where $E_{d}^{1}$ is uncertain.

The samples of \textit{SH4} with $c_v > 20\%$ for $E_{d}^{1}$ cover a similar region to that found on \textit{SH2} (Fig. \ref{fig:HU_CV_0.2}), since both datasets were generated with the same procedure from \textit{TD1}. This region extends over the dripping domain and is denser near critical values of \Cout{} and \Win{}, where uncertainty is higher. After enrichment, the density of these samples decreases and the large clusters disappear (Fig. \ref{fig:CoutWin_pre+post} in the Supplementary Material). The new VOF simulations therefore provided training data around and in the sub-regions where $E_{d}^{1}$ had higher uncertainty (see yellow regions in Fig. \ref{fig:HU_CV_0.2}).

On panel II, the tail of $E_{d}^{1}$ is substantially longer than on \textit{SH4}. The 95th percentile reaches 24.4\%, and the maximum reaches 337.8\%. \textit{SH5} was generated from \textit{T2}, which includes the new cases located in the regions where $E_{d}^{1}$ had greater uncertainty, and its candidate samples were pre-screened with $E_{r}^{2}$. As a result, \textit{SH5} contains a larger proportion of samples outside the regions covered by \textit{TD1}, where $E_{d}^{1}$ extrapolates. The distribution of $E_{d}^{2}$ shows no such tail. Its 95th percentile is 11.5\%, and its maximum is around 21.7\%. The number of samples above the threshold of 20\% decreased from 166658 to 68, confirming that even the most challenging population is effectively stabilized. Since both ensembles predict on the same samples, the samples in the tail of $E_{d}^{1}$ move approximately under 15\% for $E_{d}^{2}$ and produce a high double peak of the red histogram in panel II.

The spatial distribution of \textit{SH5} samples with $c_v > 20\%$ for $E_{d}^{1}$ covers regions of the dripping domain comparable to those identified for \textit{SH4}. A higher sample density is concentrated near the critical \Cout{} and \Win{} values where initial uncertainty peaks. Following enrichment, as observed for \textit{SH4}, the density of these high-uncertainty samples decreases markedly and the prominent clusters disappear (Fig. \ref{fig:CoutWin_pre+post_2} in the Supplementary Material).

The 95th percentile and the maximum of $E_{d}^{2}$ are similar on both datasets, \textit{SH4} and \textit{SH5}, although \textit{SH5} is considerably challenging for $E_{d}^{1}$. As it is expected, the uncertainty of the retrained diameter ensemble is therefore low both on the original samples population and on the samples selected by the updated pipeline. The comparison on \textit{SH5} in a way favors $E_{d}^{2}$, because \textit{SH5} samples were interpolated from the training data of $E_{d}^{2}$. The result on \textit{SH4}, where the candidates come from data used to train both ensembles, shows that the reduction does not depend on that advantage. In both datasets, the mode of the distribution changes little, and the reduction of uncertainty is concentrated in the tail.

Three results summarize the effect of the enrichment. First, the disagreement of the regime ensemble is lower and less concentrated in the sub-ranges associated with the dripping regime. Second, the diameter ensemble no longer produces predictions with very high $c_v$, either on \textit{SH4} or on \textit{SH5}, and its maximum $c_v$ is notably close to the threshold of 20\%. Third, on both \textit{SH4} and \textit{SH5}, the candidates with high $c_v$ disappeared from the regions identified before enrichment. The reduction of uncertainty is thus largest in the regions selected for the new VOF simulations. The accuracy of the retrained ensembles is evaluated on simulated data in Section \ref{physical_impact}. 
\subsubsection{Physical predictions impact}
\label{physical_impact}

This section evaluates whether the lower uncertainty of the retrained ensembles corresponds to a lower prediction error. The predictions of the baseline ($E^{1}$) and retrained ($E^{2}$) ensembles are compared with the results of VOF simulations on two additional datasets that were not used for training.

The regime ensemble requires this evaluation because the disagreement between predictions of its members does not measure the classification error. In addition, the enrichment process described in Section \ref{section:manual_enrich} revealed misclassifications of $E_{r}^{1}$. Of the 200 configurations predicted as dripping by $E_{r}^{1}$, about 50 produced widening jets in the simulations. Of the 250 configurations predicted as narrowing jets, only about 130 were confirmed. These configurations were obtained from regions of high disagreement or from areas adjacent to them, so these error rates do not represent the overall accuracy of $E_{r}^{1}$. However, they provide more evidence that the disagreement can be associated with classification errors. They also show that the accuracy of both ensembles must be validated against independent simulations. The same applies to the diameter ensemble, where the $c_v$ measures the spread of the predictions of the members, but not their error with respect to the simulated droplet diameter. A low $c_v$ indicates that the members agree, although they may be agreeing on an incorrect value. In the inverse design workflow, operating parameters are chosen to obtain a target droplet diameter, so the relevant quantity is the error of the predicted diameter. This error can only be measured by comparison with simulations.

The comparison uses two test datasets (Section \ref{vof_datasets} and Fig. \ref{fig:dataset_diag}). \textit{GR} contains configurations of all regimes, labeled with the regime observed in their VOF simulation. \textit{GD} contains dripping configurations, labeled with their simulated droplet diameter. Particularly, \textit{GD} contains only those configurations for which at least 17 of the 20 members of $E_{r}^{2}$ predicted dripping ($c(\mathbf{x}) \geq 0.85$). 
Neither dataset shares configurations with \textit{T2}, and both were used only for evaluation.

The configurations of \textit{GR} and \textit{GD} were sampled mainly in the regions of high disagreement and high $c_v$ identified in Sections \ref{regime_pred_unc} and \ref{diam_pred_unc}. The new cases introduced during enrichment were located in those regions, and consequently, the retrained ensembles had training data close to the test configurations (\textit{GR} and \textit{GD}), whereas the baseline ensembles did not. This design aims to principally measure the improvement in the regions targeted by the enrichment. 
Both versions of each ensemble predict on the same configurations, so the baseline and retrained ensembles were compared with paired statistical tests (McNemar's test  \citep{mcnemar1947} for regime classification and a paired t-test for diameter prediction). In both tests, the significance level was 0.05 and the alternative hypothesis was two-sided.

\begin{figure}[ht]
\centering
\includegraphics[trim={0cm 0cm 0cm 0.53cm}, clip, scale=0.5]{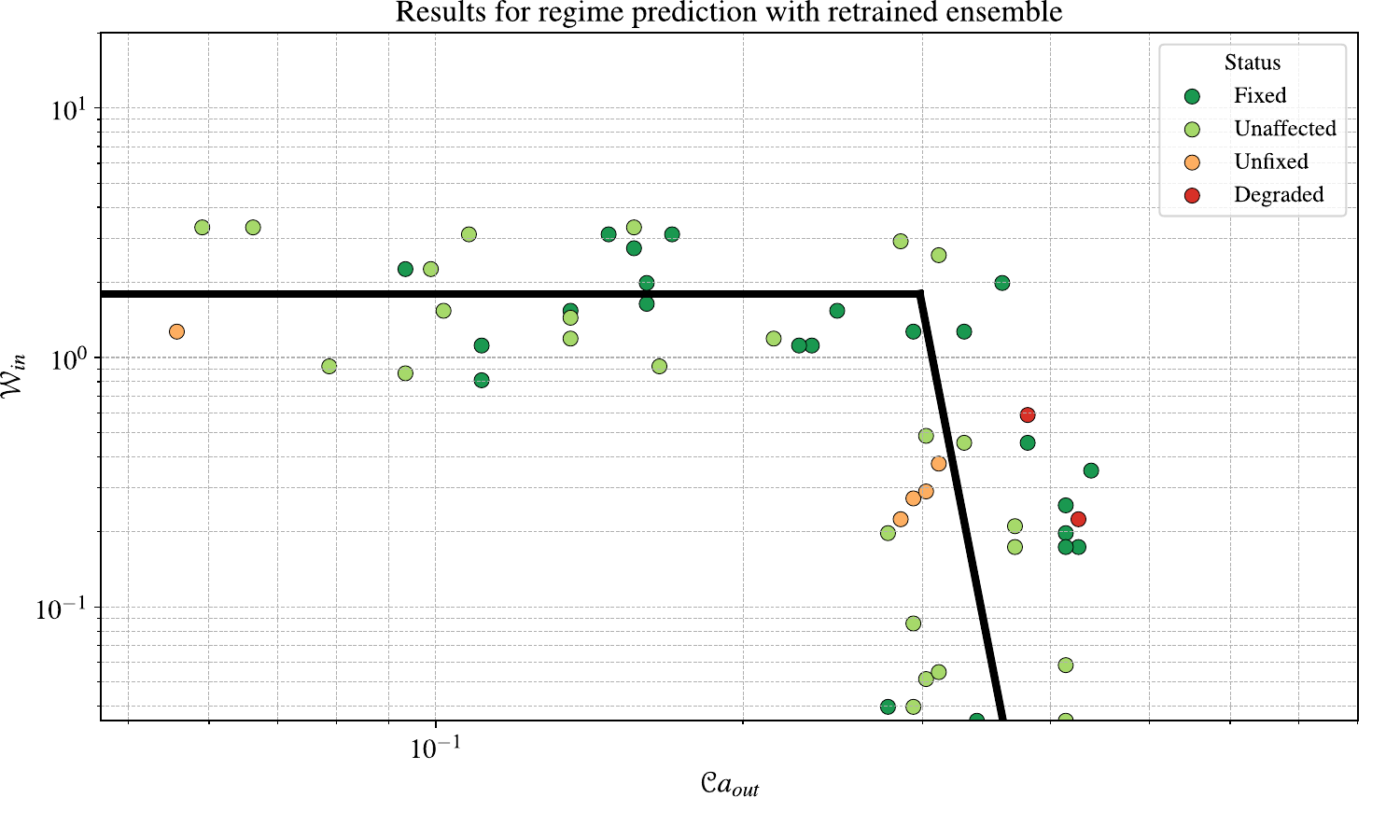}
\caption{Classification of the configurations of \textit{GR} on the (\Cout{}, \Win{}) plane by the baseline ($E_{r}^{1}$) and retrained ($E_{r}^{2}$) regime ensembles. Fixed configurations were misclassified by $E_{r}^{1}$ and correctly classified by $E_{r}^{2}$. Unaffected configurations were correctly classified by both ensembles. Unfixed configurations were misclassified by both ensembles. Degraded configurations were correctly classified by $E_{r}^{1}$ and misclassified by $E_{r}^{2}$. The black line marks the regime-transition boundaries between dripping regime and narrowing-jet or widening-jet regimes \citep{Utada2007}.}
\label{fig:regPredCompare}
\end{figure}

Figure \ref{fig:regPredCompare} shows the 55 configurations of \textit{GR} on the (\Cout{}, \Win{}) plane, colored according to the classification of both ensembles. The parameters of each configuration are given in Table \ref{tab:GR_Dataset} of the Supplementary Material. $E_{r}^{1}$ misclassified 28 configurations, and $E_{r}^{2}$ misclassified 7. The accuracy therefore increased from 49.1\% to 87.3\%. 
Of the 28 errors of $E_{r}^{1}$, 23 were corrected by $E_{r}^{2}$ (Fixed) and 5 remained (Unfixed). In addition, $E_{r}^{2}$ misclassified 2 configurations that $E_{r}^{1}$ classified correctly (Degraded). The exact McNemar's test evaluates those 23 corrected and 2 degraded configurations, considering that each prediction yields a binary outcome of correct or incorrect. This net improvement in classification accuracy is statistically significant ($p \approx 2 \times 10^{-5}$), confirming that the enrichment genuinely reduced misclassifications rather than reflecting random variation.

The corrected configurations lie mainly in two zones. The first is near the horizontal boundary at \Win{}~$\approx 2$, where the disagreement curve of $E_{r}^{1}$ had its maximum (Fig. \ref{fig:newKDEReg}). The second is at \Cout{}~$\approx 0.4$ and \Win{}~$< 0.5$, to the right of the oblique boundary. Six of the seven errors of $E_{r}^{2}$ lie close to the oblique boundary, at \Cout{} between 0.27 and 0.43. This mostly agrees with the location of the remaining disagreement of $E_{r}^{2}$ along \Cout{} (Fig. \ref{fig:newKDEReg}). Near the regime boundaries, the regime observed in a simulation can be ambiguous, even in high-fidelity simulations \citep{Harispe2026}. Part of these remaining errors may therefore reflect this ambiguity rather than a failure of the ensemble.

The error of each diameter drop prediction was computed as the signed relative error
\begin{equation}
\varepsilon_{d} = \frac{d_{\mathrm{sim}} - d_{\mathrm{pred}}}{d_{\mathrm{sim}}} \times 100\%,
\end{equation}
where $d_{\mathrm{sim}}$ is the simulated diameter and $d_{\mathrm{pred}}$ the diameter predicted by the ensemble. The change of the error magnitude after enrichment is $\Delta = |\varepsilon_{d}^{2}| - |\varepsilon_{d}^{1}|$, where the superscripts indicate the ensemble version. A negative value of $\Delta$ indicates that $E_{d}^{2}$ is more accurate than $E_{d}^{1}$ for that configuration. 

\begin{figure}[ht!]
\centering
\includegraphics[scale=0.6]
{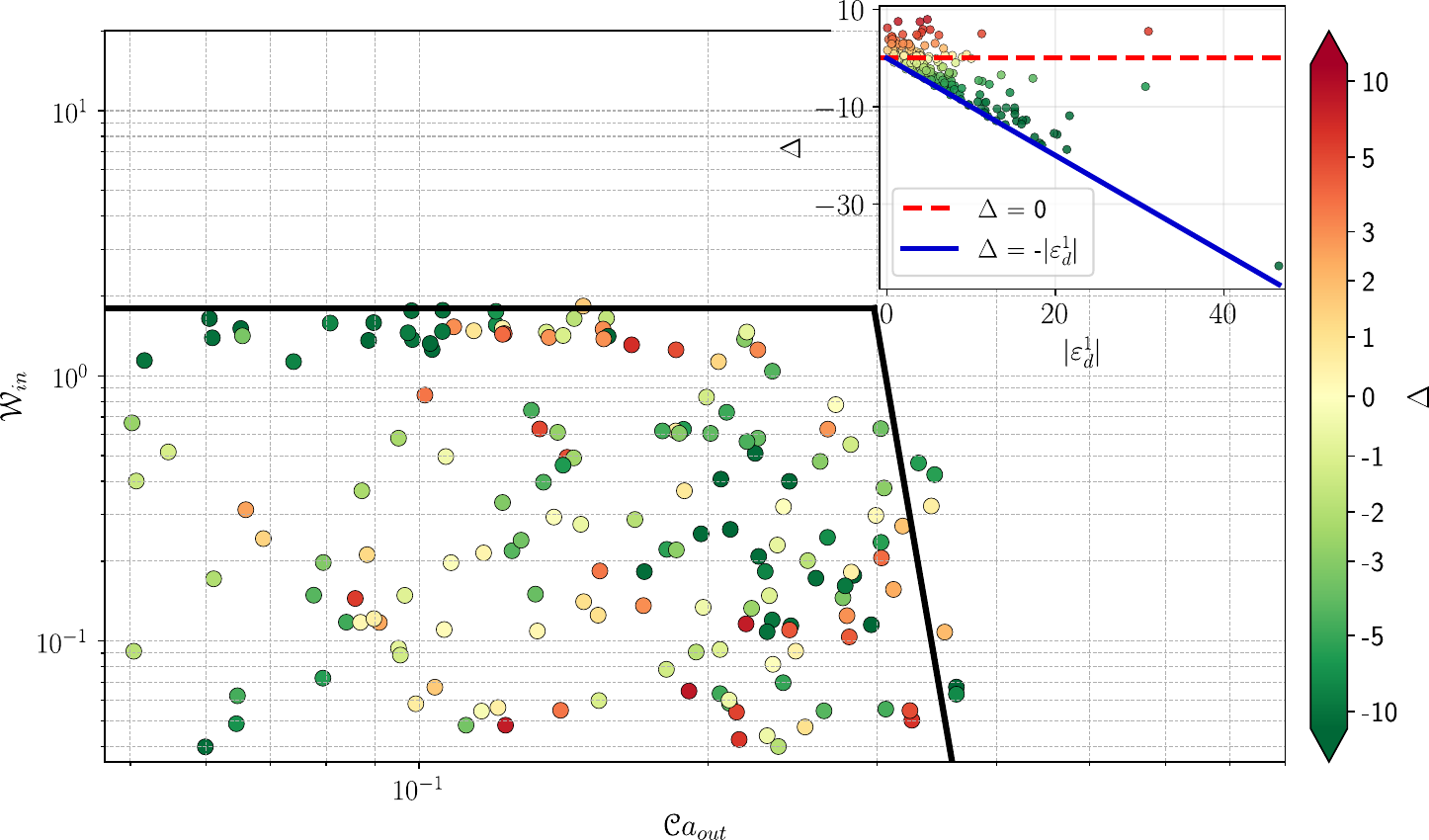}
\caption{Change in relative diameter error $\Delta$ for \textit{GD} configurations on the (\Cout{}, \Win{}) plane. Colors denote error reduction (green-hues) or increase (red-hues), saturating at $\pm10$ percentage points. The black line marks the regime-transition boundary. Inset: $\Delta$ versus the initial error magnitude $|\varepsilon_{d}^{1}|$. The dashed and solid lines mark the no-change threshold ($\Delta = 0$) and the theoretical maximum error reduction ($\Delta = -|\varepsilon_{d}^{1}|$), respectively.
}
\label{fig:CW_Diam_err_comparison}
\end{figure}

Figure \ref{fig:CW_Diam_err_comparison} shows $\Delta$ for the 171 configurations of \textit{GD} on the (\Cout{}, \Win{}) plane. As mentioned earlier, these configurations were sampled preferentially in the high-$c_v$ regions identified in Section \ref{diam_pred_unc}, and their exact parameters are listed in Table \ref{tab:GD_Dataset} of the Supplementary Material. A decrease in error (yellow-to-green hues) is observed in 115 cases, while an increase (yellow-to-red hues) occurs in 56. The largest reductions appear on two main locations. The first, near the upper boundary of the expected dripping regime area, at \Win{} between 1.1 and 1.8 and for \Cout{} below 0.13. The second, close to the right boundary, at \Cout{} between 0.19 and 0.29 and for \Win{} between 0.1 and 0.55. The increases in error are distributed and not concentrated in a particular zone.

The inset in Figure \ref{fig:CW_Diam_err_comparison} plots $\Delta$ against the baseline error magnitude $|\varepsilon_{d}^{1}|$, providing a direct view of how prediction accuracy evolves. The dashed red line marks the no-change threshold ($\Delta = 0$), while the solid black line indicates the theoretical maximum error reduction ($\Delta = -|\varepsilon_{d}^{1}|$). The distribution of points confirms that the enrichment strategy primarily reduced errors for configurations that initially exhibited moderate and large discrepancies, with points falling between these boundaries. Minor error increases were mostly concentrated where the baseline ensemble had small initial error. This trend, dominated by points falling below the no-change threshold, indicates that targeted simulations effectively reduced epistemic uncertainty in critical parameter sub-ranges, thereby validating the uncertainty-guided aspect of the process. Furthermore, the plot shows cases with low initial error (near the origin) where exists minimal degradation, clustering near or slightly above the zero line. This suggests that the enrichment process focused on correcting the challenging cases without significantly compromising the ensemble performance in well-sampled regions.

Table \ref{tab:cases_in_range} gives the number of configurations in each range of $|\varepsilon_{d}|$. The number of configurations with an error above 10\% decreased from 38 for $E_{d}^{1}$ to 11 for $E_{d}^{2}$. The number of configurations with an error below 3\% increased from 47 to 91. The number of configurations between 5\% and 10\% decreased from 47 to 29, and the number between 3\% and 5\% remained almost unchanged (39 and 40). As a result, the fraction of configurations with an error below 5\% increased from 50.3\% (86 of 171) to 76.6\% (131 of 171).

\begin{table}[ht]
\centering
\begin{tabular}{lcc}
\toprule
$|\varepsilon_{d}|$ (\%) & $E_{d}^{1}$ & $E_{d}^{2}$ \\
\midrule
$[0, 3]$  & 47 & 91 \\
$(3, 5]$  & 39 & 40 \\
$(5, 10]$ & 47 & 29 \\
$(10, \infty)$ & 38 & 11 \\
\midrule
Total & 171 & 171 \\
\bottomrule
\end{tabular}
\caption{Number of configurations of \textit{GD} in each range of the magnitude of the relative diameter error, $|\varepsilon_{d}|$, for the baseline ($E_{d}^{1}$) and retrained ($E_{d}^{2}$) diameter ensembles. Square brackets denote inclusive boundaries, while parentheses denote exclusive boundaries.}
\label{tab:cases_in_range}
\end{table}

The magnitudes $|\varepsilon_{d}^{1}|$ and $|\varepsilon_{d}^{2}|$ were compared with the Wilcoxon signed-rank test \citep{wilcoxon1945}, because the errors are paired and their distribution is asymmetric. 
The median error decreased from 4.98\% for $E_{d}^{1}$ to 2.71\% for $E_{d}^{2}$, and this difference is significant ($p \approx 3.05\times 10^{-8}$) \citep{kerby2014}. 
Despite an increase in error in 56 configurations, $|\varepsilon_{d}^{2}|$ remains below 5\% in 28 of them. These increases occur predominantly in configurations where $E_{d}^{1}$ already had a low error, as illustrated in the inset of Figure \ref{fig:CW_Diam_err_comparison}. Furthermore, these increases fall mostly within the 3\%--5\% size dispersion typically accepted for monodisperse droplets in coaxial and standard microfluidic devices across practical applications \citep{Xu2021, Chung2019, Wang2017}.

Ultimately, these results confirm that within the regions targeted by the enrichment, the reduction of uncertainty observed in Sections \ref{sec:uncertainty on regime ensembles} and \ref{sec:uncertainty on diameter ensembles} is accompanied by a significant improvement in prediction accuracy, both for the regime and for the droplet diameter. 
These are the specific regions where the baseline ensembles 
were least reliable, so the retrained ensembles provide a more robust and trustworthy basis for the inverse design procedure whose results are {described in the following section}.

\subsection{Application and validation of the inverse design algorithm}
\label{subsec:inverse_design_validation}
The inverse design method was applied to a set of case studies, each corresponding to a design request with a given target diameter. Ten target diameters were selected, logarithmically spaced between 21.75 and 248.8~$\mu$m. This interval covers the range of droplet diameters obtained in all the simulations of this work. For each target, three independent runs of PSO were performed with the retrained ensembles $E_{r}^{2}$ and $E_{d}^{2}$. Additionally, in six of the ten targets, the radius of the inner capillary $r_{\mathrm{in}}$ was fixed in the third run to represent a design constraint imposed by the user. The configuration obtained in each run was then simulated with VOF in Basilisk. These simulations were not used at any stage of training or optimization and constitute an independent validation of the inverse design method.

Table \ref{tab:inverse_design} presents the results of this validation. Each row corresponds to one PSO run, and the rows are grouped by target diameter, with three runs per target. The first two columns give the number of the case and the target diameter. The next five columns give the configuration found by PSO, defined by the values of \Cout{}, \Win{}, $\mu_{\mathrm{out}}$, $\mu_{\mathrm{in}}$, and $r_{\mathrm{in}}$. The values in bold were fixed by the user, so those were not optimized. The last three columns compare the diameter predicted by $E_{d}^{2}$ for each configuration with the diameter obtained in its VOF simulation and give the relative error between them. A negative error indicates that the simulated droplet is larger than predicted. 
All the configurations found through the PSO algorithm and listed in Table  \ref{tab:inverse_design} exhibited a dripping regime on the VOF simulations.

\begin{table}[ht!]
\centering
\caption{Configurations obtained by inverse design and their validation with VOF simulations. For each target diameter $d_{\mathrm{drop}}^{\mathrm{target}}$, three independent PSO runs were performed. Values in bold were fixed during the search. $d_{\mathrm{pred}}$ is the diameter predicted by $E_{d}^{2}$, and $d_{\mathrm{sim}}$ the simulated diameter. The error is computed as 
$(d_{\mathrm{sim}} - d_{\mathrm{pred}})/d_{\mathrm{sim}} \times 100\%$. Simulations exhibited dripping regimes in all the configurations.}

\label{tab:inverse_design}
\resizebox{0.8\textwidth}{!}{%
\begin{tabular}{@{}rccccccccr@{}}
\toprule
Case & $d_{\mathrm{drop}}^{\mathrm{target}}$ & \Cout{} & \Win{} & $\mu_{\mathrm{out}}$ & $\mu_{\mathrm{in}}$ & $r_{\mathrm{in}}$ & $d_{\mathrm{pred}}$ & $d_{\mathrm{sim}}$ & Error \\
& ($\mu$m) & & & (mPa$\cdot$s) & (mPa$\cdot$s) & ($\mu$m) & ($\mu$m) & ($\mu$m) & (\%) \\
\midrule
1 & 21.75 & 0.323 & 0.164 & 22.6 & 0.890 & 5.01 & 21.75 & 21.44 & -1.46 \\
2 &  & 0.275 & 0.235 & 15.1 & 1.60 & 5.00 & 21.75 & 20.81 & -4.53 \\
3 &  & 0.289 & 0.0356 & 9.21 & 1.16 & \textbf{8.00} & 21.75 & 22.22 & 2.10 \\
\midrule
4 & 28.52 & 0.337 & 0.110 & 43.6 & 0.893 & 6.61 & 28.52 & 29.22 & 2.38 \\
5 &  & 0.363 & 0.0600 & 56.0 & 0.890 & 6.89 & 28.52 & 29.15 & 2.15 \\
6 &  & 0.282 & 0.0350 & 10.6 & 1.37 & \textbf{10.0} & 28.52 & 29.11 & 2.02 \\
\midrule
7 & 37.39 & 0.331 & 0.0776 & 87.9 & 0.890 & 7.94 & 37.39 & 41.00 & 8.80 \\
8 &  & 0.265 & 0.748 & 18.2 & 0.890 & 7.59 & 37.39 & 37.85 & 1.22 \\
9 &  & 0.301 & 0.0355 & 13.2 & 1.47 & \textbf{13.0} & 37.39 & 36.73 & -1.80 \\
\midrule
10 & 49.01 & 0.267 & 0.113 & 87.1 & 0.964 & 9.47 & 49.01 & 53.08 & 7.66 \\
11 &  & 0.277 & 0.764 & 24.5 & 0.973 & 9.78 & 49.01 & 49.30 & 0.58 \\
12 &  & 0.181 & 0.238 & 95.0 & 1.34 & \textbf{5.00} & 49.01 & 47.96 & -2.18 \\
\midrule
13 & 64.26 & 0.202 & 0.113 & 94.8 & 0.897 & 10.8 & 64.26 & 68.88 & 6.71 \\
14 &  & 0.170 & 0.0696 & 90.4 & 0.958 & 10.6 & 64.27 & 68.09 & 5.62 \\
15 &  & 0.211 & 0.121 & 90.3 & 0.892 & 11.1 & 64.26 & 69.27 & 7.23 \\
\midrule
16 & 84.24 & 0.212 & 0.254 & 8.91 & 1.17 & 24.2 & 84.24 & 83.07 & -1.41 \\
17 &  & 0.226 & 0.101 & 11.7 & 0.894 & 25.0 & 84.04 & 87.68 & 4.15 \\
18 &  & 0.208 & 0.158 & 95.0 & 0.892 & 13.4 & 84.24 & 82.68 & -1.89 \\
\midrule
19 & 110.4 & 0.349 & 0.104 & 95.0 & 1.23 & 24.4 & 110.4 & 113.9 & 3.06 \\
20 &  & 0.162 & 1.23 & 8.90 & 0.890 & 25.0 & 110.4 & 113.4 & 2.61 \\
21 &  & 0.172 & 1.11 & 9.84 & 1.06 & 25.0 & 110.4 & 112.3 & 1.64 \\
\midrule
22 & 144.8 & 0.270 & 0.509 & 95.0 & 1.28 & 25.0 & 144.8 & 152.9 & 5.32 \\
23 &  & 0.272 & 0.634 & 94.7 & 0.901 & 25.0 & 144.8 & 152.6 & 5.15 \\
24 &  & 0.0672 & 0.647 & 92.2 & 0.929 & \textbf{10.0} & 144.8 & 144.8 & -0.01 \\
\midrule
25 & 189.8 & 0.165 & 0.662 & 93.7 & 1.61 & 25.0 & 189.8 & 227.5 & 16.58 \\
26 &  & 0.118 & 0.230 & 94.2 & 1.01 & 24.5 & 189.8 & 194.0 & 2.13 \\
27 &  & 0.0470 & 0.726 & 95.0 & 2.26 & \textbf{10.0} & 189.8 & 201.3 & 5.71 \\
\midrule
28 & 248.8 & 0.0984 & 0.713 & 90.1 & 1.77 & 23.3 & 248.8 & 255.9 & 2.79 \\
29 &  & 0.0774 & 0.185 & 94.0 & 0.890 & 25.0 & 248.8 & 238.2 & -4.48 \\
30 &  & 0.107 & 0.665 & 84.9 & 2.09 & 24.9 & 248.8 & 259.4 & 4.06 \\
\bottomrule
\end{tabular}%
}
\end{table}


The predicted diameter $d_{\mathrm{pred}}$ differs from the target by less than 0.25\% in all cases. PSO therefore found configurations that satisfy the target according to $E_{d}^{2}$, and the error with respect to the simulation measures how well this prediction holds in practice. The median absolute error is 2.7\%
, and the mean absolute error is 3.91\%. 
The absolute error is below 5\% in 21 of the 30 cases and below 10\% in 29 of them. These values are consistent with those obtained on \textit{GD}, where 76.6\% of the predictions of $E_{d}^{2}$ had an error below 5\% (Table \ref{tab:cases_in_range}).

The three runs for each target produced different configurations. For example, the three configurations for 110.4~$\mu$m (cases 19 to 21) have values of \Win{} between 0.10 and 1.23 and errors between 1.64\% and 3.06\%. The inverse problem therefore has several solutions, and the method provides alternative configurations among which the user can choose according to other criteria. The six runs with a fixed value of $r_{\mathrm{in}}$ have a mean absolute error of 2.30\%. Imposing a constraint on one parameter did not reduce the accuracy of the designs.

The errors are mostly positive. In 22 of the 30 cases, the simulated diameter is smaller than the predicted one. An exact two-sided sign test gives $p \approx 0.016$, so $E_{d}^{2}$ tends to underestimate the droplet diameter in the configurations selected by the optimizer. The three cases for 64.26~$\mu$m (cases 13 to 15) show this behavior most clearly, with errors between 5.62\% and 7.23\%.

The largest error, 16.58\%, corresponds to case 25. In this configuration, $\mu_{\mathrm{out}}$ and $r_{\mathrm{in}}$ are close to the upper bounds of their ranges, where training data could be scarce. However, cases 22 and 23 have similar values of $\mu_{\mathrm{out}}$ and $r_{\mathrm{in}}$, and errors close to 5.5\%. Proximity to those bounds is therefore not sufficient to explain this error. Since the penalty $t_4$ acts only on \Cout{} and \Win{}, the optimizer can select values of the other parameters at their bounds and
several configurations of Table \ref{tab:inverse_design} have $\mu_{\mathrm{out}}$ or $r_{\mathrm{in}}$ at or near their upper bounds.
At the same time, in case 25, the $c_v$ value obtained by the diameter ensemble was 3.14\% proving that the ensemble members can agree on an incorrect value, as mentioned in Sections \ref{diam_pred_unc} and \ref{physical_impact}. In this case, the fitness function terms $t_0$, $t_1$ and $t_2$ had a value of approximately zero, while $t_3$ and $t_4$ where around 0.237 and 0.574, respectively. $t_0$, $t_1$ and $t_2$ correctly reflect what was observed from the prediction, as the predicted diameter matches the target diameter, and the simulation, since dripping regime was obtained. The value of $t_3$ indicates some slight predictions spread of the members of $E_{d}^{2}$ ($c_v$ value), while $t_4$ suggests that the boundary proximity may be influencing the selected configuration.


Excluding case 25, all prediction errors relative to the simulated droplet diameters remain below 10\%, with the vast majority at 5\% or less. This represents an acceptable tolerance for the experimental design of droplet generators for different applications, falling within the inherent variability caused by syringe pump velocity fluctuations and microfabrication tolerances \citep{Azzini2025, Mottaghi2020, bi2025}.

\section{Discussion}
\label{sec:discussion}

A central question is whether the disagreement of the ensembles is a reliable indicator of their prediction error. Before enrichment, the regions of highest disagreement were located around and near the regime-transition boundaries reported by \citet{Utada2007}. The disagreement curves of $E_{r}^{1}$ peaked at \Win{}~$\approx 2$ and at \Cout{}~$\approx 0.42$, in the ranges where these transitions generally occurred on the literature experiments. Near these boundaries, the regime changes under small variations of the parameters, and the mapping from the inputs to the outputs is strongly nonlinear. The high disagreement in these regions is therefore consistent with the difficulty of the prediction task, and not only with a lack of capacity of the models. The simulations performed during the enrichment confirmed this association. About a quarter of the configurations predicted as dripping by $E_{r}^{1}$ produced widening jets. The disagreement of the ensembles thus identified, without prior physical information, the regions where the models were least reliable and where new simulations were needed.

As with any ensemble-based estimate, prediction disagreement and error are different quantities. The disagreement indicates where the members are least confident, but it cannot detect a systematic bias shared by all members. The inverse design results show a bias of this kind. In 22 of the 30 cases of Table \ref{tab:inverse_design}, $E_{d}^{2}$ underestimated the simulated diameter (by 5\% or less in most cases). Specifically, for case 25, where the error reached 16.58\%, the value of $c_v$ demonstrated that the ensemble members may agree on an incorrect prediction. The ensembles also do not capture the uncertainty associated with the choice of architecture, since all members share the architecture selected in Section \ref{kFoldOpt}, nor the discretization error of the VOF simulations used as reference. These effects are nevertheless included in the comparison with independent simulations, which measures the total prediction error. In this comparison, the bias remained small, with 29 of the 30 designs falling within 10\% of the simulated diameter. 
Moreover, a bias with a consistent sign could be reduced by calibrating the ensemble with a few validation simulations, and the framework could incorporate ensembles of different architectures without changes to the rest of the method.

The enrichment strategy combined two criteria. New simulations were placed where the ensembles disagreed and where the physics is known to be sensitive, near the regime boundaries. The ensembles indicate where the uncertainty is high, but not why. The regime map provides this physical context and helps to select, among the uncertain regions, those that matter for the dripping regime. After enrichment, the disagreement of $E_{r}^{2}$ decreased and became less concentrated in the sub-ranges associated with the dripping regime. Furthermore, the long tail of the $c_v$ distribution observed in the baseline ensemble is entirely absent in $E_{d}^{2}$ on both \textit{SH4} and \textit{SH5}. The concentration of non-unanimous samples of $E_{r}^{2}$ moved away from the enriched regions towards expected jetting regime regions and away from dripping regime regions. Part of the improvement of $E_{r}^{2}$ may also be due to the configurations of \textit{O1} that were excluded from \textit{TR1} by the balancing and later included in \textit{T2}. 

Although an increase in the overall fraction of non-unanimous samples was observed in the jetting regimes, this does not affect the diameter predictions, which are restricted to the dripping regime. It also indicates where new simulations would be needed to extend the method to the other regimes. The same diagnostics therefore guide both the current enrichment and any future extension of the training data.

The comparison with independent VOF simulations in Section \ref{physical_impact} showed that the reduction of disagreement was accompanied by a reduction of the prediction error. On \textit{GR}, the number of misclassified configurations decreased from 28 to 7 of 55 ($p \approx 2 \times 10^{-5}$). On \textit{GD}, the number of predictions with an error below 3\% increased from 47 to 91 of 171, and the number with an error above 10\% decreased from 38 to 11. These test configurations were sampled in the regions targeted by the enrichment, so the results describe the improvement in those regions and not over the whole parameter space. These are, however, the regions where the baseline ensembles were least reliable and where the design becomes more challenging. The results show that the uncertainty diagnostics directed the simulations to regions where they substantially reduced the prediction error. 

The inverse design method successfully identified configurations with the target diameter across the entire range of values studied, achieving a median error of 2.7\% relative to the VOF simulations. Each design required only the evaluation of the ensembles and a single simulation for validation. The three runs for each target produced different configurations, which gives the user alternative designs, and the imposed constraints on $r_{\mathrm{in}}$ did not reduce the accuracy. Two key aspects of the method offer room for improvement. The fitness function currently penalizes only the disagreement between ensemble members through $t_3$ and therefore cannot detect the systematic underestimation of the diameter. Furthermore, the penalty $t_4$ acts solely on \Cout{} and \Win{}, which allows some solutions to settle at the bounds of their ranges where training data is often scarce. Both of these issues can be effectively addressed within the proposed framework by extending $t_4$ to all free parameters and by feeding the simulations of the configurations found by the optimizer back into the training data. This second extension effectively closes the active learning loop, as the configurations proposed by the optimizer automatically become new training data that enables the ensembles to improve precisely in the regions they are utilized for design.

\section{Conclusions}
\label{sec:conclusions}

This work couples adaptive surrogate enhancement with simulation-driven inverse design to optimize microfluidic droplet generators. By leveraging neural network ensembles for instantaneous evaluations, the proposed methodology efficiently navigates a multi-dimensional design space while replacing computationally prohibitive VOF simulations that can take several hours with evaluations that require mere seconds. Uncertainty quantification via ensemble disagreement reliably isolates epistemic uncertainty and reveals that predictive confidence drops in certain regions, such as near regime transitions where flow physics is highly sensitive to parameter perturbations.

A uncertainty- and physics-informed enrichment strategy specifically targets these critical zones and significantly corrects baseline models inaccuracies. Validation on independent test datasets demonstrates the pipeline effectiveness. The enriched regime ensemble increased classification accuracy from 49.1\% to 87.3\%. For droplet diameter predictions, the fraction of cases with an error below 3\% nearly doubled and the number of high-error predictions above 10\% dropped substantially.

The integration of these refined surrogates into a Particle Swarm Optimization loop is guided by a multi-criteria fitness function that balances target diameter accuracy with predictive confidence and physical consistency. This approach prevents the optimizer from selecting impractical configurations or landing in high-uncertainty regions and ensures that all optimized designs strictly adhere to the dripping regime. The inverse design workflow successfully identified configurations with a median absolute error of less than 3\% relative to the VOF ground truth while maintaining high accuracy even when strict user-defined geometric constraints were imposed.

Looking forward, two natural extensions are possible. First, the enrichment loop could be iterated and the residual disagreement parameter-ranges identified in the post-enrichment diagnostics can pinpoint where a second round of targeted simulations would yield the greatest additional information gain. Second, if simulation costs were reduced through solver optimizations, the enrichment criterion could shift from inter-model disagreement to direct residual between VOF outputs and ensemble mean predictions to close the loop between uncertainty estimation and physical error measurement.

Ultimately, the practical value of this work lies in the predictive fidelity achieved with a modest dataset of fewer than 800 VOF simulations. This efficiency is highly significant for microfluidic modeling where high-fidelity data acquisition is often the limiting factor. Beyond the coaxial geometry presented here, the stategy and framework is fundamentally geometry-agnostic. The core logic of using ensemble disagreement to identify where the model lacks confidence, paired with domain knowledge to identify why, can be directly adapted to other microfluidic architectures such as T-junction or flow-focusing devices by redefining the relevant parameters and operational boundaries. More broadly, this strategy of leveraging physics-informed uncertainty to optimize simulation budgets is applicable to any complex multiphase CFD problem characterized by high-dimensional design spaces and sensitive physical transitions.

\section*{Acknowledgements}
This work was supported by the following institutions: Universidad Tecnológica Nacional, Argentina [grants PID-UTN-10015 and PID-UTN-2026-
I-C-TC842]; and Consejo Nacional de Investigaciones Científicas y Técnicas (CONICET) [grant PIP 11220200100819CO], Argentina. 
The present work utilised computational resources from the Piray\'{u} cluster, acquired with funds from the Santa Fe Agency for Science, Technology and Innovation (ASACTEI), Government of the Province of Santa Fe, through Project AC-00010-18, Resolution No. 117/14. This equipment is part of the National High-Performance Computing System (SNCAD) of the Ministry of Science and Technology of the Argentine Republic.

\section*{Conflict of interest}
All authors disclose any financial and personal relationships with other people or organizations that could inappropriately influence or bias this work.
\section*{Data availability}
Codes, datasets and other information related to this work can be found at\\\url{https://github.com/DGHarispe/microdroplet-surrogates-and-inverse-design}
\section*{Supplementary Material}
This work has related supplementary material available at the journal website.
\section*{Declaration of generative AI and AI-assisted technologies}

During the preparation of this work, the authors used Claude (Anthropic) for language and style refinement, as well as assistance with code refactoring and review. After using this tool, the authors carefully reviewed and edited the content as necessary and take full responsibility for the content of the published article.


\bibliographystyle{cas-model2-names}
\bibliography{sample}

\newpage

\setcounter{figure}{0}
\renewcommand{\thefigure}{S\arabic{figure}}
\renewcommand{\theHfigure}{S\arabic{figure}}
\setcounter{table}{0}
\renewcommand{\thetable}{S\arabic{table}}
\renewcommand{\theHtable}{S\arabic{table}}

\section*{Supplementary Material - Microdroplet Generator Surrogate Enhancement and Inverse Design Through an Uncertainty-Guided}

\begin{figure} [h!]
\centering
\includegraphics[scale=0.5]{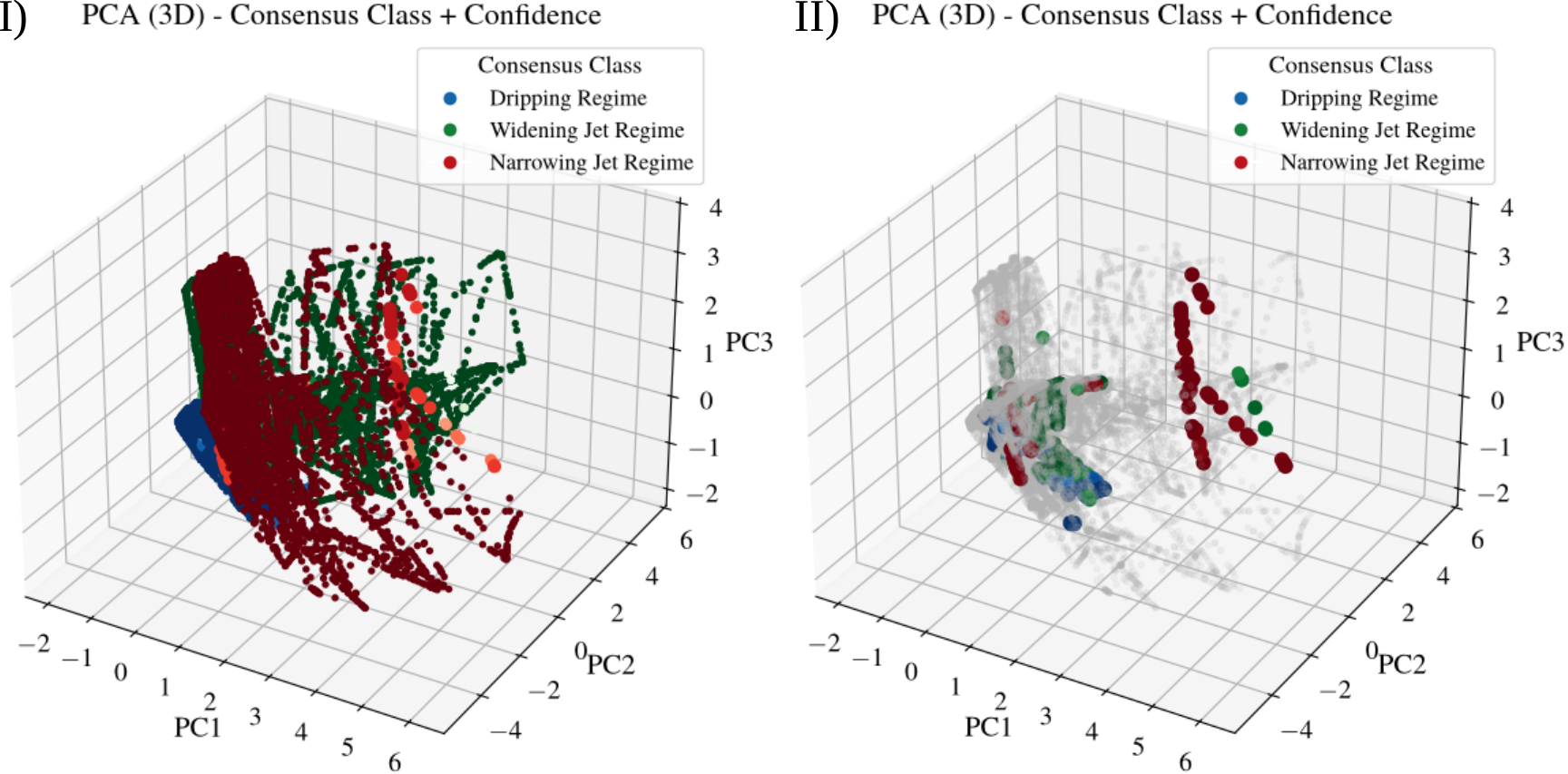}
\caption{Projection of the synthetic samples in \textit{S1} onto the first three principal components. (I) All samples, where the hue indicates the regime assigned by majority vote and the saturation indicates the consensus level, from a saturated shade for unanimous votes to a lighter tone for low consensus. (II) Same projection with the unanimous samples shown in gray, to highlight the spatial distribution of the samples with non-unanimous votes ($c<1$).}
\label{fig:3DPCA_Reg}
\end{figure}

\begin{figure} [h!]
\centering
\includegraphics[scale=0.4]{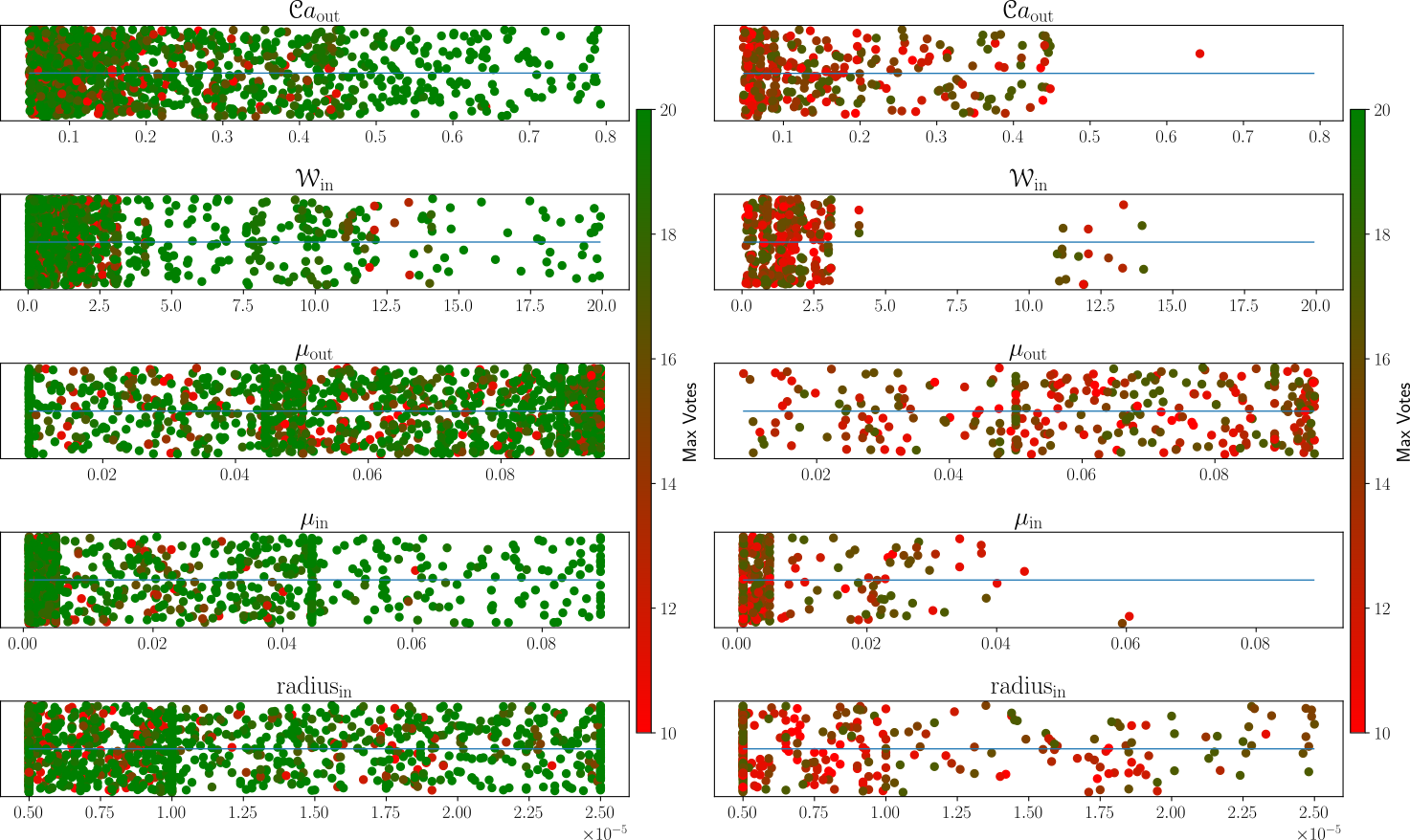}
\caption{Per-variable exploratory analysis of ensemble consensus for regime prediction. Each panel maps the synthetic sample space onto one input parameter (\Cout, \Win, $\mu_{out}$, $\mu_{in}$, $radius_{in}$). The color of the points encodes the maximum inter-member vote count (Max Votes). Left panel shows 5\% of the samples for which absolute consensus was reached, while the right panel shows only samples for which the maximum votes were 17 or less.}
\label{fig:VarRangeRegFull}
\end{figure}

\begin{figure} [h!]
\centering
\includegraphics[scale=0.5]{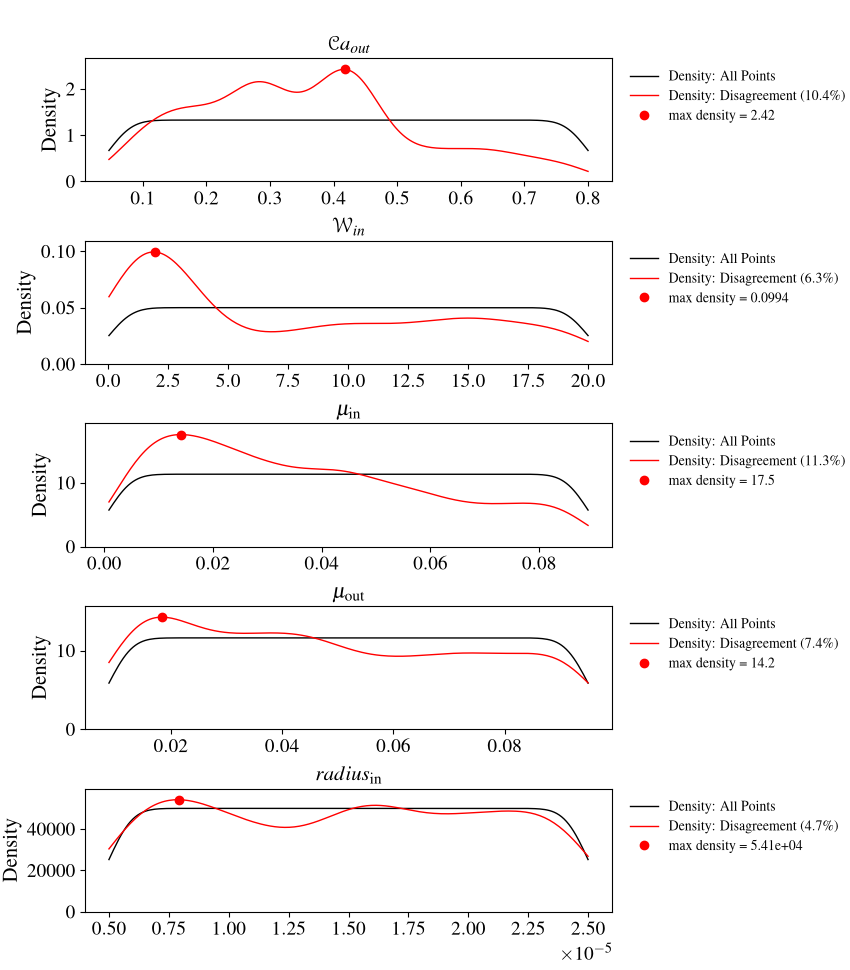}
\caption{Per-variable KDE analysis of regime prediction uncertainty. Black curves represent the marginal probability density of the uniform synthetic set, while red curves denote the conditional density of ensemble disagreement. The percentage of cases where the models fail to reach full agreement is indicated on the legend. Red peaks highlight concentrated epistemic uncertainty, and particularly for \Cout and \Win these are located near the regime transition boundaries documented in the literature.}
\label{fig:KDE_Reg_AllVar_v2}
\end{figure}

\begin{figure} [h!]
\centering
\includegraphics[scale=0.6]{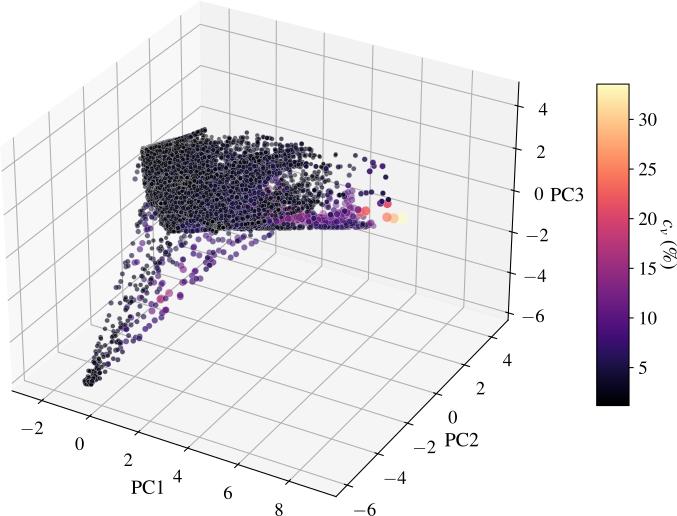}
\caption{3D PCA projection of S2's synthetic dripping samples, brightness-coded by $c_v$ level.}
\label{fig:3DPCA_Dia}
\end{figure}

\begin{figure} [h!]
\centering
\includegraphics[trim={0cm 0cm 0cm 0.5cm}, clip, scale=0.55]{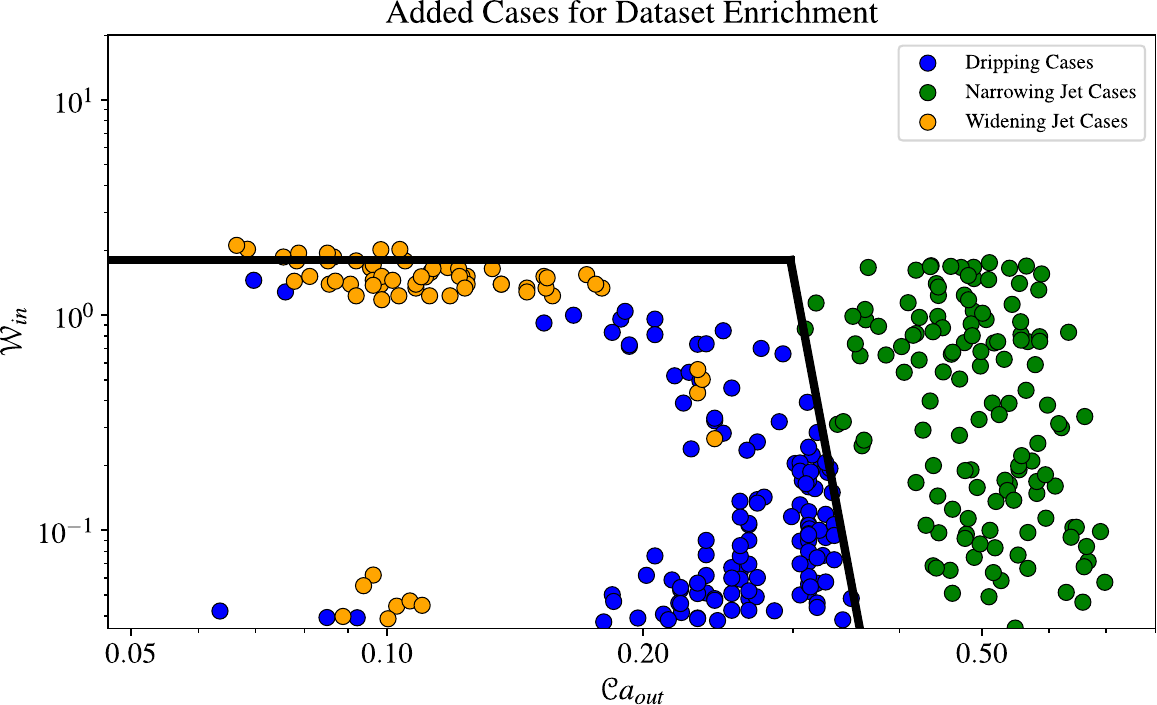}
\caption{Additional simulated cases added to the training dataset, found through the proposed enrichment strategy, shown in the (\Cout, \Win) plane. Points are color-coded by the flow regime exhibited during simulations (dripping, narrowing jet, widening jet), and the solid black lines mark the corresponding regime transition boundaries. The new cases are concentrated near the boundaries to reinforce ensembles confidence where's uncertainty is highest.}
\label{fig:datasetAddedCases}
\end{figure}

\begin{figure} [h!]
\centering
\includegraphics[scale=0.42]{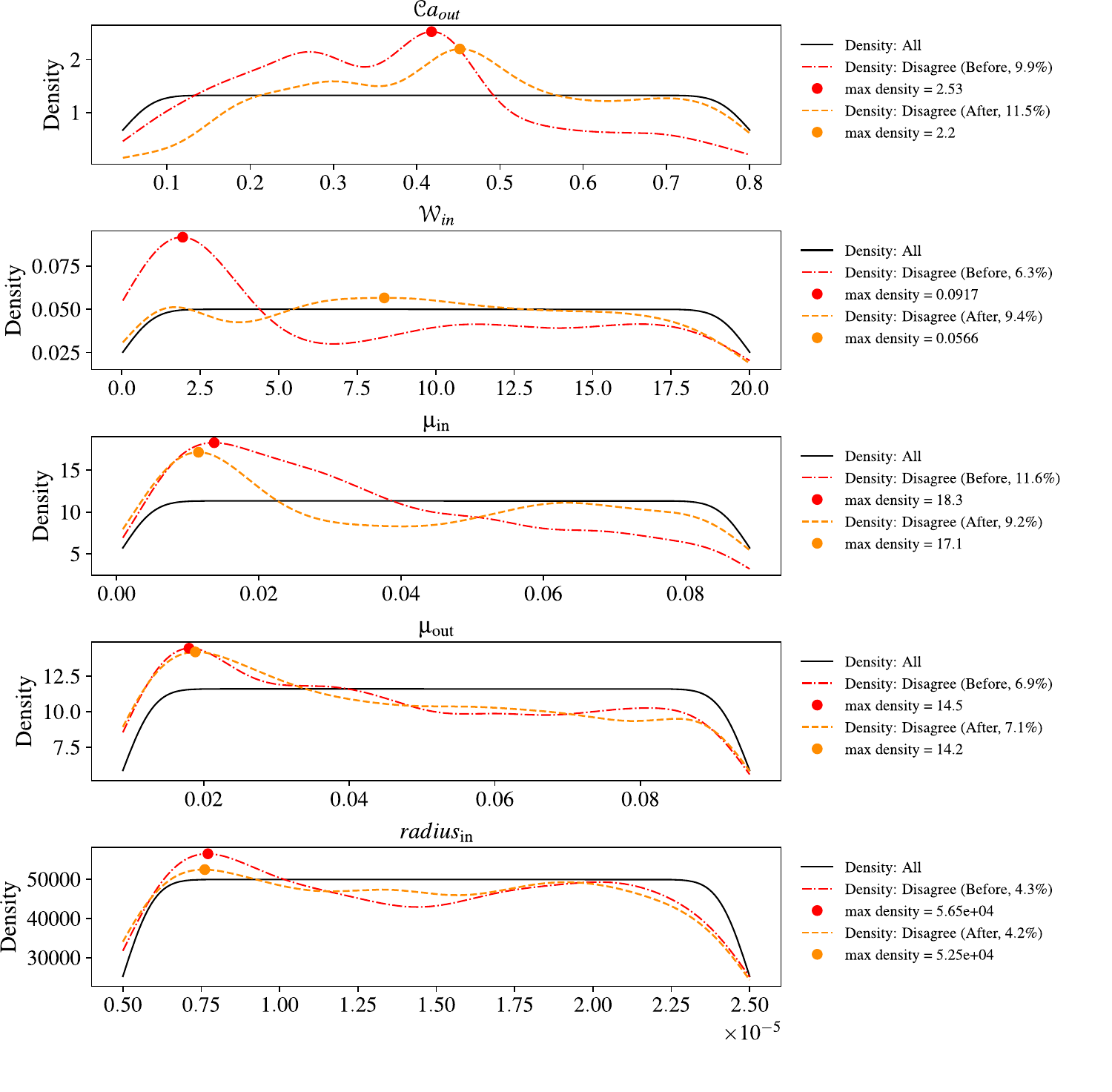}
\caption{Kernel density estimates of the samples in \textit{SH3} along all input parameters, before and after enrichment. Each panel corresponds to the subset of \textit{SH3} in which only the indicated parameter is sampled uniformly. The black curve corresponds to all samples in the subset, and the red and orange curves to the samples with non-unanimous votes ($c<1$) for the baseline ($E_{r}^{1}$) and retrained ($E_{r}^{2}$) regime ensembles, respectively. The percentage of non-unanimous samples is indicated in the legend, and the markers indicate the maximum of each curve.}
\label{fig:KDECurves_pre+post}
\end{figure}

\begin{figure} [h!]
\centering
\includegraphics[scale=0.42]{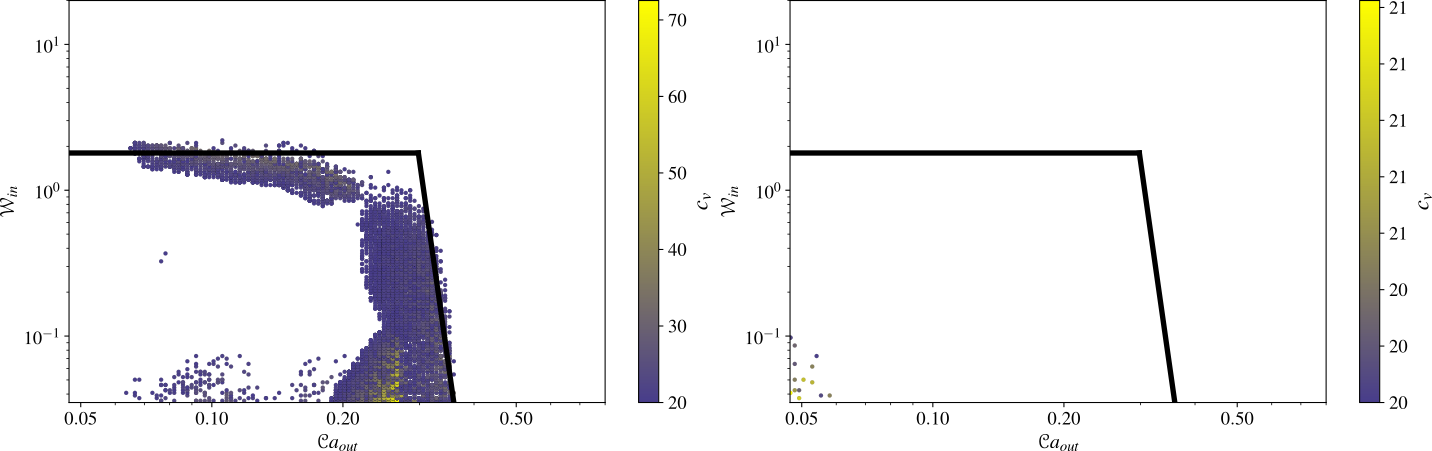}
\caption{Spatial distribution of high predictive uncertainty ($c_v > 20\%$) on the $(C_{out}, W_{in})$ plane, comparing pre-enrichment (left) and post-enrichment (right) states of the diameter-predicting ensemble. The transition demonstrates a dramatic reduction in both the density and spatial extent of high-uncertainty samples near regime boundaries, signifying the successful reduction of epistemic gaps through targeted enrichment.} 
\label{fig:CoutWin_pre+post}
\end{figure}

\begin{figure} [h!]
\centering
\includegraphics[scale=0.42]{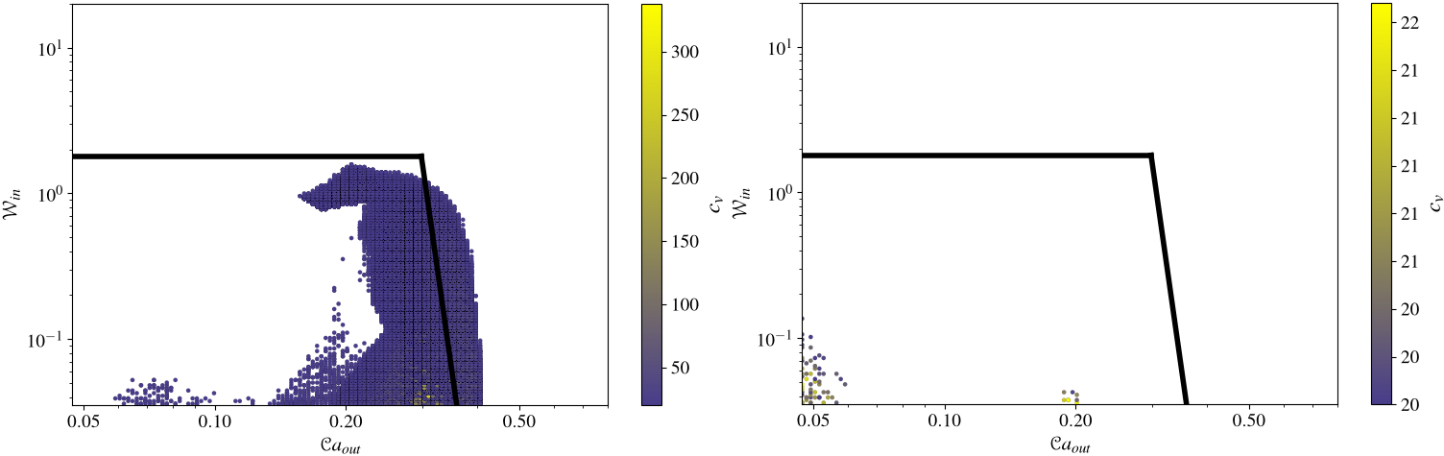}
\caption{Spatial distribution of high predictive uncertainty ($c_v > 20\%$) on the $(C_{out}, W_{in})$ plane, comparing pre-enrichment (left) and post-enrichment (right) states of the diameter-predicting ensemble. The transition demonstrates a dramatic reduction in both the density and spatial extent of high-uncertainty samples near regime boundaries, signifying the successful reduction of epistemic gaps through targeted enrichment.}
\label{fig:CoutWin_pre+post_2}
\end{figure}

\clearpage
{
\large
\begin{longtable}{@{}l c c c c c@{}}
\caption{Parametric specification and case identification for the 55 benchmark configurations comprising the $GR$ dataset.}
\label{tab:GR_Dataset}\\
\toprule
Case & \Cout & \Win & $\mu_{\mathrm{out}}$ & $\mu_{\mathrm{in}}$ & $r_{\mathrm{in}}$ \\
ID &   &   & (Pa$\cdot$s) & (Pa$\cdot$s) & (m) \\
\midrule
1&4,141E-01&1,977E-01&9,500E-04&9,500E-02&2,412E-05\\
2&1,657E-01&9,211E-01&4,450E-02&4,450E-02&1,408E-05\\
3&4,141E-01&5,846E-02&8,995E-04&8,995E-02&5,793E-06\\
4&3,800E-01&5,880E-01&9,273E-04&6,242E-02&1,000E-05\\
5&1,705E-01&3,115E+00&8,900E-04&8,900E-03&1,000E-05\\
6&2,773E-01&1,977E-01&1,324E-02&7,640E-02&2,500E-05\\
7&2,473E-01&1,539E+00&7,924E-02&8,966E-02&5,000E-06\\
8&1,564E-01&2,740E+00&1,778E-03&9,625E-03&9,898E-06\\
9&3,293E-01&4,550E-01&2,910E-03&6,983E-02&1,517E-05\\
10&6,627E-02&3,321E+00&8,900E-04&8,900E-03&6,946E-06\\
11&3,110E-01&3,754E-01&2,616E-03&7,649E-02&1,883E-05\\
12&1,564E-01&3,321E+00&2,315E-03&1,006E-02&1,049E-05\\
13&4,261E-01&1,739E-01&9,500E-04&9,499E-02&2,500E-05\\
14&1,477E-01&3,115E+00&8,900E-04&8,900E-03&1,978E-05\\
15&3,110E-01&2,570E+00&5,199E-02&9,152E-02&1,630E-05\\
16&9,344E-02&8,639E-01&1,891E-02&2,528E-02&6,023E-06\\
17&1,356E-01&1,443E+00&9,121E-04&4,062E-02&5,000E-06\\
18&3,293E-01&1,269E+00&8,900E-04&5,140E-02&1,439E-05\\
19&2,937E-01&2,724E-01&3,885E-03&6,238E-02&2,087E-05\\
20&9,894E-02&2,260E+00&8,900E-04&5,662E-02&2,500E-05\\
21&2,336E-01&1,117E+00&8,900E-04&6,450E-02&1,112E-05\\
22&4,385E-01&3,520E-01&9,468E-04&9,044E-02&2,394E-05\\
23&3,022E-01&5,142E-02&3,559E-03&6,368E-02&8,247E-06\\
24&3,693E-01&2,108E-01&1,714E-03&8,651E-02&2,123E-05\\
25&1,610E-01&1,640E+00&5,935E-03&1,302E-02&5,000E-06\\
26&2,937E-01&3,979E-02&8,900E-04&8,900E-03&7,128E-06\\
27&1,610E-01&1,988E+00&1,500E-03&1,500E-02&5,000E-06\\
28&4,141E-01&1,739E-01&9,500E-04&9,500E-02&1,000E-05\\
29&1,356E-01&1,190E+00&1,485E-02&2,029E-02&6,600E-06\\
30&5,581E-02&1,269E+00&8,900E-04&6,768E-02&5,000E-06\\
31&3,022E-01&4,851E-01&9,468E-04&9,039E-02&1,000E-05\\
32&2,143E-01&1,190E+00&9,188E-04&5,029E-02&7,596E-06\\
33&3,693E-01&1,739E-01&1,765E-03&8,595E-02&2,098E-05\\
34&4,261E-01&2,247E-01&8,900E-04&8,781E-02&2,478E-05\\
35&2,937E-01&8,589E-02&1,093E-03&1,093E-02&1,000E-05\\
36&4,141E-01&2,555E-01&8,900E-04&8,900E-02&1,167E-05\\
37&3,022E-01&2,904E-01&1,265E-03&1,265E-02&6,823E-06\\
38&1,018E-01&1,539E+00&3,204E-03&6,995E-02&1,665E-05\\
39&2,937E-01&1,269E+00&4,125E-03&4,125E-02&8,936E-06\\
40&2,773E-01&3,979E-02&2,791E-03&2,791E-02&1,806E-05\\
41&2,854E-01&2,921E+00&8,900E-04&2,258E-02&2,244E-05\\
42&1,109E-01&1,117E+00&8,104E-03&1,479E-02&2,252E-05\\
43&2,854E-01&2,247E-01&2,866E-03&2,866E-02&1,000E-05\\
44&3,110E-01&5,483E-02&3,211E-03&6,988E-02&2,500E-05\\
45&3,389E-01&3,500E-02&2,696E-03&9,488E-02&2,460E-05\\
46&3,800E-01&4,550E-01&9,185E-04&4,979E-02&5,000E-06\\
47&7,869E-02&9,211E-01&1,847E-02&2,489E-02&5,998E-06\\
48&5,910E-02&3,321E+00&8,900E-04&8,900E-03&6,119E-06\\
49&9,344E-02&2,260E+00&9,345E-04&7,275E-02&2,112E-05\\
50&2,270E-01&1,117E+00&6,804E-02&8,900E-02&8,811E-06\\
51&4,141E-01&3,500E-02&9,480E-04&9,480E-02&5,168E-06\\
52&1,078E-01&3,115E+00&8,900E-04&8,900E-03&1,000E-05\\
53&1,356E-01&1,539E+00&1,044E-02&1,669E-02&1,000E-05\\
54&1,109E-01&8,103E-01&4,480E-03&4,480E-02&2,500E-05\\
55&3,589E-01&1,988E+00&4,442E-03&5,620E-02&2,224E-05\\
\bottomrule
\end{longtable}
}

{
\large
\begin{longtable}{@{}l c c c c c@{}}
\caption{Parametric specification and case identification for the 171 benchmark configurations comprising the $GD$ dataset.}
\label{tab:GD_Dataset}\\
\toprule
Case & \Cout & \Win & $\mu_{\mathrm{out}}$ & $\mu_{\mathrm{in}}$ & $r_{\mathrm{in}}$ \\
ID &   &   & (Pa$\cdot$s) & (Pa$\cdot$s) & (m) \\
\midrule
1&2,059E-01&9,290E-02&9,494E-04&9,494E-02&1,000E-05\\
2&2,526E-01&4,740E-02&8,930E-04&1,806E-02&5,500E-06\\
3&2,059E-01&6,333E-02&1,626E-03&1,626E-02&8,611E-06\\
4&6,449E-02&4,876E-02&9,005E-04&6,103E-02&1,637E-05\\
5&1,887E-01&6,290E-01&8,944E-04&6,807E-02&1,980E-05\\
6&1,968E-01&2,536E-01&8,901E-04&1,745E-02&5,591E-06\\
7&7,937E-02&7,243E-02&9,551E-04&9,551E-03&1,534E-05\\
8&7,398E-02&1,132E+00&9,457E-04&9,457E-02&2,249E-05\\
9&9,517E-02&5,822E-01&8,900E-04&8,900E-03&5,000E-06\\
10&1,119E-01&4,814E-02&9,316E-04&7,796E-02&2,120E-05\\
11&3,029E-01&6,326E-01&8,932E-04&1,329E-02&5,000E-06\\
12&1,062E-01&1,105E-01&8,947E-04&1,689E-02&5,400E-06\\
13&1,451E-01&1,645E+00&8,909E-04&1,108E-02&2,322E-05\\
14&3,632E-01&6,702E-02&8,904E-04&3,958E-02&5,004E-06\\
15&1,014E-01&8,461E-01&9,085E-04&2,168E-02&9,961E-06\\
16&1,222E-01&3,328E-01&1,126E-03&1,131E-02&5,714E-06\\
17&2,013E-01&6,069E-01&9,202E-04&9,202E-03&5,006E-06\\
18&1,404E-01&5,480E-02&1,541E-03&2,967E-02&8,628E-06\\
19&2,643E-01&5,453E-02&9,015E-04&2,549E-02&6,000E-06\\
20&1,336E-01&6,307E-01&8,955E-04&5,955E-02&5,000E-06\\
21&6,465E-02&6,217E-02&1,655E-03&1,655E-02&6,189E-06\\
22&8,397E-02&1,178E-01&9,090E-04&1,049E-02&8,553E-06\\
23&5,985E-02&3,989E-02&8,900E-04&8,900E-03&2,450E-05\\
24&1,426E-01&4,930E-01&9,080E-04&4,837E-02&1,392E-05\\
25&2,052E-01&1,132E+00&8,934E-04&1,372E-02&8,084E-06\\
26&1,679E-01&2,873E-01&9,497E-04&9,493E-02&1,000E-05\\
27&1,231E-01&4,814E-02&1,574E-03&2,537E-02&1,090E-05\\
28&1,537E-01&1,252E-01&8,964E-04&6,404E-02&8,021E-06\\
29&5,041E-02&9,149E-02&8,915E-04&1,753E-02&5,055E-06\\
30&1,167E-01&2,148E-01&8,900E-04&8,900E-03&5,500E-06\\
31&1,717E-01&1,827E-01&8,956E-04&1,706E-02&5,410E-06\\
32&1,250E-01&2,189E-01&8,945E-04&2,900E-02&5,110E-06\\
33&6,880E-02&2,433E-01&9,089E-04&3,703E-02&9,946E-06\\
34&1,890E-01&3,686E-01&9,237E-04&6,879E-02&5,346E-06\\
35&9,517E-02&9,391E-02&9,007E-04&2,769E-02&6,000E-06\\
36&3,632E-01&6,301E-02&8,939E-04&5,608E-02&7,914E-06\\
37&8,715E-02&3,686E-01&1,452E-03&1,452E-02&5,700E-06\\
38&7,945E-02&1,977E-01&8,972E-04&2,135E-02&7,896E-06\\
39&1,812E-01&2,216E-01&9,042E-04&2,922E-02&7,473E-06\\
40&9,088E-02&1,175E-01&4,764E-03&1,206E-02&7,616E-06\\
41&2,335E-01&1,041E+00&1,002E-03&1,002E-02&5,546E-06\\
42&1,208E-01&5,611E-02&8,900E-04&8,900E-03&1,780E-05\\
43&1,993E-01&8,312E-01&1,238E-03&1,258E-02&8,413E-06\\
44&2,106E-01&5,824E-02&1,696E-03&1,696E-02&8,915E-06\\
45&2,255E-01&5,822E-01&8,926E-04&1,314E-02&5,000E-06\\
46&2,719E-01&7,787E-01&8,935E-04&2,720E-02&9,352E-06\\
47&2,961E-01&1,152E-01&8,900E-04&8,916E-03&5,000E-06\\
48&8,578E-02&1,447E-01&2,828E-03&7,107E-02&7,314E-06\\
49&1,474E-01&2,756E-01&8,900E-04&6,126E-02&8,300E-06\\
50&5,019E-02&6,648E-01&9,115E-04&4,043E-02&9,638E-06\\
51&1,449E-01&4,905E-01&1,046E-03&3,090E-02&9,094E-06\\
52&1,328E-01&1,091E-01&1,187E-03&4,071E-02&6,283E-06\\
53&1,852E-01&6,195E-01&9,160E-04&5,454E-02&1,598E-05\\
54&2,338E-01&8,167E-02&1,877E-03&3,229E-02&6,719E-06\\
55&1,484E-01&1,407E-01&1,111E-03&7,311E-02&1,155E-05\\
56&7,764E-02&1,489E-01&8,932E-04&1,393E-02&6,129E-06\\
57&1,543E-01&1,838E-01&9,366E-04&7,573E-02&2,029E-05\\
58&8,826E-02&2,116E-01&9,047E-04&9,047E-03&5,084E-06\\
59&3,034E-01&2,058E-01&8,912E-04&1,566E-02&9,658E-06\\
60&9,560E-02&8,849E-02&2,079E-03&3,803E-02&1,182E-05\\
61&6,106E-02&1,716E-01&2,773E-03&6,762E-02&9,753E-06\\
62&1,277E-01&2,400E-01&8,963E-04&2,400E-02&5,662E-06\\
63&1,911E-01&6,477E-02&1,215E-03&3,865E-02&1,138E-05\\
64&3,053E-01&3,776E-01&9,030E-04&2,779E-02&7,042E-06\\
65&2,092E-01&7,286E-01&1,304E-03&2,376E-02&9,267E-06\\
66&8,686E-02&1,176E-01&1,280E-03&3,720E-02&6,463E-06\\
67&2,063E-01&4,078E-01&1,817E-03&2,210E-02&5,700E-06\\
68&1,038E-01&6,696E-02&9,316E-04&6,528E-02&8,697E-06\\
69&1,348E-01&3,971E-01&1,494E-03&1,549E-02&8,948E-06\\
70&2,363E-01&2,301E-01&1,090E-03&3,120E-02&1,099E-05\\
71&1,867E-01&6,074E-01&1,089E-03&1,284E-02&7,454E-06\\
72&1,382E-01&2,934E-01&8,900E-04&3,272E-02&8,972E-06\\
73&1,413E-01&4,603E-01&9,015E-04&3,342E-02&5,595E-06\\
74&5,171E-02&1,142E+00&9,191E-04&8,998E-02&1,514E-05\\
75&2,441E-01&1,141E-01&8,965E-04&2,159E-02&5,410E-06\\
76&2,295E-01&1,829E-01&9,066E-04&2,728E-02&6,426E-06\\
77&1,079E-01&1,974E-01&4,054E-03&2,967E-02&6,404E-06\\
78&5,476E-02&5,165E-01&8,958E-04&2,916E-02&8,687E-06\\
79&5,071E-02&4,011E-01&9,079E-04&8,802E-02&9,719E-06\\
80&1,540E-01&5,965E-02&2,552E-03&3,402E-02&7,014E-06\\
81&1,793E-01&6,199E-01&8,989E-04&3,153E-02&6,492E-06\\
82&8,971E-02&1,214E-01&2,474E-03&3,058E-02&1,005E-05\\
83&3,420E-01&3,229E-01&9,039E-04&2,890E-02&9,276E-06\\
84&9,920E-02&5,799E-02&2,088E-03&2,519E-02&6,010E-06\\
85&3,531E-01&1,081E-01&9,444E-04&7,782E-02&1,514E-05\\
86&1,322E-01&1,502E-01&1,787E-03&2,755E-02&8,055E-06\\
87&1,811E-01&7,820E-02&9,174E-04&4,979E-02&7,237E-06\\
88&1,395E-01&6,124E-01&9,247E-04&6,130E-02&1,201E-05\\
89&2,766E-01&1,455E-01&1,036E-03&4,303E-02&6,274E-06\\
90&1,978E-01&1,340E-01&9,289E-04&6,958E-02&7,059E-06\\
91&1,854E-01&2,206E-01&8,954E-04&4,824E-02&1,364E-05\\
92&1,066E-01&4,962E-01&8,922E-04&1,224E-02&6,283E-06\\
93&3,030E-01&2,356E-01&9,059E-04&3,236E-02&5,600E-06\\
94&1,309E-01&7,416E-01&9,338E-04&7,314E-02&8,356E-06\\
95&6,596E-02&3,128E-01&8,900E-04&2,107E-02&2,078E-05\\
96&1,945E-01&9,087E-02&8,919E-04&6,498E-02&1,847E-05\\
97&2,666E-01&2,460E-01&1,262E-03&2,401E-02&7,611E-06\\
98&1,162E-01&5,442E-02&9,079E-04&4,380E-02&6,016E-06\\
99&1,713E-01&1,360E-01&8,903E-04&9,266E-03&1,642E-05\\
100&9,662E-02&1,487E-01&9,166E-04&9,166E-03&1,558E-05\\
101&1,568E-01&1,650E+00&8,900E-04&9,823E-03&2,355E-05\\
102&1,357E-01&1,463E+00&8,900E-04&8,970E-03&2,259E-05\\
103&8,856E-02&1,361E+00&9,537E-04&1,323E-02&2,215E-05\\
104&2,184E-01&1,374E+00&8,953E-04&1,873E-02&2,168E-05\\
105&1,032E-01&1,257E+00&9,132E-04&4,414E-02&2,083E-05\\
106&2,255E-01&1,254E+00&9,001E-04&1,182E-02&1,423E-05\\
107&1,666E-01&1,309E+00&8,989E-04&1,192E-02&1,573E-05\\
108&2,196E-01&1,463E+00&8,964E-04&1,129E-02&1,233E-05\\
109&1,853E-01&1,254E+00&1,032E-03&1,032E-02&1,504E-05\\
110&1,027E-01&1,325E+00&8,928E-04&1,692E-02&2,205E-05\\
111&6,517E-02&1,509E+00&8,900E-04&8,910E-03&2,253E-05\\
112&1,058E-01&1,767E+00&8,900E-04&9,404E-03&2,389E-05\\
113&6,087E-02&1,392E+00&8,913E-04&1,075E-02&2,161E-05\\
114&6,543E-02&1,416E+00&8,946E-04&8,946E-03&1,977E-05\\
115&6,044E-02&1,644E+00&9,055E-04&9,243E-03&2,373E-05\\
115&2,241E-01&5,101E-01&1,272E-03&2,115E-02&5,855E-06\\
116&2,839E-01&1,768E-01&8,900E-04&8,900E-03&5,934E-06\\
117&2,793E-01&1,246E-01&1,527E-03&6,202E-02&1,645E-05\\
118&2,156E-01&4,251E-02&4,124E-03&4,176E-02&2,079E-05\\
119&3,264E-01&5,027E-02&9,031E-04&5,881E-02&1,566E-05\\
120&2,433E-01&1,101E-01&1,310E-03&1,637E-02&7,255E-06\\
121&2,143E-01&5,393E-02&9,075E-04&1,071E-02&1,973E-05\\
122&2,334E-01&1,201E-01&1,594E-03&1,932E-02&5,193E-06\\
123&2,197E-01&5,653E-01&8,935E-04&2,981E-02&9,153E-06\\
124&3,123E-01&1,565E-01&9,101E-04&1,952E-02&5,107E-06\\
125&2,619E-01&4,757E-01&9,137E-04&6,885E-02&2,101E-05\\
126&2,432E-01&4,000E-01&8,940E-04&2,212E-02&5,000E-06\\
127&2,318E-01&1,483E-01&1,058E-03&3,448E-02&8,200E-06\\
128&3,066E-01&5,536E-02&8,980E-04&2,451E-02&5,764E-06\\
129&2,822E-01&1,819E-01&9,231E-04&8,343E-02&2,258E-05\\
130&3,447E-01&4,242E-01&9,891E-04&6,602E-02&2,128E-05\\
131&2,111E-01&2,637E-01&1,562E-03&1,565E-02&6,569E-06\\
132&2,818E-01&5,509E-01&8,959E-04&1,759E-02&5,526E-06\\
133&2,306E-01&1,084E-01&8,912E-04&2,099E-02&5,000E-06\\
134&2,369E-01&4,000E-02&8,951E-04&1,628E-02&6,835E-06\\
135&2,808E-01&1,036E-01&8,919E-04&1,954E-02&8,226E-06\\
136&2,667E-01&6,297E-01&8,951E-04&2,638E-02&1,086E-05\\
137&2,781E-01&1,615E-01&8,905E-04&2,078E-02&5,700E-06\\
138&2,259E-01&2,085E-01&8,941E-04&1,510E-02&5,366E-06\\
139&2,192E-01&1,160E-01&1,210E-03&4,114E-02&1,495E-05\\
140&2,306E-01&4,393E-02&8,987E-04&2,340E-02&5,794E-06\\
141&3,249E-01&5,467E-02&9,043E-04&3,504E-02&5,169E-06\\
142&3,317E-01&4,694E-01&8,905E-04&6,567E-02&2,067E-05\\
143&2,398E-01&3,209E-01&9,154E-04&4,535E-02&1,346E-05\\
144&2,104E-01&5,992E-02&1,001E-03&1,422E-02&5,524E-06\\
145&2,396E-01&6,962E-02&8,923E-04&1,990E-02&6,819E-06\\
146&2,994E-01&2,977E-01&9,340E-04&7,306E-02&1,353E-05\\
147&2,469E-01&9,159E-02&8,952E-04&1,878E-02&5,809E-06\\
148&3,192E-01&2,714E-01&1,032E-03&5,548E-02&1,549E-05\\
149&2,542E-01&2,009E-01&1,512E-03&8,331E-02&1,259E-05\\
150&2,221E-01&1,328E-01&8,914E-04&1,091E-02&5,355E-06\\
151&2,592E-01&1,726E-01&1,226E-03&2,957E-02&6,219E-06\\
152&1,203E-01&1,557E+00&8,902E-04&1,028E-02&2,362E-05\\
153&1,366E-01&1,397E+00&8,900E-04&8,900E-03&1,446E-05\\
154&1,555E-01&1,500E+00&8,905E-04&1,051E-02&1,347E-05\\
155&8,077E-02&1,581E+00&8,904E-04&9,488E-03&2,026E-05\\
156&1,389E-01&1,378E+00&8,900E-04&1,370E-02&1,516E-05\\
157&9,821E-02&1,761E+00&8,910E-04&1,045E-02&2,114E-05\\
158&1,087E-01&1,531E+00&8,909E-04&1,102E-02&1,866E-05\\
159&1,140E-01&1,479E+00&8,900E-04&8,900E-03&2,065E-05\\
160&9,839E-02&1,371E+00&8,900E-04&1,519E-02&1,872E-05\\
161&1,202E-01&1,751E+00&8,909E-04&1,023E-02&2,199E-05\\
162&1,574E-01&1,410E+00&8,984E-04&2,100E-02&2,464E-05\\
163&9,736E-02&1,453E+00&9,002E-04&1,330E-02&1,902E-05\\
164&1,221E-01&1,512E+00&8,900E-04&8,962E-03&1,739E-05\\
165&1,057E-01&1,470E+00&9,088E-04&1,059E-02&2,332E-05\\
166&1,227E-01&1,443E+00&9,405E-04&1,491E-02&1,596E-05\\
167&8,965E-02&1,587E+00&8,921E-04&1,184E-02&2,029E-05\\
168&1,413E-01&1,420E+00&8,924E-04&9,221E-03&2,322E-05\\
169&1,222E-01&1,435E+00&1,037E-03&1,352E-02&1,665E-05\\
170&1,483E-01&1,834E+00&8,900E-04&9,089E-03&2,386E-05\\
171&1,557E-01&1,379E+00&8,924E-04&1,519E-02&1,878E-05\\
\bottomrule
\end{longtable}
}

\end{document}